\documentclass[12pt, nofootinbib]{article}
\usepackage[left=0.2in,right=0.2in,top=0.05in,bottom=0.05in]{geometry}

\usepackage{amssymb}
\usepackage{amsmath,bm}
\usepackage{amssymb}
\usepackage{graphicx}
\usepackage{amsfonts}         
\usepackage{fancybox}

\usepackage[usenames,dvipsnames]{xcolor}%http://en.wikibooks.org/wiki/LaTeX/Colors

\definecolor{Mathematica1}{rgb}{0.368417, 0.506779, 0.709798}
\definecolor{Mathematica2}{rgb}{0.880722, 0.611041, 0.142051}
\definecolor{Mathematica3}{rgb}{0.560181, 0.691569, 0.194885}

\usepackage[pagebackref,  % this puts links to the page numbers where refs appear
bookmarks={false}, pdfauthor={
}, 
pdftitle={Yay, physics!}]{hyperref}
\hypersetup{colorlinks=true, 
linkcolor=Mathematica1,
citecolor=Mathematica3,
filecolor=OliveGreen, 
urlcolor=Mathematica2,
filebordercolor={.8 .8 1}, 
urlbordercolor={.8 .8 0}}%http://en.wikibooks.org/wiki/LaTeX/Hyperlinks
\usepackage{soul}%strike through text \st{} .  not in math mode.
\setstcolor{Red}
\definecolor{darkgreen}{rgb}{0,0.4,0}
\definecolor{darkred}{rgb}{0.4,0,0}
\definecolor{darkblue}{rgb}{0,0,0.4}
\definecolor{lightblue}{rgb}{.6,.6,0.9}

\definecolor{uglybrown}{rgb}{0.8,  0.7,  0.5}

\definecolor{palatinatepurple}{rgb}{0.41, 0.16, 0.38}
\definecolor{celebrationcolor}{rgb}{0.75,  0.0,  0.9}

\definecolor{shadecolor}{rgb}{0.90,0.90,0.90}
\definecolor{DVcolor}{rgb}{0.95,  0.5,  0.2}
\definecolor{lightbluemuons}{rgb}{0.0,.65,1.0}

\definecolor{chartreuse}{rgb}{0.70, 1.00, 0.00}

\usepackage{tikz}
\usetikzlibrary{shapes}
\usetikzlibrary{trees}
\usetikzlibrary{matrix,arrows} 				% For commutative diagram
\usetikzlibrary{positioning}				% For "above of=" commands
\usetikzlibrary{calc,through}				% For coordinates
\usetikzlibrary{decorations.pathreplacing}  % For curly braces
\usepackage[tikz]{bclogo} 					% For cute logo boxes
\usepackage{pgffor}							% For repeating patterns

\usetikzlibrary{decorations.markings}
\usetikzlibrary{intersections}				% For interesections

\usetikzlibrary{arrows,decorations.pathmorphing,backgrounds,positioning,fit,petri,automata,shadows,calendar,mindmap, graphs}
\usetikzlibrary{arrows.meta,bending}

\tikzset{
    vector/.style={decorate, decoration={snake}, draw},
    fermion/.style={postaction={decorate},
        decoration={markings,mark=at position .55 with {\arrow{>}}}},
    fermionbar/.style={draw, postaction={decorate},
        decoration={markings,mark=at position .55 with {\arrow{<}}}},
    fermionnoarrow/.style={},
    gluon/.style={decorate,
        decoration={coil,amplitude=4pt, segment length=5pt}},
    scalar/.style={dashed, postaction={decorate},
        decoration={markings,mark=at position .55 with {\arrow{>}}}},
    scalarbar/.style={dashed, postaction={decorate},
        decoration={markings,mark=at position .55 with {\arrow{<}}}},
    scalarnoarrow/.style={dashed,draw},
	vectorscalar/.style={loosely dotted,draw=black, postaction={decorate}},
}

\def\centerarc[#1](#2)(#3:#4:#5)% Syntax: [draw options] (center) (initial angle:final angle:radius)
    { \draw[#1] ($(#2)+({#5*cos(#3)},{#5*sin(#3)})$) arc (#3:#4:#5); }

\usepackage{enumitem}

\usepackage{slashed}

\usepackage[font=small,labelfont=bf]{caption}
\DeclareCaptionFont{tiny}{\tiny}
\usepackage{epstopdf}
\usepackage{wasysym}

\usepackage[yyyymmdd,hhmmss]{datetime}   
\usepackage{framed}  % needed for \begin{shaded}
 \usepackage{mdframed}
 \usepackage{wrapfig}
 \usepackage{yfonts}  
\newmdenv[%
        backgroundcolor=lightgray,
    linecolor=black,
    outerlinewidth=2pt,
]{boxedandshaded}

\numberwithin{equation}{section}

\renewcommand{\theequation}{\arabic{section}.\arabic{equation}}

\newlength{\extraspace}
\newlength{\extraspaces}
\def\be{\begin{equation}}
\def\ee{\end{equation}}

\newcommand{\bea}{\begin{eqnarray}}
\newcommand{\eea}{\end{eqnarray}}

\def\Tr{{{\rm Tr}}}
\def\tr{{\rm tr}}

\def\bra#1{\left\langle#1\right|}
\def\ket#1{\left|#1\right\rangle}

\def\CO{{\cal O}}%AEL

\def\II{\relax{I\kern-.10em I}}

\def\IB{\relax{\rm I\kern-.18em B}}

\def\ID{\relax{\rm I\kern-.18em D}}
\def\IE{\relax{\rm I\kern-.18em E}}
\def\IF{\relax{\rm I\kern-.18em F}}
\def\IG{\relax\hbox{$\inbar\kern-.3em{\rm G}$}}
\def\IGa{\relax\hbox{${\rm I}\kern-.18em\Gamma$}}
\def\IH{\relax{\rm I\kern-.18em H}}
\def\II{\relax{\rm I\kern-.18em I}}
\def\IK{\relax{\rm I\kern-.18em K}}
\def\inbar{\,\vrule height1.5ex width.4pt depth0pt}

\def\simgt{\hskip0.05in\relax{ 
\raise3.0pt\hbox{ $>$
{\lower5.0pt\hbox{\kern-1.05em $\sim$}} }} \hskip0.05in}

\def\lp10{\ell_p^{10}}
\def\lp11{\ell_p^{11}}
\def\R11{R_{11}}

\def\frac#1#2{{#1 \over #2}}

\def\up{\uparrow}

\def\Ione{\hbox{$1\hskip -1.2pt\vrule depth 0pt height 1.53ex width 0.7pt
                  \vrule depth 0pt height 0.3pt width 0.12em$}}

\newdimen\tableauside\tableauside=1.0ex
\newdimen\tableaurule\tableaurule=0.4pt
\newdimen\tableaustep
\def\phantomhrule#1{\hbox{\vbox to0pt{\hrule height\tableaurule width#1\vss}}}
\def\phantomvrule#1{\vbox{\hbox to0pt{\vrule width\tableaurule height#1\hss}}}
\def\sqr{\vbox{%
  \phantomhrule\tableaustep
  \hbox{\phantomvrule\tableaustep\kern\tableaustep\phantomvrule\tableaustep}%
  \hbox{\vbox{\phantomhrule\tableauside}\kern-\tableaurule}}}
\def\squares#1{\hbox{\count0=#1\noindent\loop\sqr
  \advance\count0 by-1 \ifnum\count0>0\repeat}}
\def\tableau#1{\vcenter{\offinterlineskip
  \tableaustep=\tableauside\advance\tableaustep by-\tableaurule
  \kern\normallineskip\hbox
    {\kern\normallineskip\vbox
      {\gettableau#1 0 }%
     \kern\normallineskip\kern\tableaurule}%
  \kern\normallineskip\kern\tableaurule}}
\def\gettableau#1 {\ifnum#1=0\let\next=\null\else
  \squares{#1}\let\next=\gettableau\fi\next}

\def\({\left(}
\def\){\right)}

\def\ii{{\bf i}}

\def\lsim{\mathrel{\mathstrut\smash{\ooalign{\raise2.5pt\hbox{$<$}\cr\lower2.5pt\hbox{$\sim$}}}}}
\def\gsim{\mathrel{\mathstrut\smash{\ooalign{\raise2.5pt\hbox{$>$}\cr\lower2.5pt\hbox{$\sim$}}}}}

\def\overleftrightarrow#1{\vbox{\ialign{##\crcr
     $\leftrightarrow$\crcr\noalign{\kern-0pt\nointerlineskip}
     $\hfil\displaystyle{#1}\hfil$\crcr}}}
     
     \def\overleftarrow#1{\vbox{\ialign{##\crcr
     $\leftarrow$\crcr\noalign{\kern-0pt\nointerlineskip}
     $\hfil\displaystyle{#1}\hfil$\crcr}}}

\def\eg{{\it e.g.}}

\def\gU{\textsf{U}}

\newif{\ifeq}           % defines a new condition @eq tested by the conditional \ifeq
\eqtrue                 % if uncommented, declares @eq to be true
\newcounter{lecturecounter}
\def\baselinestretch{1.1}
\renewcommand{\title}[1]{\vbox{\center\LARGE{#1}}\vspace{5mm}}
\renewcommand{\author}[1]{\vbox{\center#1}\vspace{5mm}}
\newcommand{\address}[1]{\vbox{\center\em#1}}

\renewcommand{\date}[1]{\vbox{\center#1}}

\usepackage[most]{tcolorbox}
\usepackage[export]{adjustbox}% for the second solution
\usepackage{tikz}
\usetikzlibrary{shapes.geometric}
\usetikzlibrary{arrows}

\usepackage{caption}
\usepackage{subcaption} 

\usepackage{duckuments}% for the images

\usepackage{comment}
\usepackage{geometry}
\usepackage{pdflscape}

\usepackage{amsmath} 
\usepackage{mathtools}
\usepackage{amssymb}

\usepackage{amsthm}
\usepackage{thmtools}

\usepackage{thm-restate}

\usepackage{diagbox}
\usepackage{boldline}

\usepackage{bbold}
\usepackage{makecell}
\makeatother

\theoremstyle{plain}
\newtheorem{theorem}{Theorem}[section] % reset theorem numbering for each chapter

\theoremstyle{assumption}

\theoremstyle{definition}
\newtheorem{definition}[theorem]{Definition} % definition numbers are dependent on theorem numbers
\newtheorem*{remark}{Remark}

\theoremstyle{plain}

\theoremstyle{plain} 
\newtheorem{lemma}[theorem]{Lemma}

\theoremstyle{plain}

\theoremstyle{plain}
\newcounter{hypothesisCounter}
\newtheorem{hypothesis}[hypothesisCounter]{Hypothesis}

\theoremstyle{plain}

\newcommand{\calB}{{\mathcal B}}

\newcommand{\calE}{{\mathcal E}}

\newcommand{\calH}{{\mathcal H}}

\newcommand{\calK}{{\mathcal K}}

\newcommand{\calO}{{\mathcal O}}
\newcommand{\calS}{{\mathcal S}}

\def\KD{\calK_{\mathbb{D}}}

\definecolor{JMblue}{RGB}{25,25,125}
\definecolor{BSorange}{RGB}{140,50,0}

\definecolor{XLgreen}{RGB}{34,139,34}

\allowbreak
\newcolumntype{M}[1]{>{\centering\arraybackslash}m{#1}}
\usepackage{pifont}
\usepackage{leftindex}
\usepackage[normalem]{ulem}
\usepackage{float}
\usepackage{multirow}

\def\veq{\stackrel{\mathbb{V}}{=}}
\def\vapprox{\stackrel{\mathbb{V}}{\approx}}
\def\Lb{\mathbb{L}}
\def\Lbe{\mathbb{L}^{(e)}}
\def\Lbo{\mathbb{L}^{(o)}}
\usepackage[T1]{fontenc}
\usepackage[scr = zapfc]{mathalpha}
\def\Leff{\widetilde{L}}
\def\Kcft{\mathscr{K}}
\def\Lcft{\mathscr{L}}
\def\Tcft{\mathscr{T}}

\def\ctot{c_{\mathrm{tot}}}

\def\({\left(} 
\def\){\right)}
\def\[{\left[} 
\def\]{\right]}
\def\betae{\beta^{(e)}}
\def\betao{\beta^{(o)}}
\def\g{\mathfrak{g}} %% be careful redefining such short symbols!
\def\gc{\mathfrak{g}^{\bullet}}
\def\calHc{\calH^{\bullet}_{\mathrm{edge}}}
\def\CFT{\mathrm{CFT}}

\def\hhh{{\mathfrak{h}}}
\begin{document}

\title{Chiral Virasoro algebra from a single wavefunction}

\author{
Isaac H.~Kim${}^{1}$, Xiang Li${}^{2}$, Ting-Chun Lin${}^{2}$,
%{\setmainfont{Noto Serif CJK TC}(林鼎鈞)}
John McGreevy${}^{2}$, Bowen Shi${}^{2}$
%\end{CJK*}
}

\address{${}^{1}$Department of Computer Science, University of California, Davis, CA 95616, USA\\
${}^{2}$Department of Physics, University of California San Diego, La Jolla, CA 92093, USA}

% =====================
% abstract
% =====================
\begin{abstract}
Chiral edges of 2+1D systems can have very robust emergent conformal symmetry. When the edge is purely chiral, the Hilbert space of low-energy edge excitations can form a representation of a single Virasoro algebra. We propose a method to systematically extract the generators of the Virasoro algebra from a single ground state wavefunction, using entanglement bootstrap and an input from the edge conformal field theory. We corroborate our construction by numerically verifying the commutation relations of the generators. We also study the unitary flows generated by these operators, whose properties (such as energy and state overlap) are shown numerically 
to agree with our analytical predictions. 
\end{abstract}
% =====================
% end of abstract
% =====================

\vfill
%\today

\eject

%\maketitle

\setcounter{tocdepth}{2}    % =2 means don't show subsubsections in toc

\renewcommand{\baselinestretch}{0.75}\normalsize
\tableofcontents
\renewcommand{\baselinestretch}{1.1}\normalsize

\vfill\eject

% =====================
% introduction 
% =====================
\newpage 
\section{Introduction}
\label{sec:introduction}

A remarkable fact about many-body quantum systems is that a variety of simple and elegant mathematical structures can emerge as a description of their low-energy physics. For instance, the low-energy physics of gapped quantum many-body systems --- aside from the notable example of fracton phases~\cite{Chamon:2004lew, Haah2011,nandkishore2019fractons,pretko2020fracton} ---
are expected to be described by topological quantum field theory (TQFT)~\cite{Witten1988}, which successfully predicts many universal features of the underlying system, such as ground state degeneracy~\cite{Wen1990}, low-energy excitations~\cite{Wilczek1982}, and entanglement properties~\cite{Kitaev2006,Levin2006}.

Over the past few decades, it was realized that fingerprints of these mathematical structures often manifest themselves in the entanglement properties of ground state wavefunctions. Early studies discovered that the entanglement entropy often encodes universal information about the underlying quantum phases, both in gapped~\cite{Kitaev2006,Levin2006} and gapless systems~\cite{Calabrese:2009qy,Ryu2006}.

More recently, new progress is being made under the moniker of ``entanglement bootstrap,'' which aims to recover the universal properties of the underlying phase of matter from ground state entanglement. For instance, the list of anyon types, their fusion rules \cite{shi2020fusion} and braiding data \cite{Shi:2019ngw}
(including the data associated with gapped interfaces \cite{Shi:2020rne}) can be extracted from the ground state wavefunction; similar statements apply in general dimensions \cite{knots-paper,Shi:2023kwr}. Moreover, a new entanglement measure known as the modular commutator was used to extract the chiral central charge of the edge theory~\cite{Kim2021}; in the presence of a $\gU(1)$ symmetry, a similar formula exists for the Hall conductivity \cite{Fan:2022ipl}. These results are strongly suggestive of the basic dogma that underlies all these studies: that all of the universal information about a state of matter is encoded in the local reduced density matrix of a representative wavefunction.

In fact, there is nontrivial evidence for this dogma even in more general physical contexts.  As a basic example, equal-time correlators of local operators in gapped ground states decay exponentially \cite{Hastings2005-decay-cor}, whereas in gapless ground states they do not. 
In conformal field theory (CFT) ground states, renormalization group monotones (called $c, a, $ or $F$ depending on the dimension) can be extracted from the entanglement entropy of a round ball \cite{Holzhey:1994we,Calabrese:2009qy,Myers:2010tj}. 

More ambitiously, one may posit that structural features, that is, mathematical relationships between these quantities, can emerge from the universal entanglement properties of the ground state wavefunctions. In systems with liquid topological order, there is already strong evidence that supports this expectation. For instance, in two~\cite{shi2020fusion} and higher~\cite{knots-paper,Shi:2023kwr} dimensions, various identities relating superselection sectors, their quantum dimensions, and fusion multiplicities emerge from simple universal properties of gapped ground states. In the presence of a gapped boundary or domain wall, similar progress can be made~\cite{Shi:2020rne}. In fact, in the context of three space dimensions, fusion spaces associated with knotted excitations (and partons for domain wall excitations) are identified, making the existing puzzle of ``what is the minimal set of data?'' more pronounced in such physical contexts.

The focus of this paper is a simple but nontrivial physical setup that differs from these previous studies, namely the chiral gapless edge of a 2+1D system. From the field theory perspective, the key mathematical structure that should be present in this system is the Virasoro algebra:
\begin{equation}\label{eq:vir}
[\Lcft_m, \Lcft_n] = (m-n) \Lcft_{m+n} + \frac{c}{12} (m^3-m) \delta_{m+n,0},
\end{equation}
where $c$ is the central charge and $\{ \Lcft_n\}_{n \in \mathbb{Z}}$ is the set of Virasoro generators.
The Virasoro algebra and its unitary representations appear in 1+1D CFT, 1+1D critical lattice models\footnote{See \cite{Koo:1993wz} for a concrete example of how Virasoro algebra could appear in critical lattice models.}, such as the critical transverse field Ising model, and the edge theories of 2+1D topological orders, such as the $\nu=2/3$ quantum Hall state \cite{Levin2013} and critical points describing phase transitions between two gapped edges \cite{Chatterjee2022}. 
In general, these theories contain both the left and right moving parts which are the representations of two separate copies of Virasoro algebra.  
However, in a 2+1D system with a \emph{purely chiral} edge, such as the edge of $\nu = 1/3$ Laughlin state \cite{Laughlin1983} or the most natural ($c = 8$) edge of the $E_8$ bosonic invertible phase \cite{Kitaev2005}, there is only one copy of the Virasoro algebra, simplifying the analysis. 

Compared to the case of liquid gapped ground states, 
for gapless systems, 
precisely how these mathematical objects (such as the Virasoro algebra) emerge from entanglement is much more wide open. Before diving into this problem directly, we need to do exploratory work and test various conjectures. This is the main aim of this paper. 

In this paper, we identify generators of such a Virasoro algebra in terms of linear combinations of modular Hamiltonians (entanglement Hamiltonian) of a purely chiral ground state supported near the edge. These form a part of 
a larger class of modular flows called {\it good modular flows} 
that can create excitations at the edge whilst preserving the bulk local density matrices. 
(By `bulk local density matrices' here we mean reduced density matrices of regions that do not touch the gapless edge.  Throughout the paper, we will refer to such regions as `bulk regions'.) 
We provide analytical and numerical arguments that support our claim in Section~\ref{sec:good-modular-flows}. The arguments presented in this Section are independent of any CFT assumption and heavily rely on the techniques from entanglement bootstrap~\cite{shi2020fusion}. 

In Section~\ref{sec:chiral-cft}, we put forward a hypothesis [Hypothesis~\ref{assump:chiral-CFT}] that relates the generators of certain subsets of the good modular flows to quantities in CFT. This hypothesis will be the foundation upon which the results of the ensuing sections rely. While proving this hypothesis is beyond the scope of this paper, we provide several pieces of nontrivial evidence in support of it. First, we provide several locally checkable consequences of the hypothesis in Section~\ref{sec:locally_checkable}, 
including a generalization of 
the recently-discovered vector fixed point equation of \cite{Lin2023}.
We verify these numerically in Section \ref{sec:numerics}.  Secondly, we prove a special case of this conjecture under a reasonable physical assumption: that a 2+1D gapped ground state with a chiral gapless edge on a strip can be formally viewed as a 1+1D CFT ground state under dimensional reduction [Section~\ref{subsec:cylinder-argument}].

In Section \ref{sec:edge-subspace-coherent}, we define a carrier space for our representation of the chiral Virasoro algebra from purely chiral states on a two-dimensional disk.  It is swept out by the generators of good modular flow.  

We then construct the Virasoro generators modulo our hypothesis and explore their properties in Section \ref{sec:virasoro-construction} and \ref{sec:numerics}. 
In Section~\ref{sec:virasoro-construction}, we devise a systematic method to extract the Virasoro generators, which are shown to be a special class of good modular flow generators. We numerically verify their commutation relations in Section~\ref{sec:numerics} using $p+ip$ superconductor and chiral semion lattice ground states, which both show good agreement with the predictions. 

We remark that there is a large literature on the construction of Virasoro algebras in 1+1D systems, starting from \cite{Koo:1993wz} and more recently in \cite{Hu:2020suv,Osborne:2021ppp,Wang:2022qxf,Zeng:2022swq}. Here, we make some comparisons: Firstly, the idea of these works is to perform an exact Fourier transformation of the Hamiltonian density. The starting point of their construction is very different from ours: \cite{Koo:1993wz} starts from the partition function, \cite{Hu:2020suv} directly starts from the local Hamiltonian and \cite{Wang:2022qxf,Zeng:2022swq} starts from a tensor network representation of the partition function. So all of these begin with a time evolution operator, either transfer matrix or Hamiltonian.  Our construction, instead, is directly from a single quantum state. Taking advantage of the form of a certain linear combination of modular Hamiltonians near the edge, designed in such a way that it acts nontrivially only on the edge, we can perform an approximate\footnote{We also design a systematic process to make the approximation converge to the exact Fourier transformation.} Fourier transformation to the stress-energy tensor, even without the construction of the Hamiltonian density beforehand. Secondly, previous work focuses on 1+1D systems, which involve two copies of Virasoro algebra, while our construction is for 2+1D purely chiral systems, where there is only one copy of Virasoro algebra. Therefore, there is no need to separate the two copies of Virasoro algebra as in the 1+1D construction. 

% =====================
% end of the introduction 
% =====================

\section{Primer on entanglement bootstrap}
\label{sec:primer_eb}

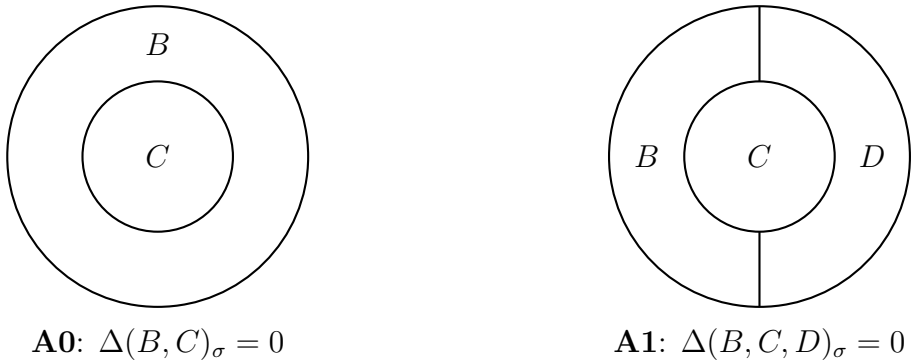
\begin{figure}[h]
    \centering
    \begin{tikzpicture}
        \draw[ thick] (0,0) circle (1cm);
        \draw[ thick] (0,0) circle (2cm);
        \node[] () at (0,0) {$C$};
        \node[] () at (0, 1.5) {$B$};
        \node[] () at (0, -2.5) {\textbf{A0}: $\Delta(B,C)_{\sigma}=0$};

        \begin{scope}[xshift=8cm]
        \draw[ thick] (0,0) circle (1cm);
        \draw[ thick] (0,0) circle (2cm);
        \draw[ thick] (0,1) -- ++ (0,1);
        \draw[ thick] (0,-1) -- ++ (0, -1);
        \node[] () at (-1.5, 0) {$B$};
        \node[] () at (0, 0) {$C$};
        \node[] () at (1.5, 0) {$D$};
        \node[] () at (0, -2.5) {\textbf{A1}: $\Delta(B, C,D)_{\sigma}=0$};
        \end{scope}
    \end{tikzpicture}
    \caption{Axioms of the entanglement bootstrap. Here $\sigma$ is the global state, often referred to as the \emph{reference state}. Here $\Delta(B,C)_{\sigma} := S(\sigma_{BC}) + S(\sigma_C) - S(\sigma_B)$ and $\Delta(B,C,D)_{\sigma} := S(\sigma_{BC}) + S(\sigma_{CD}) - S(\sigma_B) - S(\sigma_D)$.}
    \label{fig:eb_axioms}
\end{figure}

In this Section, we provide a brief review of the basics of entanglement bootstrap~\cite{shi2020fusion}. In the absence of edges, there are two axioms (\textbf{A0} and \textbf{A1}) that constitute the core of entanglement bootstrap. These axioms pertain to the entanglement property of the global state, often referred to as the \emph{reference state}. These are described in Figure~\ref{fig:eb_axioms}. Here we use the convention that the subscript $\sigma$ appearing after the parenthesis represents the underlying global state used for the evaluation of the entanglement entropies; for instance, in expressions like $(S_{AB} + \ldots)_{\sigma}$, one can simply replace $S_{AB}$ with $S(\sigma_{AB})$. While the only assumptions we make are that \textbf{A0} and \textbf{A1} hold true on every constant-sized ball, as a logical consequence, one can show that the analogous set of assumptions holds true at larger scales~\cite{shi2020fusion}. Therefore, without loss of generality, we can assume that these axioms hold on any disks, independent of their size.

The axioms \textbf{A0} and \textbf{A1} are intimately related to quantum information-theoretic measures of correlations, such as mutual information and conditional mutual information. Mutual information with respect to a state $\rho$ is defined as 
\begin{equation}
    I(A:B)_{\rho} := S(\rho_A)+ S(\rho_B) - S(\rho_{AB}).
\end{equation} 
Conditional mutual information is a generalization of mutual information, defined as 
\begin{equation}
    I(A:C|B)_{\rho} := S(\rho_{AB})+ S(\rho_{BC}) - S(\rho_B) - S(\rho_{ABC}).
\end{equation}
If the global state is pure, one can verify that \textbf{A0} and \textbf{A1} reduce to $I(A:C)_{\sigma}=0$ (choosing $A$ as a purifying system) and $I(A:C|B)_{\sigma}=0$ (choosing $A$ as a purifying system). Therefore, \textbf{A0} and \textbf{A1}  imply that certain mutual information and conditional mutual information are zero.

Intuitively, the fact that the mutual and conditional mutual information are zero can be understood as follows. Because mutual information is a measure of the correlation between two subsystems, in gapped systems we can expect $I(A:C)$ to approach $0$ as the distance between $A$ and $C$ diverges. Therefore, \textbf{A0} can be thought of as a condition that we expect to hold at a long-distance scale, in particular, at the fixed point of some coarse-graining procedure. We can arrive at the same conclusion for \textbf{A1}, but using a different reasoning. At long distance, the entanglement entropies of the subsystems are expected to be approximated well by a leading ultraviolet-divergent term, proportional to the area, and a subleading universal term, known as the topological entanglement entropy~\cite{Kitaev2006,Levin2006}. It is an easy exercise to show that this scaling law implies \textbf{A1}. Therefore, \textbf{A0} and \textbf{A1} are the conditions that we can expect to hold in gapped systems at long distances.
\footnote{Note that for chiral systems with finite-dimensional Hilbert spaces, we expect the axioms to only hold approximately.}

In the context of this paper, the main power of \textbf{A0} and \textbf{A1} comes from the fact that they let us decompose the \emph{modular Hamiltonian} \cite{Casini2008,Li2008} (also known as the entanglement Hamiltonian, defined as $-\ln \rho_A$ for the subsystem $A$) of any region into terms that act locally within that region. For any quantum state, the following relations are known:
\begin{equation}
    \begin{aligned}
        I(A:B)_{\rho}=0 &\iff  \ln  \rho_{AB} = \ln  \rho_A + \ln  \rho_B, \\
        I(A:C|B)_{\rho}=0 &\iff \ln \rho_{ABC} = \ln \rho_{AB} + \ln \rho_{BC} - \ln  \rho_B.
    \end{aligned}
    \label{eq:logrdm_decomposition}
\end{equation}
Therefore, by iteratively applying these decompositions, one can often decompose the modular Hamiltonian of a region to a linear combination of modular Hamiltonians acting on smaller regions. 
In the presence of edges, one should not expect these axioms to hold. However, one can still assume that the axioms hold at any location that is sufficiently far away from the edge; we numerically verified this fact, the result of which is presented in Section~\ref{sec:fidelity_tests}. The fact that the axioms are still satisfied in the bulk comes into play in our analysis, letting us decompose the linear combination of modular Hamiltonians over regions that are anchored on the edge. However, the remaining piece of the puzzle is what kind of entanglement-based assumptions one ought to make at the edge. This is what we discuss and explore in the rest of the paper. 

% =====================
% good modular flows
% =====================
\section{Good modular flows}
\label{sec:good-modular-flows}

In this Section, we introduce and study a special class of unitary flows\footnote{i.e. flows generated by unitary operators}, referred to as \emph{good modular flows}.
A good modular flow creates only edge excitations, 
and this is the 
intuitive reason that their study can be related to 
an (anomalous) 1+1D system.
The central hypothesis of this paper [Hypothesis~\ref{assump:chiral-CFT}] is that certain generators of the good modular flows can be identified as the Virasoro generators, in the sense we make precise in Section~\ref{sec:virasoro-construction}.

\subsection{Definition and properties of good modular flows}
\label{sec:def-good-modular-flows}
In order to motivate the definition of good modular flows, it will be instructive to consider a general linear combination of modular Hamiltonians:  
\begin{equation}
    \mathbb{L} = \sum_{A\in \calS} \lambda_A K_A, \quad \lambda_A \in \mathbb{R}, 
    \label{eq:modular_generator_general}
\end{equation}
where $\calS$ is a set of regions,  $K_A = -\ln \rho_A$ is the modular Hamiltonian of the subsystem on the region $A$, $\rho_A = \text{Tr}_{\bar{A}}\left( {|\Psi\rangle}{\langle \Psi|}\right)$ is the reduced density matrix over $A$, and $\lambda_A$ is a real number. Upon evolving the global state by a unitary flow $e^{-\ii \mathbb{L} t}, t\in \mathbb{R}$, (such that $|\Psi(t)\rangle = e^{-\ii \mathbb{L} t}|\Psi\rangle$), we can ask how the reduced density matrices of different subsystems change. 

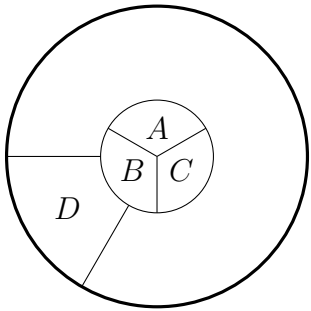
\begin{figure}[h]
    \centering
    \begin{tikzpicture}
        \draw[] (0,0) circle (0.75cm);
        \draw[very thick] (0,0) circle (2cm);
        \draw[] (0,0) -- (30:0.75cm);
        \draw[] (0,0) -- (150:0.75cm);
        \draw[] (0,0) -- (270:0.75cm);
        \node[] () at (90:0.375cm) {$A$};
        \node[] () at (210:0.375cm) {$B$};
        \node[] () at (330:0.375cm) {$C$};
        \draw[] (180:0.75cm) -- (180:2cm);
        \draw[] (240:0.75cm) -- (240:2cm);
        \node[] () at (210:1.375cm) {$D$};
    \end{tikzpicture}
    \caption{A modular flow generated by $K_{BCD}$ changes the density matrix on $AB$. The outermost boundary in this diagram is the physical edge.}
    \label{fig:flow_changing_rdm}
\end{figure}

In general, such a flow can change the reduced density matrices in the bulk. For instance, consider a flow generated by $\mathbb{L} = K_{BCD}$ for the subsystems shown in Figure~\ref{fig:flow_changing_rdm}. For small $\delta t$, $e^{-\ii K_{BCD}\delta t}$ changes the entanglement entropy of the state on $AB$ by an amount $\frac{\pi}{3}c_-\delta t$, where $c_-$ is the chiral central charge~\cite{Kim2021,Kim:2021tse}, which is generally nonzero. Therefore, this particular flow can change reduced density matrices in general. Indeed, for a chiral state\footnote{Throughout this paper we use ``chiral'' to mean $c_- \neq 0$ and ``purely chiral'' to mean $|c_-| = c_\text{tot}$, where $c_\text{tot} \equiv c_L + c_R$ is the total central charge.}, by which we mean $c_- \ne 0$, the density matrix on $AB$ must change in the first order of $\delta t$.

However, unlike these more general modular flows, good modular flows are constructed in such a way that they do not change the bulk reduced density matrices. Here is a simple rule that we advocate.

\begin{definition}[Good modular flows]
Let $\calS$ be a set of regions and $\mathbb{L} = \sum_{A\in\calS} \lambda_A K_A$ (with $\lambda_A \in \mathbb{R}$) be a modular flow generator. 
This generator is \emph{good} if for any line segment $e$
\begin{equation} 
    \sum_{A\in \calS \text { with } \partial A\supset e } \lambda_A=0, \label{eq:good_constraint}
\end{equation}
where $\partial A\supset e $ means $\partial A$, the entanglement boundary of $A$\footnote{By entanglement boundary $\partial A$ (sometimes called entanglement cut) we mean the boundary of $A$ inside the bulk, away from the boundary of the sample.}, fully contains $e$.
Throughout this paper, we denote the class of good modular flow generators as $\mathfrak{g}$. We call the modular flow $e^{-\ii \Lb t}$ \emph{good} if $\Lb \in\mathfrak{g}$. 
\label{definition:good_modular_flow_succinct}
\end{definition}

Any linear combination of modular Hamiltonians that does not satisfy Eq.~\eqref{eq:good_constraint} shall be referred to as a \emph{bad} modular flow generator. 

\begin{figure}[!h]
    \centering
    \includegraphics[width=0.5\columnwidth]{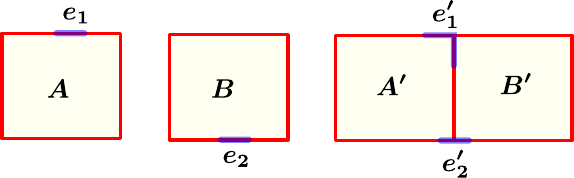}
    \caption{A good and a bad modular flow generators $\mathbb{L} = K_{A}+K_{B}-K_{AB}$ and $\mathbb{L}' = K_{A'}+K_{B'}-K_{A'B'}$. $A,B,A',B'$ are regions in the bulk and $e_1,e_2,e_1',e_2'$ are examples of line segments.}
    \label{fig:good-bad-examples}
\end{figure}

We now explicitly explain the Definition~\ref{definition:good_modular_flow_succinct} with the following examples of good and bad modular flow generators. More examples are given in Table~\ref{tab:flow-examples}. Consider two modular flow generators $\mathbb{L} = K_{A}+K_{B}-K_{AB}$ and $\mathbb{L}' = K_{A'}+K_{B'}-K_{A'B'}$, where $A,B,A',B'$ are regions in the bulk shown in Figure~\ref{fig:good-bad-examples}. Based on the definition, $\mathbb{L}$ is a good modular flow generator while $\mathbb{L}'$ is bad. For $\mathbb{L}$, we can consider an arbitrary line segment, say $e_1$ shown in Figure~\ref{fig:good-bad-examples}. Since $\partial A \supset e_1, \partial(AB)\supset e_1$ and $\lambda_A = - \lambda_{AB} = 1$, the sum in Eq.~\eqref{eq:good_constraint} for $e_1$ is indeed zero: $\lambda_{A} + \lambda_{AB} = 0$. One can examine that for any line segment inside $\partial A,\partial B,\partial(AB)$\footnote{One only needs to consider the line segments on the entanglement boundaries of the regions involved in the sum since those are not parts of any entanglement boundaries and will not give any constraints.}, Eq.~\eqref{eq:good_constraint} condition is satisfied, and therefore $\mathbb{L}$ is a good modular flow generator. For $\mathbb{L}'$, notice for the line segment $e'_1$, it is only fully contained in $\partial A'$, therefore the sum in Eq.~\eqref{eq:good_constraint} results in $\lambda_{A'} = 1$, which makes $\mathbb{L}'$ a bad modular flow generator. 

%See Table~\ref{tab:flow-examples} for examples of good and bad modular flow generators. 

\begin{table}[!h]
    \centering
    \begin{tabular}{|l|}
    \hline 
    Modular flow generators and their types (Good/Bad) \\ 
    \hline
    $\includegraphics[width=3cm,valign=c,margin=0.2cm]{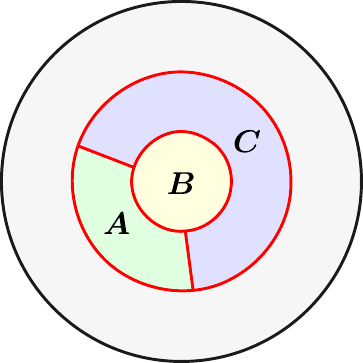} \; \; 
    \begin{aligned}
        &K_{ABC}+K_{B}+K_{AC} && \text{Bad} \\ 
        & K_{AB}+K_{BC}+K_{A}+K_{C} && \text{Bad}\\
        &\hat{\Delta}(AC,B,\emptyset) = K_{ABC}+K_{B}-K_{AC}&&\text{Good} \\ 
        & \hat{\Delta}(A,B,C) = K_{AB}+K_{BC}-K_{A}-K_{C} && \text{Good}
    \end{aligned}$ \\ \hline 

    $\includegraphics[width=3cm,valign=c,margin=0.2cm]{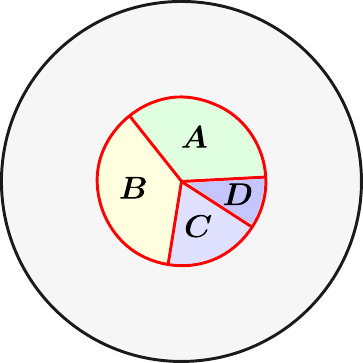}\; \; 
    \begin{aligned}
        &\hat{I}(A:C|B) = K_{AB}+K_{BC}-K_{ABC}-K_B && \text{Bad}\\
        &K_{AB}+K_{BCD}+K_{CDA}-K_{A}-K_{B}-K_{CD}-K_{ABCD} && \text{Good}
    \end{aligned}$  \\ \hline 

    $\includegraphics[width=3cm,valign=c,margin=0.2cm]{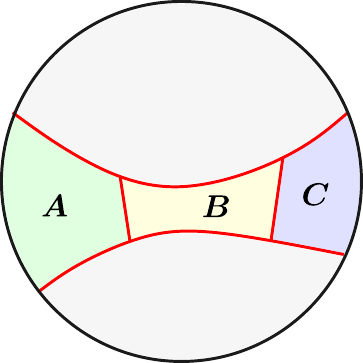}\; \; 
    \begin{aligned}
        &K_A+K_C+K_{AC} && \text{Bad}\\
        &K_{AB}+K_{BC}+K_{ABC}+K_{B} && \text{Bad}\\
        &\hat{I}(A:C) = K_{A}+K_{C}-K_{AC} && \text{Good} \\ 
        &\hat{I}(A:C|B) = K_{AB}+K_{BC} -K_{ABC}-K_{B} && \text{Good}
    \end{aligned}$ \\ \hline

    $\includegraphics[width=3cm,valign=c,margin=0.2cm]{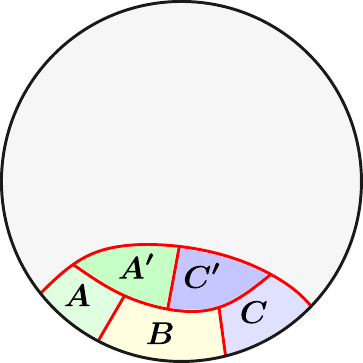}\; \; 
    \begin{aligned}
        &K_{AA'B}+K_{BCC'}+K_{AA'}+K_{CC'} && \text{Bad}\\ 
        &K_{AB}+K_{BC}+K_{ABC}+K_{B} && \text{Bad}\\
        &\hat{\Delta}(AA',B,CC') = K_{AA'B}+K_{BCC'}-K_{AA'}-K_{CC'} && \text{Good}\\ 
        &\hat{I}(A:C|B) = K_{AB}+K_{BC}-K_{ABC}-K_{B} && \text{Good}
    \end{aligned}$  \\ \hline 

    $\includegraphics[width=6cm,valign=c,margin=0.2cm]{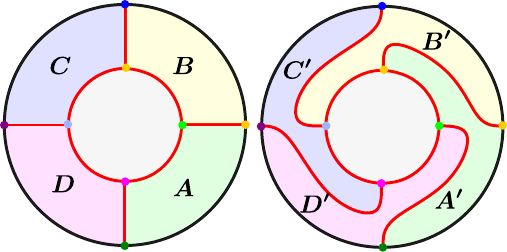}\quad \begin{aligned}
        &L = K_{A}-K_B+K_C-K_D && \text{Bad}\\ 
        &L' = K_{A'}-K_{B'}+K_{C'}-K_{D'} && \text{Bad}\\
        &L+L' && \text{Good}
    \end{aligned}$ \\ \hline
\end{tabular}
\caption{Examples of good and bad modular flow generators. The setups are shown in the figure on the left. Each line of the equation is a modular flow generator, with its type (good or bad) marked at the end. The two figures in the last row are two ways to partition the same disk, where the colored dots are identified. }
\label{tab:flow-examples}
\end{table}

While we do not have a rigorous proof for the invariance of the bulk reduced density matrices under good modular flows, we provide an argument that supports the validity of this claim. As a starter, we prove this claim for an infinitesimally small flow. Let $\mathbb{L}$ be a good modular flow generator. We compute the reduced density matrices in the bulk under a flow generated by $\exp(-\ii \mathbb{L} \delta t)$, for an infinitesimally small $\delta t$. We show that the linear-order contribution in $\delta t$ vanishes. 

Without loss of generality, consider a local bulk region $S$, by which we mean $S$ is a subset of a bulk disk. 
Under the good modular flow generated by $\mathbb{L}$, the first order contribution can be computed as 
\begin{equation}
    \rho_S(\delta t) = \rho_S + \ii \delta t\text{Tr}_{\bar{S}}\left( [{|\Psi\rangle}{\langle\Psi|}, \mathbb{L}]\right).
\end{equation}
The generator $\mathbb{L}$ is a linear combination of modular Hamiltonians. Our goal is to show that the linear order contribution vanishes, $\text{Tr}_{\bar{S}}\left( [{|\Psi\rangle}{\langle\Psi|}, \mathbb{L}]\right) = 0$.

By the cyclicity of the trace, for any modular Hamiltonian associated with a region included in $\bar{S}$, its contribution is identically zero. Similarly, by noting the identity $K_X|\Psi\rangle = K_{\overline{X}}|\Psi\rangle$ for any subsystem $X$, any contribution coming from the region that includes $S$ is also identically zero. Therefore, for computing the linear-order contribution, it suffices to compute the contributions from the modular Hamiltonians associated with regions that have nontrivial intersections with both $S$ and $\overline{S}$. Put another way, only the modular Hamiltonians whose entanglement cut passes through $S$ contribute.

To show that these contributions vanish, we will deform the cuts away from the local bulk region $S$. In particular, we will show that 
\begin{equation}
    \mathbb{L}|\Psi\rangle = \mathbb{L}'|\Psi\rangle, \label{eq:deformation_generator}
\end{equation}
where $\mathbb{L}'$ is a good modular flow generator whose entanglement cut does not pass through $S$.
This implies
\begin{equation}
\begin{aligned}
    \Tr_{\overline{S}}\left( [{|\Psi\rangle}{\langle\Psi|}, \mathbb{L}]\right) = \Tr_{\overline{S}}\left( [{|\Psi\rangle}{\langle\Psi|}, \mathbb{L}']\right) 
    =0,
\end{aligned}
\label{eq:first_order_contribution_zero}
\end{equation}
where the second equality applies the argument from the previous paragraph.

Therefore, if we were to show that the entanglement cuts present in the definition of $\mathbb{L}$ can be deformed away from the local bulk region $S$, we would be done. More precisely, if we can show that 
\begin{equation}
    \mathbb{L}|\Psi\rangle = \mathbb{L}'|\Psi\rangle, \label{eq:deformation_generator}
\end{equation}
where $\mathbb{L}'$ is a good modular flow generator whose entanglement cut does not pass through $S$, we have
\begin{equation}
\begin{aligned}
    \Tr_{\overline{S}}\left( [{|\Psi\rangle}{\langle\Psi|}, \mathbb{L}]\right) &= \Tr_{\overline{S}}\left( [{|\Psi\rangle}{\langle\Psi|}, \mathbb{L}']\right) \\
    &=0.
\end{aligned}
\label{eq:first_order_contribution_zero}
\end{equation}

\begin{figure}[h]
    \centering
    \begin{tikzpicture}[scale=0.8]
        \draw[very thick] (0,0) circle (2cm);
        \draw[] (0,-2cm) --++ (0, 4cm);
        \node[] () at (-1cm, 0) {\footnotesize $A$};
        \node[] () at (1cm, 0) {\footnotesize $B$};

        \begin{scope}[xshift=6cm]
        \draw[very thick] (0,0) circle (2cm);
        \draw[] (0,-2cm) -- (0, -0.5cm);
        \draw[] (0,2cm) -- (0, 0.5cm);
        \draw[] (0, -0.5cm) arc (-90:90:0.5cm);
        \node[] () at (-0.75cm, 0) {\footnotesize $A\cup c$};
        \node[] () at (1.25cm, 0) {\footnotesize $B\setminus c$};
        \end{scope}
    \end{tikzpicture}
    \caption{Deformation of the subsystem $A$ and $B$ 
(left) by a half-disk $c$ (right).}
    \label{fig:deformation_example}
\end{figure}
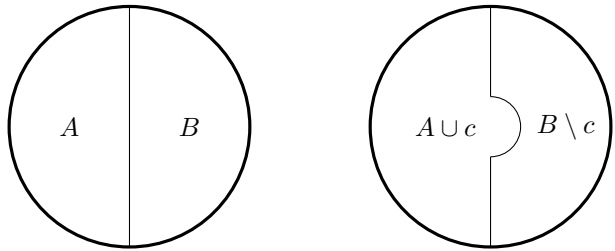

To prove the deformability of the entanglement cut [Eq.~\eqref{eq:deformation_generator}], let us consider a special toy example, which will be an instructive exercise that ultimately leads to the proof of this claim. Consider the bipartition of the entire system into two disks, $A$ and $B$ [Figure~\ref{fig:deformation_example}]. Our claim is that the action of $K_A - K_B$ on the reference state is invariant under the deformation of the entanglement cut [Eq.~\eqref{eq:deformation_generator}]. Upon deforming $A$ to $A\cup c$ and $B$ to $B\setminus c$, the change in the good modular flow generator is $K_{A\cup c} - K_{B\setminus c} - K_A + K_B$. It is known that this quantity is zero if and only if $S(\rho_{A\cup c}) + S(\rho_B) - S(\rho_{B\setminus c}) - S(\rho_A)=0$~\cite{Lin2022-op}. Because of strong subadditivity (SSA) \cite{Lieb1973},
\begin{equation}
    S(\rho_{A\cup c}) + S(\rho_c) - S(\rho_{B\setminus c}) - S(\rho_A) \leq  S(\rho_{a\cup c}) + S(\rho_c) - S(\rho_{b\setminus c}) - S(\rho_a).\label{eq:sandwich_bound_example}
\end{equation}
The right hand side of Eq.~\eqref{eq:sandwich_bound_example} is zero by \textbf{A1} [Section~\ref{sec:primer_eb}]. 
On the other hand, the left-hand side of Eq.~\eqref{eq:sandwich_bound_example} is at least zero by the weak monotonocity~\cite{Lieb1973}. Therefore, $S(\rho_{A\cup c}) + S(\rho_c) - S(\rho_{B\setminus c}) - S(\rho_A)=0$, yielding 
\begin{equation}
    (K_{A\cup c} - K_{B\setminus c} - K_A + K_B)|\Psi\rangle=0, \label{eq:operator_identity_wm}
\end{equation}
where $|\Psi\rangle$ is the global state.

This argument can be straightforwardly extended to a more general setup in which $A$ and $B$ do not necessarily partition the entire system. As long as the subsystem $c$ is a disk in the bulk lying in a small neighborhood of an entanglement cut shared by $A$ and $B$, our argument works exactly the same way. Furthermore, we remark that $A$ and $B$ need not lie on opposite sides of the entanglement cut. A similar argument applies to the case in which the subsystems $A$ and $B$ lie on the same side of the cut. In this case, while the overall structure of the argument is similar, strong subadditivity plays the role of weak monotonicity. 

Now we are in a position to prove our main claim: that the action of good modular flow generators on the reference state are invariant under deformations of the entanglement cut. Without loss of generality, consider a modular flow generator $\Lb$ and a small line segment $e$ along an entanglement cut. We shall show that one can always deform the entanglement cut $e$ in a way similar to Figure~\ref{fig:deformation_example}. Let $K_{A_i}, i=1,\cdots,n$ be the modular Hamiltonians in $\Lb$ whose entanglement boundaries contain $e$, i.e. $e\subset \partial A_i$, and let $\lambda_i$ be the coefficient of $K_{A_i}$ in $\Lb$. The idea behind this proof is simply to ``pair up'' these modular Hamiltonians in $\Lb$ as
\begin{equation}\begin{split}
    \sum_{i=1}^n \lambda_{A_i}K_{A_i} &= \sum_{i=1}^n \lambda_{A_i}(K_{A_i}-K_{A_1}) + \left(\sum_{i=1}^n \lambda_{A_i}\right) K_{A_1} \\ 
    &= \sum_{i=1}^n \lambda_{A_i}(K_{A_i}-K_{A_1}),
\end{split}
\end{equation}
where the first line is simply a rearrangement of the terms in the left hand side and the second line follows because $\sum_{i=1}^{n}\lambda_{A_i}=0$ by the definition of the good modular flow generators [Definition~\ref{definition:good_modular_flow_succinct}]. Now we can see that the action of $\Lb$ on $e$ is simply the action of a sum of $n$ pairs $K_{A_i}-K_{A_1}$. Notice each pair is just the simple case we described in the previous paragraph (or the version with both regions on the same side of the cut). Therefore we can deform the entanglement cut $e$ for each pair as in Figure~\ref{fig:deformation_example}, without changing the action on the reference state. Thus, the action of a good modular flow generator on the reference state is invariant under deformations near any segment of an entanglement cut. Repeatedly applying this argument, we conclude that the action of a good modular flow generator remains invariant under any smooth deformation of the entanglement cuts. 

Therefore, we have shown that for any good modular flow generator, its entanglement cuts can be deformed [Eq.~\eqref{eq:deformation_generator}]. In particular, for any local bulk region, we can deform the generator so that none of its entanglement cuts pass through it. Therefore, for an infinitesimally small flow, the reduced density matrix of any local bulk region\footnote{An example of what we are excluding is a cylinder, 
where the modular flow generators involve regions touching both boundaries, and the bulk region in question wraps around the cylinder.} remains invariant [Eq.~\eqref{eq:first_order_contribution_zero}]. 
We will comment elsewhere on the effects of good modular flow on density matrices of more general regions.

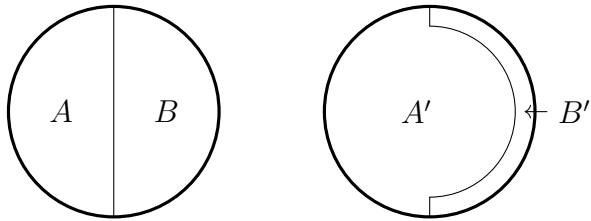
\begin{figure}[h]
    \centering
    \begin{tikzpicture}[scale=0.7]
        \draw[very thick] (0,0) circle (2cm);
        \node[] () at (-1cm, 0) {$A$};
        \node[] () at (1cm, 0) {$B$};
        \draw[] (0, -2cm) -- ++ (0, 4cm);

        \begin{scope}[xshift=6cm]
        \draw[very thick] (0,0) circle (2cm);
        \node[] () at (-0.25cm, 0) {$A'$};
        \node[] (Bp) at (2.75cm, 0) {$B'$};
        \draw[] (0, -2cm) -- (0, -1.625cm) arc (-90:90:1.625cm) -- (0, 2cm);
        \draw[->] (Bp) -- (1.8cm, 0);
        \end{scope}
    \end{tikzpicture}
    \caption{Here we consider a simple example of a good modular flow generated by $K_A - K_B$, where the relevant subsystems are shown in the left figure. The entanglement cut of this good modular flow generator can be deformed close to the physical edge (right). In particular, $(K_A - K_B)|\Psi\rangle = (K_{A'} - K_{B'})|\Psi\rangle,$ where $|\Psi\rangle$ is the state from which the modular Hamiltonians are defined.} 
    \label{fig:cut_deformation}
\end{figure}

What if the flow is no longer infinitesimal? In this case, we can heuristically argue for the invariance of the bulk reduced density matrices by making a nontrivial assumption: that Eq.~\eqref{eq:operator_identity_wm} can be promoted to an operator identity acting on an appropriate subspace. Specifically, let $\mathbb{L}$ be a good modular flow generator. Let $\mathbb{L}'$ be a deformation of $\mathbb{L}$ such that $(\mathbb{L} - \mathbb{L}')|\Psi\rangle=0$. (For instance, one may consider $\mathbb{L} = K_A - K_B$ and $\mathbb{L}' = K_{A'} - K_{B'} $ in Figure~\ref{fig:cut_deformation}.) We posit that
\begin{equation}
    (\mathbb{L} - \mathbb{L}')|\phi_{\text{edge}}\rangle = 0
    \label{eq:actual_op_identity_wm}
\end{equation}
for any $|\phi_{\text{edge}}\rangle$ which is indistinguishable from the reference state in the bulk. This includes the reference state $|\Psi\rangle$ but also includes states that may have excitations at the boundary.

We claim that $\mathbb{L}^n|\Psi\rangle = {\mathbb{L}'}^{n}|\Psi\rangle$ for any $n\in \mathbb{N}$. For $n=1$, the claim is what we showed above  [Eq.~\eqref{eq:deformation_generator}]. For $\mathbb{L}^2 |\Psi\rangle$, note that $\mathbb{L}^2|\Psi\rangle = \mathbb{L} \mathbb{L}'|\Psi\rangle$, where $\mathbb{L}'|\Psi\rangle$ is a state with excitations only at the edge. Therefore, using Eq.~\eqref{eq:actual_op_identity_wm}, we get $\mathbb{L}\mathbb{L}' |\Psi\rangle = {\mathbb{L}'}^2|\Psi\rangle$. The same argument can be repeated recursively, establishing $\mathbb{L}^n|\Psi\rangle = {\mathbb{L}'}^n|\Psi\rangle$. In particular, using Taylor expansion we conclude that $e^{-\ii \mathbb{L} t} |\Psi\rangle = e^{-\ii \mathbb{L}' t} |\Psi\rangle$ for any $t\in \mathbb{R}$. 

\begin{figure}[h]
    \centering
    \begin{tikzpicture}
        \draw[very thick] (0,0) circle (2cm);
        \draw[] (0, -2cm) -- (0, -1.625cm) arc (-90:90:1.625cm) -- (0, 2cm);
        \draw[fill=blue!20!white] (0,0) circle (1.5cm);
        \draw[fill=red!20!white] (0,0) circle (1.25cm);
        \node[] () at (0,0) {$C$};
        \node[] (D) at (-2.5, 0) {$D$};
        \draw[->]  (D) -- (-1.375cm, 0);
    \end{tikzpicture}
    \caption{For any bulk disk ($C$) there is a region surrounding $C$ (denoted as $D$) on which a disentangling unitary exists. Upon applying this disentangling unitary, we obtain a state in which the disk $C$ is fully entangled with part of $D$.}
    \label{fig:bulk_disentangling}
\end{figure}
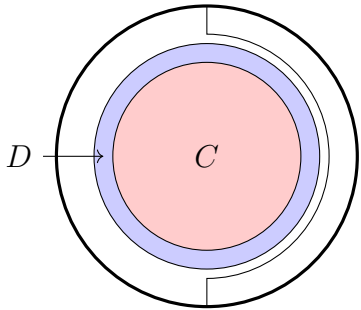

Now we claim that the reduced density matrix of a disk in the local bulk region (denoted as $C$) is invariant under $e^{-\ii\mathbb{L}'t}$ provided that the entanglement cut associated with $\mathbb{L}'$ does not pass through $C$. This is due to the existence of a disentangling unitary acting on a region surrounding $C$ (denoted as $D$) [Figure~\ref{fig:bulk_disentangling}]. More concretely, the entanglement bootstrap axioms imply that \textbf{A0} is satisfied on this disk; it is known that~\cite{shi2020fusion}
\begin{equation}
    \left(S_{CD} + S_C - S_D \right)_{\sigma}=0.
\end{equation}
Therefore, for any extension of $\sigma_{CD}$ to a larger system, say, $\rho_{BCD}$, there is a disentangling unitary acting on $D$ (denoted as $U_D$) such that 
\begin{equation}
    U_D \rho_{BCD} U_D^{\dagger} = {|\phi\rangle_{CD_L}}{\langle \phi|} \otimes \rho_{D_R B},
\end{equation}
for some pure state $|\phi\rangle_{CD_L}$, where $\mathcal{H}_D \cong \mathcal{H}_{D_L}\otimes \mathcal{H}_{D_R}$. What is more, $|\phi\rangle_{CD_L}$ depends only on $\sigma_{CD}$, independent of the choice of extension $\rho_{BCD}$. In particular, the \emph{same} $U_D$ disentangles the global state $|\Psi\rangle$ as well as its reduced density matrices over $A'$. 

Therefore, all the modular Hamiltonians appearing in $\mathbb{L}$ associated with supersets of $CD$ can be disentangled using $U_D$. Without loss of generality, let $X$ be such a subsystem.
\begin{equation}
    K_X = \ln  \left(|\phi\rangle_{CD_L}\langle \phi| \right) + \ln  \rho_{D_R\cup(X\setminus CD)}. \label{eq:kx_logarithm}
\end{equation}
Therefore, $\mathbb{L}$ can be expressed as a sum of terms that act nontrivially on the complement of $CD_L$ and the terms that act on $CD_L$. The latter term is a multiple of $\ln  \left(|\phi\rangle_{CD_L}\langle \phi| \right)$, which acts trivially on $|\Psi\rangle$. Therefore, the reduced density matrix of $|\Psi\rangle$ over $CD_L$ is invariant under the good modular flow. In particular, the reduced density matrix over $C$ is invariant under the flow. Because $e^{-\ii\mathbb{L} t}|\Psi\rangle = e^{-\ii\mathbb{L}' t}|\Psi\rangle$ and we showed that the bulk reduced density matrix of $e^{-\ii\mathbb{L}' t}|\Psi\rangle$ is independent of $t$, we can also conclude that the bulk reduced density matrix of $e^{-\ii\mathbb{L} t}|\Psi\rangle$ is invariant. 

We remark that our argument for the invariance of the bulk reduced density matrices for non-infinitesimal flow is not rigorous. There are two key mathematical issues that need to be dealt with. First, Eq.~\eqref{eq:actual_op_identity_wm} may not hold even if Eq.~\eqref{eq:operator_identity_wm} is true. Second, Eq.~\eqref{eq:kx_logarithm} is unbounded because we are taking the logarithm of the rank-one projector $|\phi\rangle_{CD_L}\langle \phi|$. Making our argument into a rigorous argument that takes care of these subtleties is left for future work. We complement this shortcoming of our argument by providing strong numerical evidence for our claim in Section~\ref{sec:fidelity_tests}.

In summary, the argument we present in this section says that if we apply a good modular flow to the ground state, the resulting state
\begin{equation}\label{eq:invariant-subspace?}
    \ket{\Psi(t)} = e^{-\ii\Lb t}\ket{\Psi},\quad \Lb \in \mathfrak{g}
\end{equation}
will have only edge excitations, without any bulk excitations. This is because, as we argued, the bulk density matrix should not change. We will denote the subspace spanned by these states as $\mathcal{H}_{\text{edge}}$ from now on. 

\subsection{Fidelity tests}
\label{sec:fidelity_tests}
In this subsection, we provide numerical evidence for our claim that the bulk reduced density matrices are invariant under good modular flow. 

\subsubsection{Fidelity as a diagnostic of change under modular flow}\label{sec:fid-general} 

In order to assess how close two density matrices are, we employed the fidelity $F(\rho,\sigma)\equiv \Tr\left( \sqrt{\sqrt{\rho}\sigma\sqrt{\rho}} \right)$.  
For the ensuing discussion, recall the basic properties of fidelity: that $0\leq F(\rho,\sigma)\leq 1$ and $F(\rho,\sigma) = 1$ if and only if $\rho=\sigma$.  

Consider a flow (one-parameter family of unitary) $U(t) = e^{\ii Gt}$ on a state $\ket{\Psi}$ generated by a Hermitian operator $G$. We compute $F(\rho_X(0),\rho_X(t))$, where $\rho_X(t) = \Tr_{\overline{X}}\ket{\Psi(t)}\bra{\Psi(t)}$ and $\ket{\Psi(t)} = U(t)\ket{\Psi}$ to quantify the extent to which the local reduced density matrix of a region $X$ has been changed under the flow. 

We will also be interested in the change of the state under the flow $U(t) = e^{\ii Gt}$ for infinitesimal $t$. 
Provided that $F(\rho_X(0),\rho_X(t))$ is twice differentiable around $t=0$, we define 
\begin{equation}\label{eq:alpha}
    \alpha(G,X) \equiv  -\left. \frac{d^2}{dt^2} F(\rho_X(0),\rho_X(t)) \right|_{t=0}.
\end{equation}
The Taylor expansion around $t$ becomes: 
\begin{equation}\label{eq:f-smallt}
    F(\rho_X(0),\rho_X(t)) = 1 - \frac{1}{2}\alpha(G,X)t^2 + O(t^3).  
\end{equation}
The absence of the linear term comes from the fact that $F(\rho_X(0),\rho_X(t))\leq 1$. Otherwise, under an infinitesimal time $t$, $F(\rho_X(0),\rho_X(t))$ can exceed $1$. (For a similar reason, we can also conclude that $\alpha(G,X) \geq 0$.) Thus the single number $\alpha(G,X)$ accurately captures the small $t$ behavior of the fidelity\footnote{This is related to the fact that $\sqrt{1-F}$ is a metric; see Ref.~\cite{Fidelity-2008} and \emph{Bures distance}~\cite{Bures1969} for related discussions.} under the flow. 

In the following section, we shall numerically study a specific model and utilize these quantities, namely fidelity and $\alpha(G,X)$, to quantitatively study the change of local reduced density matrices various under modular flows.

\subsubsection{Numerical tests}

The main goal of our numerical tests is to verify our prediction in Section~\ref{sec:def-good-modular-flows}: that good modular flow at finite time preserves the bulk local density matrices.\footnote{Note that this statement holds in the limit in which the size of the involved subsystems are sufficiently large compared to the correlation length. For realistic systems, the invariance of the bulk local density matrix is only approximate.} For this purpose, we shall use the fidelity [Section~\ref{sec:fid-general}] to quantify the degree to which the bulk local density matrix changes. For concreteness, we will study the ground state of a $p+\ii p$ superconductor on a square lattice:
\begin{equation}\label{eq:pipH-2}
    H = \sum_{\vec{r},\vec{a}} \left[ -ta_{\vec{r}}^{\dagger}a_{\vec{r}+\vec{a}} + \Delta a_{\vec{r}}^{\dagger}a_{\vec{r}+\vec{a}}^{\dagger} e^{\ii \vec{a}\cdot \vec{A}}  + h.c. \right] - \sum_{\vec{r}}(\mu - 4t)a_{\vec{r}}^{\dagger}a_{\vec{r}}, 
\end{equation}
where $\vec{r} = (x,y)$ represents a site on the square lattice, and $\vec{a}$ is the lattice vector, taking values of $(1,0)$ and $(0,1)$ for positive $x$ and $y$ direction respectively.  
We set $\vec{A} = (0,\pi/2)$, so that $e^{\ii \vec{a}\cdot \vec{A}}$ is either $1$ or $\ii$. 
In the following test, we choose $t = 1.0,\Delta=1.0,$ and $\mu = 1.3$; this choice of parameter yields a relatively small correlation length (approximately $1.2$ lattice spacings). We impose an anti-periodic boundary condition on the $x$-direction so that no flux is threaded. 

We have numerically observed that the ground state of this model satisfies entanglement bootstrap axioms {\bf A0} and {\bf A1} approximately with error decaying with subsystem sizes; see Section~\ref{sec:axioms-test-p+ip} for details. We thus anticipate the conclusions made in the previous section to continue to hold, at least approximately. In particular, we expect the local bulk reduced density matrices to be approximately invariant under good modular flows.

\begin{figure}[!h]
    \centering
    \includegraphics[width=0.95\columnwidth]{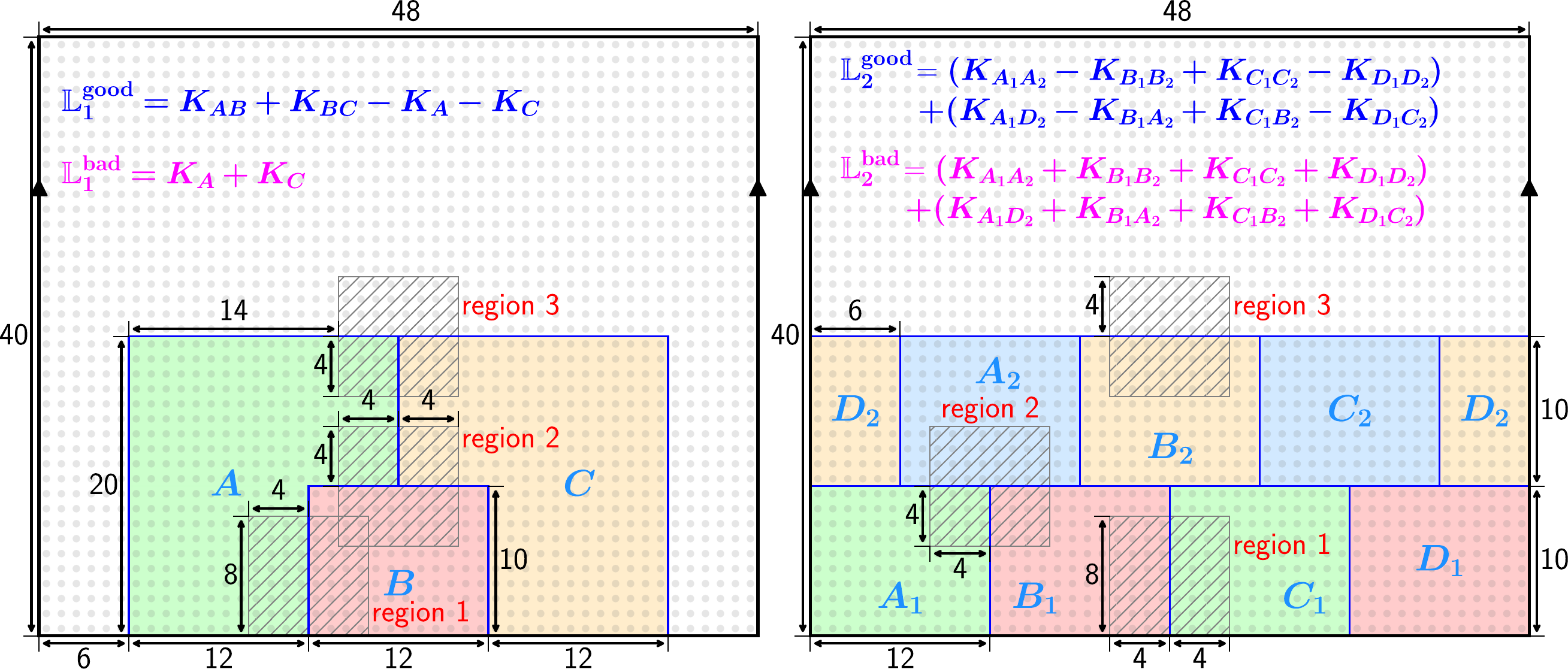}
    \caption{Setups for testing the good and bad modular flows. The left and right figure represents the setup for test 1 and test 2, respectively. Each test involves a choice of good (blue) and bad (pink) modular flow generators. The sizes and the locations of the subsystems can be inferred from the figure by noting that each dot represents a site of a square lattice.}
    \label{fig:fidelity-regions}
\end{figure}

Let us now discuss our setup [Figure~\ref{fig:fidelity-regions}]. We considered two types of good and bad modular flows, each shown on the left and the right side in Figure~\ref{fig:fidelity-regions}; we shall refer to these two different setups as ``test 1'' and ``test 2,'' respectively. In both cases, we chose three disk-like regions, indexed by $i=1, 2, 3.$ As $i$ increases, the region gets progressively farther away from the physical edge. In particular, $i=1$ corresponds to a region anchored at the physical edge. We computed $F(\rho_i(0),\rho_i(t))$ for $i=1, 2, 3$ in Figure~\ref{fig:fidelity-regions}, where $\rho_i(t)$ is the reduced density matrix of region $i$ at time $t$. 

\begin{figure}[!h]
    \centering
    \includegraphics[width=0.95\columnwidth]{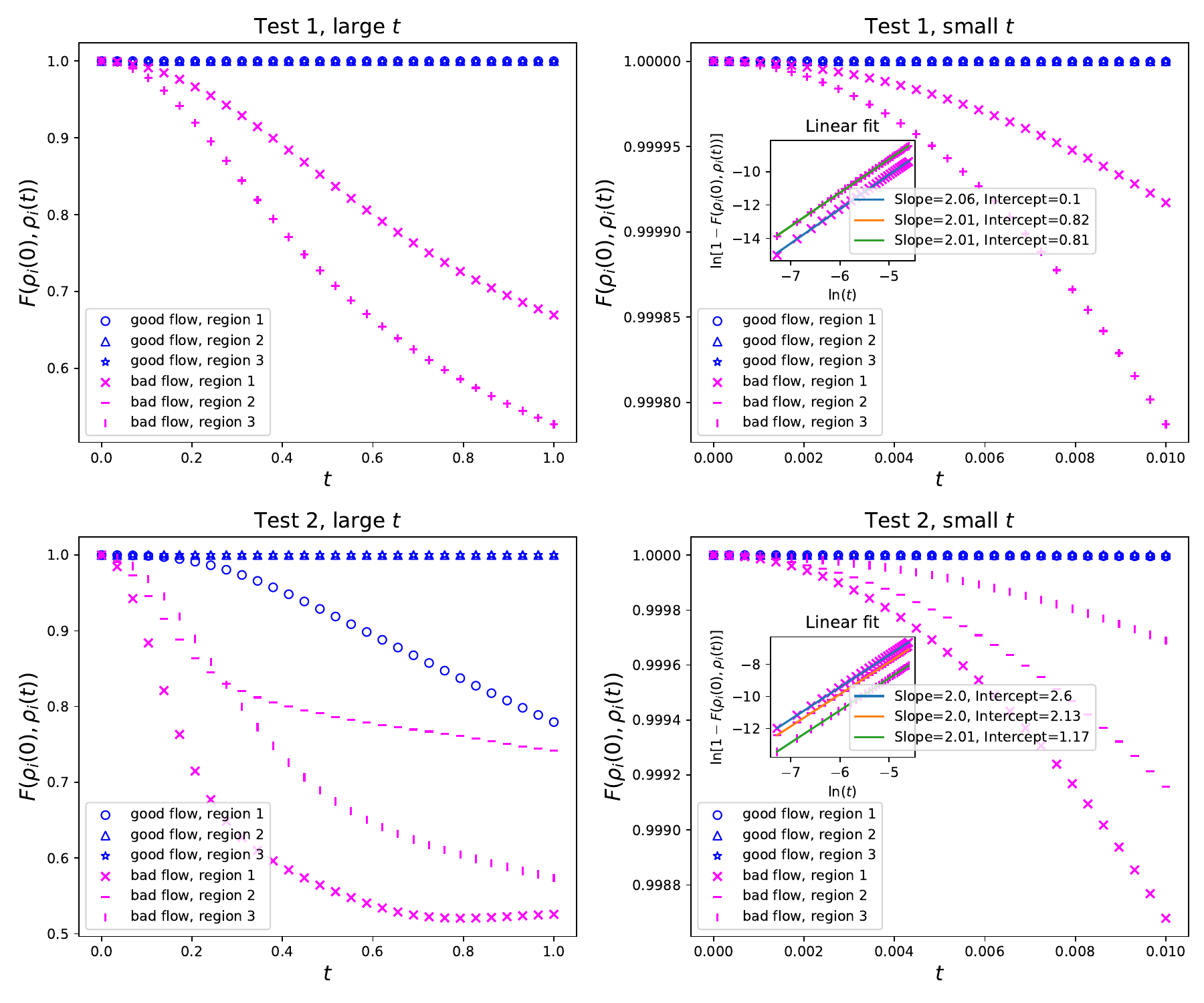}
    \caption{The fidelity tests associated with the regions shown in Fig.~\ref{fig:fidelity-regions}. For each test (each row), we have $F(\rho_i(0),\rho_i(t))$ for modular flow with a long evolution time (left) and a zoom-in to the short evolution time (right). For the short time data, we perform linear fit of $\ln\left( 1 -F(\rho_i(0),\rho_i(t)) \right)$ vs $\ln t$. From the intercepts, we compute the $\alpha$s for the associated bad modular flows: For test 1: $\alpha_1 = 1.10,\alpha_2 = 2.27,\alpha_3 = 2.25$. For test 2: $\alpha_1 = 13.41,\alpha_2 = 8.45,\alpha_3 = 3.23$.}  
    \label{fig:fidelity-nums}
\end{figure}

The fidelities are plotted in Figure~\ref{fig:fidelity-nums}. As expected, the reduced density matrices of the bulk regions (regions 2 and 3) remain invariant under the good modular flow, though they can change under the bad modular flow. On the other hand, for the reduced density matrices anchored at the physical edge (region 1), the good modular flow can change the fidelity. We remark that, for test 1, the fidelity of region 1 appears unchanged under the good modular flow. The origin of this invariance is unclear to us and we leave it as an open problem to explain this unexpected behavior. 

At small $t$, the fidelity can be approximated as 
\begin{equation}
    F(\rho_i(0),\rho_i(t)) \approx  1 - \frac{1}{2}\alpha_i \, t^2,
\end{equation} 
where $\alpha_i$ is equal to the $\alpha(G,X)$ are defined in Eq.~\eqref{eq:alpha}, with $G$ being the generators of the flow and $X$ being the disk-like region $i$. The coefficient $\alpha_i$ can be inferred from the intercept of a linear fit between $\ln (1-F(\rho_i(0), \rho_i(t)))$ and $\ln (t)$. We list the results of $\alpha_i$ in the caption of Figure~\ref{fig:fidelity-nums}. (For these tests, we estimated $\alpha_i$ only for the bad modular flows, since good modular flows did not change the density matrices for small $t$.) 
% =====================
% end of sec:good-modular-flows
% =====================

% =====================
% sec:chiral CFT assumption
% =====================
\section{Good modular flow generators and CFT}
\label{sec:chiral-cft}

The numerical results in Section~\ref{sec:fidelity_tests} showed that good modular flows do not alter the bulk reduced density matrices, corroborating our argument in Section~\ref{sec:def-good-modular-flows}. This suggests that there is an invariant subspace associated with good modular flows, which is a set of states that are locally indistinguishable from the reference state in the bulk. 
This subspace, $\mathcal{H}_{\rm edge}$, can therefore be thought of as a subspace formed by states with excitations that are localized at the edge. 

Now the natural question is how the good modular flows act on $\mathcal{H}_{\rm edge}$. To make progress, in the following, we shall employ an extra assumption (motivated by the chiral CFT description of the edge) which allows us to establish a dictionary between the good modular flow generators and the Virasoro generators of a 1+1D chiral CFT.  
More specifically, we apply these concepts to gapped ground states with purely-chiral gapless edges, laying the groundwork for the construction of Virasoro generators in Section~\ref{sec:virasoro-construction}. 

We shall study the following subclass of good modular flow generators, denoted as $\mathfrak{g}^{\bullet}$:

\begin{definition}
\label{def:gdot}
We define $\mathfrak{g}^{\bullet}$ as the subclass of good modular flow generators [Definition~\ref{definition:good_modular_flow_succinct}] of the form $\mathfrak{g}\ni\mathbb{L} = \sum_{A} \lambda_A K_A$, where every subsystem $A$ appearing in the summation is \emph{a disk-like region that intersects with at most one single interval on the edge.} 
\end{definition}
\noindent

By definition, $\mathfrak{g}^{\bullet} \subset \mathfrak{g}$. 
In Table~\ref{tab:gc-examples}, we shall see that $\mathfrak{g}^{\bullet} \ne \mathfrak{g}$. There we list examples of good modular flow generators, delineating the difference between the generators that are in $\gc$ and those that are not.

\begin{table}[!h]
    \centering
    \begin{tabular}{|l|}
        \hline 
        Good modular flow generators and their types \\ \hline 
        $\includegraphics[width=3cm,margin=0.2cm,valign=c]{flow-gen3.pdf} \quad \begin{aligned}
            &\hat{I}(A:C) = K_{A}+K_{C}-K_{AC} &&\notin \gc \\ 
            &\hat{I}(A:C|B) = K_{AB}+K_{BC}-K_{ABC}-K_{C} &&\notin \gc
        \end{aligned}$  \\ \hline 
        $\includegraphics[width=3cm,margin=0.2cm,valign=c]{flow-gen4.pdf} \quad \begin{aligned}
            &\hat{I}(A:C) = K_{A}+K_{C}-K_{AC} &&\notin \gc \\ 
            &\hat{I}(A:C|B) = K_{AB}+K_{BC}-K_{ABC}-K_{C} &&\in \gc \\ 
            &\hat{\Delta}(AA'CC',B,\emptyset) =  K_{AA'CC'B}+K_{B}-K_{AA'BCC'} &&\notin \gc \\ 
            &\hat{\Delta}(AA',B,CC') = K_{AA'B}+K_{CC'B}-K_{AA'}-K_{CC'} &&\in \gc 
        \end{aligned}$  \\ \hline 
    \end{tabular}
    \caption{Examples of good modular flow generators. Some of them are in $\gc$ and some are not. We remark that not all linear combinations of modular Hamiltonians of the form $\hat{I}(A:C)$ or $\hat{I}(A:C|B)$ are good modular flow generators; see for example Figure~\ref{fig:good-bad-examples} and Table~\ref{tab:flow-examples} second column.}
    \label{tab:gc-examples}
\end{table}

At a high level, the main motivation for defining $\gc$ is to focus on the good modular flow generators that can be identified with the modular Hamiltonians of CFT on an interval, for which we have a good understanding~\cite{Cardy:2016fqc}. For the good modular flow generators outside of $\gc$, it is less clear if such an identification holds because the corresponding subsystems in the CFT would contain more than one interval. The CFT modular Hamiltonians of such subsystems are not local~\cite{Casini2009,Klich2017,Arias2018}, posing a challenge for making such an identification. 

As we shall soon see, by hypothesizing a certain relationship between the good modular flow generators and the modular Hamiltonians of CFT, we can identify the good modular flow generators as weighted integrals of the stress-energy tensor in CFT. Moreover, this will lead to several testable predictions which we verify, thereby supporting our hypothesis.

\subsection{CFT Perspective}
We shall focus on the low-energy physics of (2+1)D systems which are gapped and chiral in the bulk. Such a system has robust gapless edge, which is expected to be described by a chiral CFT~\cite{Halperin1982,Witten:1988hf,Wen1991}.
Due to the gravitational anomaly of the chiral CFT, this theory cannot be realized on a stand-alone 1D system with a tensor product Hilbert space structure~\cite{Witten:1988hf,Hellerman:2021fla}. However, the chiral CFT is well-defined as an edge of a 2+1D system due to the cancellation of the anomaly~\cite{Callan:1984sa,Witten:1988hf}. 

From the CFT perspective, it is natural to expect the modular Hamiltonian of the edge to have a particularly simple form. It is well-known that for a 1+1D CFT (without gravitational anomaly) on a circle\footnote{The circle can have some radius $R$. The form of the modular Hamiltonian in Eq.~\eqref{eq:modH-CFT} in angular coordinates does not depend on $R$, because of conformal symmetry.}, the modular Hamiltonian of the ground state $|\Omega\rangle$ over an interval $[\theta_1, \theta_2]$ is of the following form \cite{Cardy:2016fqc}: 
\begin{equation} \label{eq:modH-CFT}
    \Kcft_{[\theta_1,\theta_2]}  = \int_0^{2\pi}d\theta\beta^{[\theta_1,\theta_2]}(\theta) \(\Tcft(\theta) + \bar{\Tcft}(\theta) \) + \alpha_{[\theta_1,\theta_2]}\mathbb{1},
\end{equation}
where $\alpha_{[\theta_1.\theta_2]}$ is a UV-dependent constant, $\beta^{[\theta_1,\theta_2]}(\theta)$ is 
\begin{equation}\label{eq:coolness-function}
    \beta^{[\theta_1,\theta_2]}(\theta) =  2 \Theta(\theta-\theta_1) \Theta(\theta_2-\theta) \frac{\sin(\frac{\theta-\theta_1}{2}) \sin(\frac{\theta_2-\theta}{2})}{\sin(\frac{\theta_2-\theta_1}{2})} ,
\end{equation}
and $\Tcft(\theta), \bar{\Tcft}(\theta)$ are the holomorphic and anti-holomorphic components of the stress-energy tensor, respectively. (Both $\Tcft(\theta)$ and $\bar{\Tcft}(\theta)$ are hermitian.)
$\Theta(x)$ is the Heaviside function, namely $\Theta(x)=1$ for $x\ge 0$, and $\Theta(x)=0$ for $x<0$.
We will refer to $\beta^{[\theta_1,\theta_2]}(\theta)$ as the \emph{coolness function}\footnote{Here we consider the CFT is on a circle. If the CFT is on an infinite line, for an interval $[x_1,x_2]$, the coolness function takes the form $\beta^{[x_1,x_2]}(x) = 2\Theta(x-x_1)\Theta(x_2-x)\frac{(x-x_1)(x_2-x)}{x_2-x_1}$.}.

For a \emph{purely-chiral} CFT, by which we mean a 1+1D CFT with only holomorphic components, we expect the modular Hamiltonian to take the form: 
\begin{equation}\label{eq:modH-chiCFT} 
\Kcft_{[\theta_1,\theta_2]}^{\chi} = \int_0^{2\pi} d\theta\beta^{[\theta_1,\theta_2]}(\theta) \Tcft(\theta) + \alpha'_{[\theta_1,\theta_2]}\mathbb{1},
\end{equation}
where the upper index $\chi$ signifies the fact that we are talking about a purely-chiral CFT.

While Eq.~\eqref{eq:modH-chiCFT} is a well-defined object in CFT, as we mentioned, there is an obstruction to defining a chiral CFT on a stand-alone 1+1D system. However, there are certain 2+1D states with an energy gap in the bulk, whose gapless edge is described by such purely-chiral CFTs. We shall also refer to these 2+1D states as ``purely-chiral''. Next, we put forward a hypothesis that quantitatively connects good modular flow generators in $\gc$ from a purely-chiral 2+1D state to the modular Hamiltonians in a purely-chiral CFT. 
For concreteness, let $\mathcal{H}_{\text{edge}}$ be the low-energy subspace which is spanned by the states obtained from good modular flows [Section~\ref{sec:good-modular-flows}].

\begin{figure}[h]
    \centering
    \begin{tikzpicture}
    \begin{scope}
    \clip (0, 0) circle (2cm);
    \fill[blue!30!white] (1.75cm, 0) circle (1cm);
  \end{scope}
    \draw[] (0,0) circle (2cm);
    \draw[very thick, red] (30:2cm) arc (30:-30:2cm);
    \node[] () at (1.35cm, 0) {$A$};
    \node[] (ba) at (3.5cm, 0) {$b(A)$};
    \draw[->] (ba) -- (2cm, 0);
        
    \end{tikzpicture}
    \caption{The edge interval $b(A)$ (red) corresponding to a disk $A$ (blue).}
    \label{fig:bulk_region_vs_boundary_region}
\end{figure}
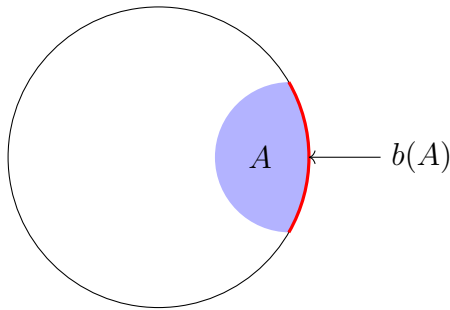

\begin{hypothesis}[Action of good modular flow generators for purely-chiral states]
\label{assump:chiral-CFT}
Consider a purely chiral reference state $\ket{\Psi}$ on a two-dimensional disk.  Let $\mathbb{L} = \sum_A \lambda_A K_A \in g^{\bullet}$. There exists an isometry $\mathbb{V}: \mathcal{H}_{\rm CFT} \to \mathcal{H}_{\rm edge}$ (i.e., $\mathbb{V}^{\dagger}\mathbb{V} = \mathbb{1}$) such that 
\begin{equation} \label{eq:conj:chiral-CFT}
    \mathbb{L} \mathbb{V} = \mathbb{V}\left( \sum_A \lambda_A  \Kcft_{b(A)}^{\chi} + \alpha''(\mathbb{L})\mathbb{1} \right) \quad \text{ and } \quad |\Psi\rangle = \mathbb{V} |\Omega\rangle%_{\rm CFT}
\end{equation}
where $\Kcft_{b(A)}^{\chi}$ is the modular Hamiltonian of the CFT ground state $\ket{\Omega}$ for the interval $b(A)$ obtained from the intersection of $\partial A$ with the edge 
[see Figure~\ref{fig:bulk_region_vs_boundary_region}]. $\alpha''(\mathbb{L})$ is a constant.
\end{hypothesis} 
\noindent
We will also use the following short-hand notation: 
\begin{equation}
    \mathbb{L} \veq  \sum_A \lambda_A  \Kcft_{b(A)}^{\chi} + \alpha''(\mathbb{L})\mathbb{1} \quad \textrm{is short for} \quad \mathbb{L} \mathbb{V} =  \mathbb{V}\left( \sum_A \lambda_A  \Kcft_{b(A)}^{\chi}+ \alpha''(\mathbb{L})\mathbb{1}\right).
\end{equation}

A few remarks are in order. (1) Firstly, although the hypothesis may seem to be limited to expressions linear in $\mathbb{L}$, this is actually not the case. Because $\mathbb{V}$ is an isometry, the hypothesis immediately implies the following identity:  
\begin{equation}\label{eq:identity_polynomial}
    \bra{\Psi} f(\{\Lb_i \in \gc\})\ket{\Psi}  = \bra{\Omega} f(\{ L^{\chi\mathrm{CFT}}_i \}) \ket{\Omega},\quad \gc \ni \Lb_i \veq L^{\chi\mathrm{CFT}}_i
\end{equation}
for any (multi-variable) polynomial $f$. 
(2) Secondly, we note that the hypothesis does not say anything about how the individual $K_A$ are related to the CFT modular Hamiltonians. One may think that the hypothesis is demanding that $K_A \buildrel{\mathbb{V}?}\over{=}\Kcft_{b(A)}^{\chi} + \alpha''(A)\mathbb{1}$ for every disk $A$, but this is not the case. Rather, the hypothesis tells us only how to map elements of $g^{\bullet}$ \footnote{A stronger hypothesis we can make is to replace $\mathfrak{g}^\bullet$ with $\mathfrak{g}$. We will not need this stronger version in this paper.} to a linear combination of CFT modular Hamiltonians. (3) Thirdly, the constant $\alpha''(\mathbb{L})$ of \eqref{eq:conj:chiral-CFT} is necessary and non-trivial in some cases. One example of a good modular flow generator with such non-zero $\alpha''$ is shown in the second row in Table~\ref{tab:flow-examples}, where the constant term $\alpha''$ comes from the topological entanglement entropy \cite{Kitaev2006,Levin2006} (see also \cite{shi2020fusion}) of bulk disks.\footnote{We suspect the only source of contribution to $\alpha''$ is topology dependent.} 
(4) Moreover, by applying the good modular flow generators in $\mathfrak{g}^\bullet$ to $|\Psi\rangle$, one can obtain states of the form of $\mathbb{V}|\Omega'\rangle$, where $|\Omega'\rangle$ is some excited state of the CFT. 
Physically, we can expect $\mathbb{V}$ to map states of the CFT to $\mathcal{H}_{\rm edge}$. In particular, because the CFT Hilbert space is infinite-dimensional, this isometry can only exist in the limit in which the $\mathcal{H}_{\rm edge}$ is an infinite-dimensional Hilbert space. If $\mathcal{H}_{\rm edge}$ is finite-dimensional, such as some purely-chiral reference state realized in a finite lattice system we shall numerically study later, we anticipate $\mathbb{V}$ to map a certain regularized subspace of $\mathcal{H}_{\rm CFT}$ to $\mathcal{H}_{\rm edge}$. (5) Lastly, we remark that due to the Stinespring dilation theorem \cite{Stinespring_1955}, the existence of isometry can be reformulated as the existence of a quantum channel that maps a set of operators acting on $\calH_{\mathrm{edge}}$ to the ones on $\calH_{\mathrm{CFT}}$.

As a consequence of this hypothesis, we can establish a relationship between good modular flow generators in $\mathfrak{g}^\bullet$ and the stress-energy tensor of the purely-chiral CFT:
\begin{equation}\label{eq:L-T-k}
    \Lb \veq \sum_A \lambda_A \, \Kcft^{\chi}_{b(A)} + \alpha''(\mathbb{L}) \mathbb{1}    
    =   \int_{0}^{2\pi}f(\theta) \Tcft(\theta) + \kappa(\mathbb{L}) \mathbb{1}, 
\end{equation}
where $\kappa(\mathbb{L})$ is a real constant (depending on $\mathbb{L}$) and $f(\theta)$, which will be called a \emph{weight function}, is a linear combination of the coolness functions $\beta^a(\theta)$ of intervals $a$ with the coefficients appearing in the good modular flow generator:  
\begin{equation}
    f(\theta) = \sum_A \lambda_A \beta^{b(A)}(\theta).
\end{equation}
Eq.~\eqref{eq:L-T-k} is the main starting point of our ensuing analysis. This is an important identity that establishes a correspondence between good modular flow generators and the CFT stress-energy tensor for purely chiral states.

In the remainder of this paper, we will study many good modular flow generators, whose corresponding weight functions $f(\theta)$ are of special interest to us. One particularly simple case is when the weight function in Eq.~\eqref{eq:L-T-k} vanishes. This will result in an equation (which we call the \emph{vector fixed-point equation}) that we can verify using means independent of Hypothesis~\ref{assump:chiral-CFT}. We discuss this equation in Section~\ref{sec:locally_checkable} and provide an alternative argument for its derivation in Section~\ref{subsec:cylinder-argument}.

\begin{remark}
    Hypothesis~\ref{assump:chiral-CFT} implies that the behavior of a good modular flow for any $t \in \mathbb{R}$ matches a certain CFT prediction. Note, however, Hypothesis~\ref{assump:chiral-CFT} is intended for systems in the thermodynamic limit.
    For systems with finite sizes (and with finite onsite Hilbert dimensions), we expect the match to hold approximately only within a finite time range $t \in (-t^\ast, t^\ast)$, where $t^\ast$ depends on the system size and parameters. At larger $t$, the modular flow behavior depends on the UV and is not universal. 
    We will not study such UV phenomena in this paper. 
\end{remark}

\subsection{Vector fixed-point equation}
\label{sec:locally_checkable}

A straightforward consequence of Hypothesis~\ref{assump:chiral-CFT} is that there is a nontrivial identity that certain good modular flow generators ought to satisfy. Consider a 1+1D CFT ground state $|\Omega\rangle$. Let $a, b, $ and $c$ be three contiguous intervals. It was shown in \cite{Lin2023} that
\begin{equation}\label{eq:op-id-CFT}
    \eta \hat{\Delta}^{\rm CFT}(a,b,c) + (1-\eta)\hat{I}^{\rm CFT}(a:c|b)  \propto \mathbb{1},
\end{equation}
where $\hat{\Delta}^{\mathrm{CFT}}(a,b,c) \equiv \Kcft_{ab}+\Kcft_{bc}- \Kcft_{a}- \Kcft_c, \hat{I}^{\CFT}(a:c|b)\equiv \Kcft_{ab}+ \Kcft_{bc}- \Kcft_{abc}- \Kcft_b$, 
and $\eta$ is the cross-ratio\footnote{Explicitly, $\eta = \frac{|a| |c|}{|ab| |bc|}$, where on an infinite line $|a|$ is the length of edge interval $a$, and on a circle $|a|$ is the chordal distance associated with arc $a$.}.
Eq.~\eqref{eq:op-id-CFT} follows from the form of the coolness function, as observed in \cite{Lin2023}. In particular, the derivation of this equation can be generalized straightforwardly to purely chiral CFTs.

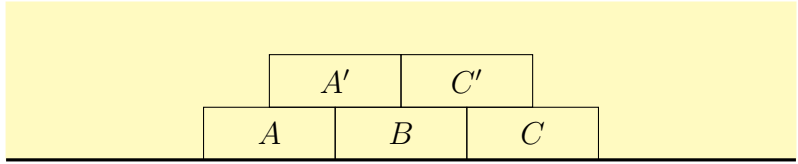
\begin{figure}[h]
    \centering
    \begin{tikzpicture}[scale=0.35]
    \filldraw[yellow!30!white] (-15cm, 0) -- (15cm ,0)  -- ++ (0, 6cm) -- (-15cm, 6cm)-- cycle;
        \draw[very thick] (-15cm, 0) -- (15cm, 0);
        \draw[] (-7.5cm, 0) -- ++ (5cm, 0) -- ++ (0, 2cm) -- ++ (-5cm, 0) -- cycle;
        \draw[] (-2.5cm, 0) -- ++ (5cm, 0) -- ++ (0, 2cm) -- ++ (-5cm, 0) -- cycle;
        \draw[] (2.5cm, 0) -- ++ (5cm, 0) -- ++ (0, 2cm) -- ++ (-5cm, 0) -- cycle;

        \draw[] (-5cm, 2cm) -- ++ (5cm, 0) -- ++ (0, 2cm) -- ++ (-5cm, 0) -- cycle;
        \draw[] (0cm, 2cm) -- ++ (5cm, 0) -- ++ (0, 2cm) -- ++ (-5cm, 0) -- cycle;

        \node[] () at (-5cm, 1cm) {$A$};
        \node[] () at (0cm, 1cm) {$B$};
        \node[] () at (5cm, 1cm) {$C$};
        \node[] () at (-2.5cm, 3cm) {$A'$};
        \node[] () at (2.5cm, 3cm) {$C'$};
    \end{tikzpicture}
    \caption{Subsystems used in the construction of $\hat{I}$ and $\hat{\Delta}$ appearing in Eq.~\eqref{eq:KD-in-I-Delta}. The yellow region represents the bulk and the thick bold line represents the physical edge. }
    \label{fig:KD_eta}
\end{figure}

Then by Hypothesis~\ref{assump:chiral-CFT}, we obtain a vector fixed-point equation for a purely chiral 2+1D state $|\Psi\rangle$; see Eq.~\eqref{eq:bdy-vector-fixed-point-equation}. Explicitly, for the regions $AA'B CC'$ shown in Figure~\ref{fig:KD_eta}, we consider
 a particular linear combination of modular Hamiltonians:
\begin{equation}\label{eq:KD-in-I-Delta}
    \KD(\eta) \equiv \eta \hat{\Delta}(AA',B,CC') + (1-\eta) \hat{I}(A:C|B), 
\end{equation} 
where $\hat{\Delta}(AA',B,CC') \equiv K_{AA'B}+K_{CC'B}-K_{AA'}-K_{CC'}$, and $\hat{I}(A:C|B)\equiv K_{AB}+K_{BC}-K_{ABC}-K_B$. Here $\eta$ is the cross-ratio computed from the length of the edge interval.  Note that 
\begin{equation}
    \hat{\Delta}, \hat{I} \in \mathfrak{g}^\bullet \quad \Rightarrow \quad  \KD(\eta) \in   \mathfrak{g}^\bullet. 
\end{equation}
In other words, $\KD(\eta)$ is a good modular flow generator.
It is a very special one, because by Hypothesis~\ref{assump:chiral-CFT} we have 
\begin{equation}
    \KD(\eta)  \overset{\mathbb{V}}{\propto}  \mathbb{1} ,
\end{equation}
implying that $\KD(\eta)$ acts trivially on the invariant subspace $\mathcal{H}_{\rm edge}$, that is,
\begin{equation}
\label{eq:bdy-vector-fixed-point-equation}
\boxed{    \KD (\eta)|\Psi\rangle \propto |\Psi\rangle. }
\end{equation}
Eq.~\eqref{eq:bdy-vector-fixed-point-equation} is the \emph{vector fixed-point equation} (for a chiral edge), which we study in the rest of this Section.

While Eq.~\eqref{eq:bdy-vector-fixed-point-equation} is a priori only a \emph{necessary condition} for Hypothesis~\ref{assump:chiral-CFT} to be true, it has the advantage of being \emph{locally checkable}. This is due to the equivalence of the following two statements:
\begin{equation}
\label{eq:locally_checkable}
    \KD(\eta)\ket{\Psi} \propto \ket{\Psi}  \quad \Leftrightarrow \quad \KD(\eta) \rho_{AA'BCC'} \propto \rho_{AA'BCC'},
\end{equation}
where $\rho_{AA'BCC'}$ is the reduced density matrix of $|\Psi\rangle$. Note that the right-hand side of Eq.~\eqref{eq:locally_checkable} is formulated entirely in terms of local reduced density matrices. As such, this condition can be verified locally. 

Now let us provide the proof of Eq.~\eqref{eq:locally_checkable}. The $\Rightarrow$ direction is obvious as one can simply trace out the complement of $AA'BCC'$ on both hand side of $\KD(\eta)\ket{\Psi}\bra{\Psi} \propto \ket{\Psi}\bra{\Psi}$. For $\Leftarrow$ direction, one can first consider a purification of $\rho_{AA'BCC'}$, denoted as  $\ket{\Psi'}$, to get $\KD(\eta)\ket{\Psi'}\propto \ket{\Psi'}$. 
By Uhlmann's theorem~\cite{Uhlmann1976}, there is an isometry acting on the complement of $AA'BCC'$ that maps $\ket{\Psi'}$ from $\ket{\Psi}$. Because this unitary commutes with $\KD(\eta)$ by construction, we obtain the original vector equation. Therefore, the vector fixed-point equation can be verified from the local reduced density matrix of $|\Psi\rangle$ over $AA'BCC'$.

We note that our Hypothesis~\ref{assump:chiral-CFT} assumes that the bulk state is purely chiral.  The vector fixed-point equation, however, applies more generally.  We expect that a suitable generalization of Hypothesis~\ref{assump:chiral-CFT} to non-purely-chiral states would also imply the vector fixed-point equation.

Currently, we do not know if the vector fixed-point equation implies Hypothesis~\ref{assump:chiral-CFT} in the case of purely-chiral states. While we leave this question for future work, we make some further remarks on it in Section~\ref{subsec:cylinder-argument}.

\subsection{Analytical evidence for Hypothesis \ref{assump:chiral-CFT}}
\label{subsec:cylinder-argument}

The purpose of this section is to give evidence for Hypothesis~\ref{assump:chiral-CFT}.  We do this by first proving the vector fixed point equation under an assumption that we call the ``cylinder assumption''.   
Then we ask about the extent to which the vector fixed point equation and its operator version imply Hypothesis~\ref{assump:chiral-CFT}. 

{\bf Analytical evidence for the vector fixed-point equation.}
In Section~\ref{sec:locally_checkable}, we introduced a straightforward consequence of Hypothesis~\ref{assump:chiral-CFT}, which we referred to as the vector fixed-point equation. In this subsection, 
we use physics arguments to derive the vector fixed-point equation.
A certain reformulation of this condition will be used centrally in our upcoming work~\cite{q-cross-ratio}. 
We then discuss the strength of such fixed-point equations, and the extent to which they in turn imply Hypothesis~\ref{assump:chiral-CFT}.  
This subsection thereby provides nontrivial evidence for Hypothesis~\ref{assump:chiral-CFT}.

Because we will use an independent physics argument to derive the vector fixed-point equation, we emphasize that the assumptions used in this Section (and its associated Appendix~\ref{appendix:cylinder-argument}) are different from the rest of the paper. Our starting point is the idea that a (possibly chiral) 2+1D topological phase on a cylinder is a (non-chiral) 1+1D CFT. This assumption lets us use the entanglement structure of a 1+1D CFT ground state and the bulk assumptions [Section~\ref{sec:primer_eb}] to learn about the entanglement structure of the 2+1D state altogether.  We will call this line of thought ``the cylinder argument''.

\begin{figure}[h]
    \centering
    \begin{tikzpicture}[scale=2]
    \begin{scope}

  \draw[thick, red, -latex] (6mm, 0) arc
  (0:180:6mm and 1.9mm);

   \draw[thick, red, -latex] (6mm, -1.25cm) arc
  (360:180:6mm and 1.9mm);
  
  \draw [very thick]
  (180:7.5mm) coordinate (A)
  -- ++(0,-12.5mm) coordinate (B) node [midway, right, inner sep=1pt] {}
  arc (180:360:7.5mm and 2.625mm) coordinate (D)
  -- (A -| D) coordinate (C) arc (0:180:7.5mm and 2.625mm);
  \draw [very thick]
  (0,0) coordinate (T) circle (7.5mm and 2.625mm);
  \draw [densely dashed] (D) arc (0:180:7.5mm and 2.625mm);
\begin{scope}[xshift=0.25cm, yshift=-0.2cm]
    \draw [thick]
  (180:7.5mm)   -- ++(0,-12.5mm);
  \node[] () at (-0.6cm, -0.625cm) {$A$};
\end{scope}
\begin{scope}[xshift=0.55cm, yshift=-0.25cm]
    \draw [thick]
  (180:7.5mm)   -- ++(0,-12.5mm);
\end{scope}
\begin{scope}[xshift=-0.25cm, yshift=-0.2cm]
    \draw [thick]
  (0:7.5mm)   -- ++(0,-12.5mm);
  \node[] () at (0.6cm, -0.625cm) {$C$};
\end{scope}
\begin{scope}[xshift=-0.55cm, yshift=-0.25cm]
    \draw [thick]
  (0:7.5mm)   -- ++(0,-12.5mm);
\end{scope}
\node[] () at (0, -0.85cm) {$B$};
    \end{scope}
\node[] () at (1.75cm, -0.625cm) {$\longrightarrow$};
    \begin{scope}[xshift=3cm, yshift=-0.625cm]
         \draw [very thick]
  (0,0) coordinate (T) circle (7.5mm and 2.625mm);
  \begin{scope}[xshift=0.25cm, yshift=-0.1375cm]
    \draw [thick]
  (180:7.5mm)   -- ++(0,-0.125cm);
  \node[] () at (-0.6cm, -0.25cm) {$A$};
\end{scope}
\begin{scope}[xshift=0.55cm, yshift=-0.1875cm]
    \draw [thick]
  (180:7.5mm)   -- ++(0,-0.125cm);
\end{scope}
\begin{scope}[xshift=-0.25cm, yshift=-0.1375cm]
    \draw [thick]
  (0:7.5mm)   -- ++(0,-0.125cm);
  \node[] () at (0.6cm, -0.25cm) {$C$};
\end{scope}
\begin{scope}[xshift=-0.55cm, yshift=-0.1875cm]
    \draw [thick]
  (0:7.5mm)   -- ++(0,-0.125cm);
\end{scope}
\node[] () at (0, -0.45cm) {$B$};
    \end{scope}
\end{tikzpicture}
    \caption{An illustration of the cylinder assumption. One can view a 2+1D gapped system with a chiral gapless edge mode (red) as a 1+1D CFT using dimensional reduction. For illustration purposes we draw a purely-chiral 2+1D system where we have either left or right moving mode for each edge (not both). The cylinder assumption works more generally.
    } 
    \label{fig:cylinder_argument}
\end{figure}
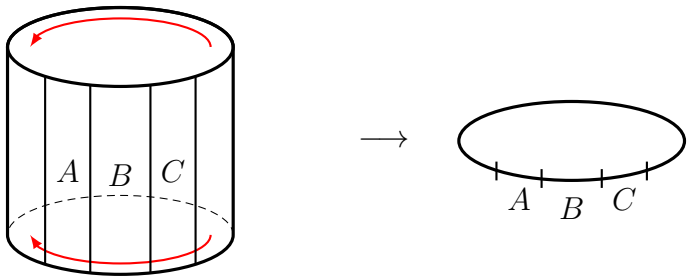

Consider a system with a large bulk gap on a cylinder.  
It is natural to regard the whole system as a 1+1D CFT [Figure~\ref{fig:cylinder_argument}]:
we dimensionally reduce along the direction normal to the boundary, meaning that we regard the degrees of freedom with the same value of the position along the boundary as lying at the same site of the 1D system.  The basic assumption is that the resulting system satisfies the vector fixed-point equation of \cite{Lin2023}, reviewed in Appendix \ref{appendix:cylinder-argument}, for regions of the form indicated on the right side of Figure~\ref{fig:cylinder_argument}.  We will call this statement (that the dimensionally reduced state satisfies the 1+1D vector fixed-point equation) the {\it cylinder assumption}. 

Taking the cylinder assumption as a starting point, with entanglement bootstrap axioms \textbf{A0} and \textbf{A1} plus two global assumptions, we indeed partially verify the ``dictionary'' provided by Hypothesis~\ref{assump:chiral-CFT}. Here we state the assumptions and results and defer proof details to Appendix \ref{appendix:cylinder-argument}.

\begin{figure}[t]
    \centering
    \begin{tikzpicture}[scale=0.5]

    \node[rectangle, draw=black] () at (5.5cm, 5.5cm) {Full boundary axioms};
        \draw[very thick, draw=black] (0,0) circle (4cm);
        \draw[fill=blue!30!white, dashed] (0,0) circle (3cm);
        \draw[fill=black!30!white, dashed] (0,0) circle (2cm);
        \draw[very thick, fill=white, draw=black] (0,0) circle (1cm);
        \node[] () at (0,1.5) {$C$};
        \node[] () at (0, 2.5) {$B$};
        \node[] () at (0, -5.5) {\textbf{A0}: $\Delta(B,C)_{\sigma}=0$};

        \begin{scope}[xshift=11cm]
        \draw[very thick] (0,0) circle (4cm);
        \draw[fill=blue!30!white, dashed] (0,0) circle (3cm);
        \draw[fill=black!30!white, dashed] (0,0) circle (2cm);
        \draw[fill=orange!30!white, draw=none] (90:2cm) -- (90:3cm) arc (90:-90:3cm) -- (-90:2cm) arc (-90:90:2cm);
        \draw[very thick, fill=white, draw=black] (0,0) circle (1cm);
        \draw[dashed] (0,2) -- ++ (0,1);
        \draw[dashed] (0,-2) -- ++ (0, -1);
        \node[] () at (-2.5, 0) {$B$};
        \node[] () at (0, 1.5) {$C$};
        \node[] () at (2.5, 0) {$D$};
        \node[] () at (0, -5.5) {\textbf{A1}: $\Delta(B, C,D)_{\sigma}=0$};
        \end{scope}
    \end{tikzpicture}
    \caption{Full-boundary versions of the axioms \textbf{A0} and \textbf{A1}. The solid lines represent the physical boundary and the dashed lines represent the entanglement cut.}
    \label{fig:full_boundary_axioms}
\end{figure}
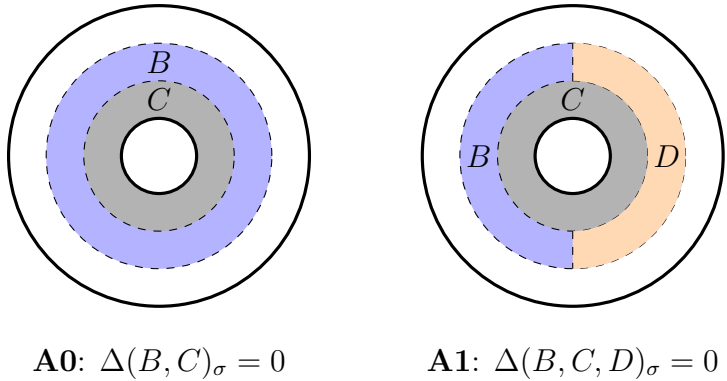

We introduce the two technical assumptions (similar to the axioms \textbf{A0} and \textbf{A1}) to guarantee that no nefarious person % nice!
has distributed a Bell pair between the two boundaries of the cylinder [Figure~\ref{fig:full_boundary_axioms}] so as to break the ensuing argument. These can be thought of as versions of \textbf{A0} and \textbf{A1} over regions that include the inner boundary of the cylinder. If the inner boundary is contracted to a point, this reduces to the usual \textbf{A0} and \textbf{A1}. Nonetheless, we note that these are extra assumptions that do not follow from \textbf{A0} and \textbf{A1}. To see why these assumptions are necessary, we note the following fact. If we impose conditions on disks that only include the bulk regions, we cannot exclude the possibility of having an end-point of a twist defect~\cite{Bombin2010} (or Majorana zero modes in fermionic models \cite{Read1999}) in the inner boundary. It is well-known that this leads to long-range mutual information~\cite{Brown2013,Kim2014-1D}. We are ruling out such pathological cases by imposing the full-boundary version of the axioms \textbf{A0} and \textbf{A1}.

The main result of this Section is the following proposition.
\begin{restatable}[Cylinder assumption gives boundary vector fixed-point equation]{Proposition}{cylindertovectorfixedpointeqn}\label{prop:cylinder-to-vector-fixed-point-eqn}
Given the cylinder assumption, bulk {\bf A1}, full-boundary {\bf A0} and full-boundary {\bf A1}, 
the 2+1D vector fixed-point equation 
\eqref{eq:bdy-vector-fixed-point-equation}
holds near each boundary independently.   
\end{restatable}
\noindent
We defer the proof of this Proposition to Appendix~\ref{appendix:cylinder-argument}.

\noindent{\bf Vector fixed-point equation hints at Hypothesis~\ref{assump:chiral-CFT}.} 
Having just explained that the vector fixed-point equation near the physical boundary follows from the cylinder assumption, we now discuss to what extent the vector fixed-point equation supports Hypothesis~\ref{assump:chiral-CFT}. We shall give a physical discussion here. Work on the question of the strength of the vector fixed-point equation (as well as its operator version) will be reported elsewhere \cite{strength-of-FPE}.

In fact, the vector fixed-point equation largely determines the form of the modular Hamiltonian of a disk $A$ that touches the gapless boundary along an interval $b(A)$, at least when it acts on the reference state $\ket{\Psi}$. Recall that a modular Hamiltonian $K_A$ only has non-trivial action on $\ket{\Psi}$ along its entanglement cut $\partial A$\footnote{Explicitly, $K_A\ket{\Psi} = O_{\partial A}\ket{\Psi}$. This only requires bulk {\bf A0}.}. 
If the reference state satisfies the vector fixed point equation for every set of three contiguous intervals, it has the following three consequences. (1) Vector fixed-point equations suggest that  
\begin{equation}\label{eq:KA-action}
    K_A\ket{\Psi} = \left( \int_{b(A)} d\theta \beta(\theta)\mathfrak{h}(\theta) + \calB_{\partial A}\right) \ket{\Psi},
\end{equation}
where $\calB_{\partial A}$ acts on $\partial A$ (the entanglement cut of $A$) inside the bulk, and will be canceled in a linear combination of modular Hamiltonians that gives a good modular flow generator.
(2) Moreover, the vector fixed-point implies that the weight function $\beta(x)$ takes the form of the coolness function \eqref{eq:coolness-function}. (3) Finally, the operator $\mathfrak{h}(\theta)$ in \eqref{eq:KA-action} is the ``reconstructed Hamiltonian density'' \cite{Lin2023} that governs the edge excitations. 

We make three further remarks regarding the reconstructed Hamiltonian density. Firstly, $\mathfrak{h}(\theta)$ is a suitable limit of a particular linear combination of modular Hamiltonians of small disks that touch the gapless edge at infinitesimal intervals near the point labelled $\theta$. The reconstructed Hamiltonian from the integral of $\mathfrak{h}(\theta)$ can be regarded as the continuous version of the reconstructed Hamiltonian in \cite{Lin2023}. Secondly, if one considers a lattice realization of $\ket{\Psi}$, then the Hamiltonian density shall be supported on the thickened edge. The thickness is inevitable at least in chiral state, since otherwise, one can obtain a lattice realization of chiral CFT, which is forbidden by gravitational anomaly. Lastly, we should point out that from the vector fixed-point equation (which only involves a linear action of modular Hamiltonians), it is not clear that $\mathfrak{h}(\theta)$ satisfies the operator product expansion (OPE) relation required for a stress-energy tensor of a CFT.  

% =====================
% end of sec:chiral CFT assumption
% =====================

\section{Coherent states}\label{sec:edge-subspace-coherent}

In the previous section, we introduced Hypothesis~\ref{assump:chiral-CFT}, relating the action of a good modular flow generator in $\gc$ on the reference state to a weighted integral of the stress-energy tensor on the CFT ground state. In this section, we discuss the general properties of states obtained from good modular flows. These states have the form $\ket{\Psi(t)} = e^{\ii \Lb t}\ket{\Psi}$ for some $t\in\mathbb{R}$, where $\Lb \in \gc$ is Hermitian. We shall refer to these states as \emph{coherent states} for the reasons we discuss in Section~\ref{subsec:virasoro_and_coherent_states}. The vector space spanned by such coherent states gives a representation of the Virasoro algebra. This representation is known as the ``identity module'' in literature \cite{francesco2012conformal}. 

\subsection{Virasoro algebra and coherent states}
\label{subsec:virasoro_and_coherent_states}

Consider a good modular flow generator $\Lb\in \gc$:
\begin{equation}
    \Lb \veq \int d\theta \,f(\theta) \Tcft (\theta).
\end{equation}
Define $|\Psi(t)\rangle \equiv e^{i\Lb t} |\Psi\rangle$. According to Hypothesis~\ref{assump:chiral-CFT}, this state is of the following form:
\begin{equation}
    \ket{\Psi(t)} = e^{\ii \Lb t}\ket{\Psi} = \mathbb{V} \exp\( \ii t\int d\theta \, f(\theta)\Tcft(\theta) \)\ket{\Omega},
\end{equation}
where $\ket{\Omega}$ is the CFT ground state. The state $\exp\left( \ii t\int d\theta \, f(\theta)\Tcft(\theta) \right)\ket{\Omega}$ is known as a \emph{coherent state} in the CFT literature \cite{Hollands2019}. As such, we shall refer to $|\Psi(t)\rangle$ also as a coherent state. (We remark that $|\Psi(t)\rangle$ is a state in 2+1 dimensions, though $\ket{\Omega}$ describes a state in 1+1 dimensions.)

In purely chiral CFTs, the group generated by $\left\{ \exp\left( \ii t\int f(\theta)\Tcft(\theta) \right)|f(\theta)\right\}$
forms a projective representation of the orientation-preserving diffeomorphisms over $S^1$, denoted as $\mathrm{Diff}_{+}(S^1)$ from now on. As such, in the context of 2+1D purely chiral state, we expect the good modular flow $e^{i\Lb t}$ to be an element of a (projective) representation of $\mathrm{Diff}_{+}(S^1)$. A natural Hilbert space to consider is $\calHc$, defined as:
\begin{equation}
    \calHc \equiv \mathrm{Span}\left\{ e^{\ii \Lb t}\ket{\Psi}| \Lb \in \gc,t\in\mathbb{R} \right\}.
\end{equation}
This is our carrier space for $\mathrm{Diff}_{+}(S^1)$. 

We now formally define the representation of the Virasoro generators with respect to $\calHc$. We define $L_n$ as an operator satisfying 
\begin{equation}\label{eq:def-Ln}
    L_n \veq \Lcft_n,\quad n\in \mathbb{Z},
\end{equation}
where $\mathbb{V}$ is the isometry appearing in Hypothesis~\ref{assump:chiral-CFT}. We shall discuss constructions of $L_n$ in Section~\ref{sec:virasoro-construction}. Following from Hypothesis~\ref{assump:chiral-CFT}, $L_n$ shall satisfy the Virasoro algebraic relation in $\calHc$, with the same charge $c$ as $\Lcft_n$.   

Let us make a few remarks. First, Eq.~\eqref{eq:def-Ln} does not define $L_n$ uniquely in the original Hilbert space of the 2+1D chiral system. This is because there can be different operators that satisfy Eq.~\eqref{eq:def-Ln} for the same value of $n$. However, within the subspace $\calHc$, these different choices act in an identical way. Specifically, consider $L_n$ and $L_n'$ such that $L_n\mathbb{V} = L_n'\mathbb{V} = \mathbb{V}\Lcft_n$. Then for any $|\Psi(t)\rangle = \mathbb{V}e^{\ii \Lcft t}|\Omega\rangle$, 
\begin{equation}
    L_n |\Psi(t)\rangle = L_n' |\Psi(t)\rangle = \mathbb{V} \Lcft_n e^{\ii \Lcft t}|\Omega\rangle.
\end{equation}
Second, $\mathbb{L}^k|\Psi\rangle \in \calHc$ for any $\mathbb{L}\in \gc$. This can be seen by taking the $n$'th order derivative of $e^{\ii \mathbb{L}t}|\Psi\rangle$ at $t=0$. In Section~\ref{sec:virasoro-construction}, we shall see that $L_n$ can be approximated arbitrarily well by a linear combination of various good modular flow generators. Therefore, $\calHc$ is closed under applications of $L_n$s.

Now, we can understand the action of a good modular flow generator in $\gc$ in terms of the Virasoro algebra (up to the isometry $\mathbb{V}$).  
Recall the mode expansion of the stress-energy tensor: 
\begin{equation}\label{eq:L-T}
    \Lcft_n = \frac{1}{2\pi} \int_0^{2\pi} d\theta e^{\ii n\theta} \Tcft(\theta) \quad \text{and} \quad \Tcft(\theta) = \sum_{n\in\mathbb{Z}} e^{-\ii n\theta}\Lcft_n. 
\end{equation}
By Hypothesis~\ref{assump:chiral-CFT}, any good modular flow generator in $\gc$ can be expressed in terms of a weighted integral of the stress-energy tensor:
\begin{equation}\label{eq:Lb-fourier}
    \Lb \veq \int_0^{2\pi} d\theta \, f(\theta)\Tcft(\theta) = \sum_{n\in\mathbb{Z}}\lambda_n \Lcft_n,\quad \lambda_n = \int_0^{2\pi} d\theta \, f(\theta)e^{-\ii n \theta}.
\end{equation}
Recalling $L_n\veq \Lcft_n$, we can write $\Lb$ in Eq.~\eqref{eq:Lb-fourier} as 
\begin{equation}\label{eq:L-expand}
    \Lb = \sum_{n\in\mathbb{Z}} \lambda_n L_n. 
\end{equation}
Therefore, the action of $\Lb$ on $\ket{\Psi}$ is essentially a linear combination of Viraoro generators acting on the gapless edge. Moreover, a coherent state can also be expressed as 
\begin{equation}\label{eq:coherent-L}
    \ket{\Psi(t)} =e^{\ii \Lb t}\ket{\Psi} = \exp\( \ii t \sum_{n\in\mathbb{Z}} \lambda_n L_n \)\ket{\Psi}.
\end{equation}

We wish to remark on an important property of the Fourier coefficient $\lambda_n$, namely, $|\lambda_n|$ decays as $O(|n|^{-3})$. This follows because the weight function $f(\theta)$ is piecewise\footnote{The precise definition of piecewise $C^m$ is given in Appendix \ref{sec:smooth-fourier}.} $C^2$ by the construction of good modular flow. By contrast, if the action of a bad modular flow on the edge could be written as a weighted integral of the stress tensor (as in Hypothesis~\ref{assump:chiral-CFT} for good modular flow), the weight function would be piecewise $C^1$, which leads to $O(|n|^{-2})$ decay of its Fourier coefficients.
(See Lemma~\ref{lemma:fourier-discontinuity} for the detailed proof.) As we shall discuss below, the Fourier coefficients of the modular flow generators play an important role in computing several quantities, such as the variance of the generator and the energy of the edge excitations. These quantities are finite for good modular flow, thanks to the $O(|n|^{-3})$ decay of the Fourier coefficients for the generators. However, these quantities diverge for bad modular flow, since the Fourier coefficients of the generators are $O(|n|^{-2})$ decay.  

Using this expression Eq~\eqref{eq:coherent-L}, one can obtain several analytic properties of the coherent states, which we  discuss in the following two subsections.

\subsection{State overlap}
\label{subsec:state-difference}

In this subsection, we compare $\ket{\Psi(t)} = e^{\ii \Lb t}\ket{\Psi},
\Lb\in\gc$ and $\ket{\Psi}$. The difference between $\ket{\Psi}$ and $\ket{\Psi(t)}$ can be characterized by the overlap $|\langle \Psi(t)|\Psi\rangle|^2$. The behavior for small $t$ is
\begin{equation}
    |\langle \Psi(t)|\Psi\rangle|^2 = 1 - \sigma(\Lb)^2 t^2 + O(t^4),
\end{equation}
where $\sigma(\Lb)^2$ is the variance of the operator $\Lb$ on $\ket{\Psi}$: 
\begin{equation}
    \sigma(\Lb)^2\equiv \bra{\Psi}\Lb^2\ket{\Psi} - \bra{\Psi}\Lb\ket{\Psi}^2. 
\end{equation}
$\sigma$ quantifies the change of $\ket{\Psi}$ under the good flow $e^{\ii \Lb t}$ for small $t$. 

Moreover, the states $\ket{\Psi(t)}$ and $\ket{\Psi}$ only differ near the edge region where $\Lb\in\gc$ has non-trivial action. To be more precise, we utilize the fidelity $F(\rho_A(0),\rho_A(t))$ to compare reduced density matrices from $\ket{\Psi}$ and $\ket{\Psi(t)}$ on a region $A$. (By fidelity, we mean $F(\rho,\lambda)=\Tr\sqrt{\sqrt{\rho}\lambda \sqrt{\rho}}$ as in Section~\ref{sec:fidelity_tests}.) Denoting the total system as $\mathcal{X}$, we can show that
\begin{equation}\label{eq:fidelity-X-edge}
    F(\rho_{\mathrm{edge}}(0),\rho_{\mathrm{edge}}(t)) = F(\rho_{\mathcal{X}}(0),\rho_{\mathcal{X}}(t)) = |\langle \Psi(t)|\Psi\rangle|,
\end{equation}
where $\rho_{\mathrm{edge}}(0)$ and $\rho_{\mathrm{edge}}(t)$ are reduced density matrices on the thickened edge region. The first equality is proven in Appendix~\ref{app:fidelity}. 
The idea is that one can deform the support of $\Lb\in\gc$ to a region near the gapless edge without changing the action on the reference state, and hence, only the reduced density matrix near the edge is changed. In the circumstances where there are multiple gapless edges, such as when the system is on a cylinder, the proof requires the full-boundary {\bf A0} condition\footnote{If the system is on a disk (i.e.~it has only one gapless edge), this full-boundary {\bf A0} condition follows from the bulk {\bf A0} condition.}, which is explained previously in Figure~\ref{fig:full_boundary_axioms} in Section~\ref{subsec:cylinder-argument}. The role of full boundary {\bf A0} can be understood as a constraint that there is no long-ranged entanglement between the gapless edges.
The second equality in Eq.~\eqref{eq:fidelity-X-edge} follows from the definition of fidelity. 

As a result of Eq.~\eqref{eq:fidelity-X-edge}, we obtain 
\begin{equation}
    F(\rho_{\mathcal{X}}(0),\rho_{\mathcal{X}}(t)) = F(\rho_{\text{edge}}(0), \rho_{\text{edge}}(t)) = 1 - \frac{1}{2} \sigma(\Lb)^2 t^2 + O(t^3). 
\end{equation}
and therefore 
\begin{equation}\label{eq:sigma-alpha}
   \boxed{ \sigma(\Lb)^2 = \alpha(\Lb,\mathcal{X}) = \alpha(\Lb,\text{edge region}),\quad \Lb\in\gc}, 
\end{equation}
where $\alpha$ is defined in Eq.~\eqref{eq:alpha} in Section~\ref{sec:fid-general}. Later, we shall numerically test Eq.~\eqref{eq:sigma-alpha} for some selected good modular flow generators. We shall also compare the results with theoretical predictions, which are obtained by the Fourier coefficients in the Virasoro mode expansion $\gc\ni \Lb = \sum_{n\in\mathbb{Z}} \lambda_n L_n$ 
\begin{equation}\label{eq:sigma-Lb}
    \sigma(\Lb)^2 = \frac{c}{12}\sum_{n\geq 0} |\lambda_n|^2 (n^3-n). 
\end{equation}
It is worth noting that $\sigma(\Lb)^2$ converges due to the decaying property $|\lambda_n|\sim |n|^{-3}$ for the good modular flow.
In contrast, for the bad modular flow, where $|\lambda_n|\sim |n|^{-2}$, $\sigma(\Lb)^2 \sim \sum_{n=1}^{\infty} 1/n \to \infty$, which diverges.

\subsection{Energy of the coherent states}
\label{subsec:schwarzian}

Since the Virasoro algebra contains a natural notion of Hamiltonian, 
we can also compute the ``energy'' of the edge excitations. In Section~\ref{subsubsec:L0}, we will construct  
\begin{equation}
    L_0 \veq \Lcft_0,
\end{equation}
and the energy of the edge excitation created by the good modular flow $e^{\ii \Lb t}$ with small $t$ is 
\begin{equation}\label{eq:leading-term}
    \bra{\Psi(t)}L_0 \ket{\Psi(t)} = \frac{c \,  t^2}{12} \sum_{m\geq 0} |\lambda_m|^2 (m^4-m^2) + O(t^4).  
\end{equation}
Interestingly, for small $t$, the leading order of this average energy is equal to the leading order (i.e. the $t^2$ order) of the Schwarzian action of an element $\varphi_t \in \rm{Diff}(S^1)$
\begin{equation}\label{eq:schwarz}
    I[\varphi_t] = {2\pi c\over 24 }  \int_0^{2\pi} d\theta 
\( \varphi'_t(\theta)^2 + \( { \varphi''_t(\theta) \over \varphi'_t(\theta)} \)^2 - 1 \) .
\end{equation}
Specifically, the transformation $\varphi_t(\theta)$ is generated by the good modular flow $e^{\ii \Lb t}$. The weight function $f(\theta)$ of $\Lb$ in Eq.~\eqref{eq:Lb-fourier} is the vector field of the transformation, i.e. $\varphi_t(\theta) = \theta + t f(\theta)$ for small $t$, with which one can show Eq.~\eqref{eq:leading-term} and Eq.~\eqref{eq:schwarz} agrees at the leading order.  

The Schwarzian action \eqref{eq:schwarz} governs the dynamics of various theories of low-dimensional quantum gravity and appears to be an effective description of certain 0+1-dimensional approximate CFTs like the low-energy limit of the SYK model (see \eg~\cite{Stanford:2017thb}).  
It also gives the relative entropy of coherent states in chiral CFT \cite{Hollands2019}. 

%==================================
% sec:virasoro-construction
%==================================

\section{Construction of approximate Virasoro generators}
\label{sec:virasoro-construction}
In the previous section, we formally defined the Virasoro generators $L_n$ by requiring $L_n \veq \Lcft_n$. We now begin to discuss their constructions from a single 2+1D purely-chiral wavefunction $\ket{\Psi}$. 

In Section~\ref{subsec:explicit-vir}, we shall use good modular flow generators to construct approximate Virasoro generators $\Leff_n$, assuming the state $\ket{\Psi}$ satisfies Hypothesis~\ref{assump:chiral-CFT}. For $n=0,\pm 1$, the construction is exact $\Leff_0 = L_0,\Leff_{\pm 1} = L_{\pm 1}$\footnote{For this reason, we will occasionally drop the ``$\sim$'' for $\Leff_0$ and $\Leff_{\pm 1}$.}, while for the other generators, $\Leff_{\pm n}$ take the form 
\begin{equation}\label{eq:Leff-intro}
    \Leff_{\pm n} = L_{\pm n} + \sum_{|k|\geq 2} \lambda_{k n} L_{k n},\quad n\geq 2.
\end{equation}
The first term $L_{\pm n}$ gives the leading-order contribution and the remaining higher-harmonic terms are sub-leading in the sense that $|\lambda_{kn}| \sim |k|^{-3}$. We shall discuss the exact version of Eq.~\eqref{eq:Leff-intro} at the end of the next subsection. 

In Section~\ref{subsec:quant-properties}, we analytically discuss several properties of $\Leff_n$. Using the result of Eq.~\eqref{eq:Leff-intro}, we compute several meaningful quantities that are related to $\Leff_n$. The purposes of studying these quantities are not only for their own physical meaningfulness but also for comparisons. Later in Section~\ref{sec:numerics}, we first compare the theoretical predictions for these quantities with numerical results. The good agreement between them supports the validity of Hypothesis~\ref{assump:chiral-CFT}. Secondly, we compare the results for these quantities from $\Leff_n$ and those from $L_n$, from which we will see that the approximation ``error'' of $\Leff_{n}$ to $L_{n}$ is quite small. 

We remark that one can further improve the construction by a procedure that reduces the discrepancy between $\Leff_n$ and $L_n$. The procedure is introduced in Appendix~\ref{app:improvement}. As the limit of such a procedure, one can obtain exact $L_n$. 

\subsection{Explicit construction of $\Leff_n$}
In this subsection, we introduce the explicit construction of $\Leff_n$ using good modular flow generators from a 2+1D purely-chiral state $\ket{\Psi}$. The general idea of the construction can be understood as an approximate spatial Fourier transformation. Recall the Virasoro generator $L_n$ is
\begin{equation}\label{eq:Ln-T}
    L_n \veq \Lcft_n = \frac{1}{2\pi}\int_0^{2\pi} d\theta e^{\ii n\theta} \Tcft(\theta). 
\end{equation}
Based on Hypothesis~\ref{assump:chiral-CFT}, a good modular flow generator in $\gc$ under the isometry $\mathbb{V}$ is a weighted integral of stress-energy tensors, i.e.~a linear combination of them. That is: 
\begin{equation}\label{eq:Lb-T-expt-vir}
   \sum_i \lambda_i \Lb_i \veq \int_0^{2\pi}f(\theta)\Tcft(\theta),\quad \Lb_i \in \gc,\quad \lambda_i \in\mathbb{C}.\footnote{Notice since $\lambda_i$ can take values of complex numbers, such a linear combination might not be Hermitian and hence might not a good modular flow generator.}
\end{equation}
Therefore, to approximate $L_n$, one can find a certain linear combination of suitable good modular flow generators in $\gc$ such that the weight function $f(\theta)$ is approximately $e^{\ii n\theta}$. 

We shall introduce the construction of $L_0$ and $L_{n\neq 0}$ separately, as their construction schemes are different. For $L_0$, it can be obtained directly from a single good modular flow generator [Section~\ref{subsubsec:L0}]. However, the construction for $L_{n},n\neq 0$ takes more steps [Section~\ref{subsubsec:Leffn}]. First of all, since $L_{n}$ for $n\neq 0$ are not Hermitian while good modular flow generators are, a certain linear combination of good modular flow generators with imaginary coefficients is necessary. Secondly, in the construction of $\Leff_{n\neq 0}$, we utilize a method to which we shall refer as ``twisting trick'' (illustrated in Figure~\ref{fig:twist}). The twisting trick enables us to obtain a certain set of good modular flow generators with desirable weight functions whose linear combinations can approximate $e^{\ii n\theta}$. 

\label{subsec:explicit-vir}

\subsubsection{Construction of $L_0$}
\label{subsubsec:L0}
We begin with the construction of the Virasoro generator $L_0$. We shall construct $L_0$ as a good modular flow generator in $\mathfrak{g}^\bullet$. Such a generator needs to give $f(\theta)\equiv \frac{1}{2\pi}$ for $f(\theta)$ 
appearing in Eq.~\eqref{eq:Lb-T-expt-vir}.
As discovered in 1+1D CFT \cite{Lin2023}, if the system is on a circle divided into $N$ equal-sized intervals (labeled by $i=1,\cdots,N$), a constant weight function can be obtained by summing over all the coolness functions on two successive intervals $i,i+1$ and then subtracting the sum over all the coolness functions on single intervals: 
\begin{equation}\label{eq:K2-K1}
    \sum_{i} \( \Kcft^{\chi}_{i,i+1} - \Kcft^{\chi}_{i}\)  \propto \int_0^{2\pi} d\theta \Tcft(\theta)+\kappa \mathbb{1},
\end{equation}
where $\Kcft^{\chi}_{a}$, as a reminder, is the modular Hamiltonian of interval $a$ from a chiral CFT ground state, defined in Eq.~\eqref{eq:modH-chiCFT}. The constant $\kappa$ can be computed by taking the expectation value on the ground state. 

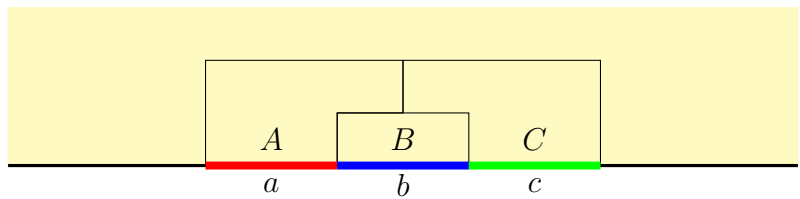
\begin{figure}[h]
    \centering
    \begin{tikzpicture}[scale=0.35]
    \filldraw[yellow!30!white] (-15cm, 0) -- (15cm ,0)  -- ++ (0, 6cm) -- (-15cm, 6cm)-- cycle;
        \draw[very thick] (-15cm, 0) -- (15cm, 0);
        \draw[] (-7.5cm, 0) -- ++ (5cm, 0) -- ++ (0, 2cm) -- ++ (2.5cm, 0) -- ++ (0, 2cm) -- ++ (-7.5cm, 0) -- cycle;
        \draw[] (-2.5cm, 0) -- ++ (5cm, 0) -- ++ (0, 2cm) -- ++ (-5cm, 0) -- cycle;
        \draw[] (2.5cm, 0) -- ++ (5cm, 0) -- ++ (0, 4cm) -- ++ (-7.5cm, 0) -- ++ (0, -2cm);

        \node[] () at (-5cm, 1cm) {$A$};
        \node[] () at (0cm, 1cm) {$B$};
        \node[] () at (5cm, 1cm) {$C$};

        \node[] () at (-5cm, -0.75cm) {$a$};
        \node[] () at (0cm, -0.75cm) {$b$};
        \node[] () at (5cm, -0.75cm) {$c$};

        \draw[line width=3pt, red] (-7.5cm, 0) --++ (5cm,0);
        \draw[line width=3pt, blue] (-2.5cm, 0) --++ (5cm,0);
        \draw[line width=3pt, green] (2.5cm, 0) --++ (5cm,0);
    \end{tikzpicture}
    \caption{Subsystems for $\hat{\Delta}(A,B,C)$. The gapless edge is indicated by the thick black line. The construction of Virasoro generator $L_0$ utilizes a similar choice of subsystems with $(A,B,C)$ taken to be $(A_i,B_i,C_i)$ that touches the gapless edge at interval $(i-1,i,i+1)$.} 
    \label{fig:L0_construction}
\end{figure}

For the 2+1D purely chiral state $\ket{\Psi}$ with a gapless edge, we will be using a linear combination of modular Hamiltonians from $\ket{\Psi}$, which takes the form of $\hat{\Delta}(A,B,C)\equiv K_{AB} + K_{BC}-K_{A}-K_{C} $ for the region $(A,B,C)$ near the edge [Figure~\ref{fig:L0_construction}]. Note that $\hat{\Delta}(A,B,C) \in \mathfrak{g}^\bullet$ is already a good modular flow generator by construction. 
We expect
\begin{equation}\label{eq:delta-and-K}
    \hat{\Delta}(A,B,C)\veq \Kcft_{ab}^{\chi} + \Kcft_{bc}^{\chi} - \Kcft_a^{\chi} - \Kcft_c^{\chi}
\end{equation}
for intervals $a, b,$ and $c$ [Figure~\ref{fig:L0_construction}]. 
While we believe equation \eqref{eq:delta-and-K} holds exactly, Hypothesis~\ref{assump:chiral-CFT} allows a possible additive multiple of the identity operator.  
This additive constant is irrelevant since (in the next paragraph) we will express $L_0$ as a sum of $\hat{\Delta}$s with their expectation values subtracted.  

We now introduce the construction from the reference state $\ket{\Psi}$ that satisfies Hypothesis~\ref{assump:chiral-CFT} near the gapless edge. 
We divide the edge into $N$ equal-sized intervals, labeled by $i=1,2,\cdots, N$. Let $A_i, B_i, C_i$ be regions touching the edge at the $i-1$, $i$, and $i+1$th intervals respectively, and they are arranged in way topologically equivalent to the $A,B,C$ shown in Figure~\ref{fig:L0_construction}. 
We see that 
\begin{equation}
\label{eq:L0}
    L_0 \equiv \frac{1}{4\pi \tan(\pi/N)}\sum_{j=1}^{N} \left( \hat{\Delta}(A_i, B_i, C_i) - \Delta(A_i, B_i, C_i) \right)
\end{equation}
satisfies the desired property $L_0 \veq  \frac{1}{2\pi}\int_0^{2\pi} d\theta \, \Tcft(\theta) =\Lcft_0$, and it is the generator we are looking for. 
Here, the lower indices of the regions should be understood modulo $N$ and $\Delta(A_i, B_i, C_i) = \bra{\Psi} \hat{\Delta}(A_i, B_i, C_i) \ket{\Psi}$. 

\subsubsection{Construction of $\Leff_n,n\neq 0$ with the twisting trick}
\label{subsubsec:Leffn}
In this subsection, we introduce a scheme to construct the approximate Virasoro generator $\Leff_n$ which is generically applicable for all $n\neq 0$. Our construction for $n = \pm 1$ shall give the exact $L_{\pm 1}$, i.e. $ \Leff_{\pm 1}=L_{\pm 1}$. On the other hand, $\Leff_n$ for $n\geq 2$ is equal to $L_{n}$ plus a sum of sub-leading higher-harmonic terms $L_{kn}, |k|\geq 2$, as we mentioned in Eq.~\eqref{eq:Leff-intro}. 

\begin{figure}[!h]
    \centering
    \includegraphics[width=0.8\columnwidth]{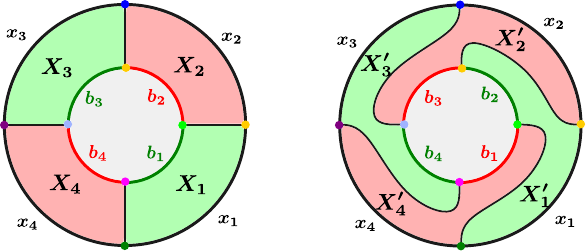}
    \caption{Illustration of the twisting trick. The color (green, red) of each region stands for the sign $(+,-)$ of the modular Hamiltonian on that region. Between the two plots, the colored dots are identified. $L = K_{X_1}-K_{X_2}+K_{X_3}-K_{X_4}$, $L^{\mathrm{twist}} = L_{X'_1}-L_{X'_2}+K_{X'_3}-L_{X'_4}$. Neither $L$ nor $L^{\mathrm{twist}}$ is a good modular flow generator, as there are ``excitations'' near the entanglement cut. Once we add them, such ``excitations'' cancel.}
    \label{fig:twist}
\end{figure}

As we mentioned at the beginning of this subsection, to construct an approximate Virasoro generator, our goal is to design a suitable set of good modular flow generators which realize approximate spatial Fourier transformations. We shall later introduce a \emph{twisting trick} [Figure~\ref{fig:twist}] to achieve this goal. This trick leads to a judicious choice of linear combination of modular Hamiltonians that only create excitations on the edge. Below we explain this idea in detail with an example, followed by the discussion on the more general cases.

The main task of the construction is to find two types of good modular flow generators, denoted as $\Lbo_n,\Lbe_n$, such that  
\begin{equation}
    \Lbo_n \vapprox \frac{1}{2\pi}\int_0^{2\pi}d\theta \sin(n\theta)\Tcft(\theta),\quad  \Lbe_n \vapprox \frac{1}{2\pi}\int_0^{2\pi}d\theta \cos(n\theta)\Tcft(\theta).  
\end{equation}
Once $\Lbo_n,\Lbe_n$ are constructed, one can obtain the approximate Virasoro generators $\Leff_{\pm n}$ as
\begin{equation}
    \Leff_{\pm n} = \Lbe_n \pm \ii \Lbo_n \vapprox \frac{1}{2\pi}\int_0^{2\pi}d\theta e^{\pm \ii n\theta} \Tcft(\theta) = \Lcft_{\pm n},\quad n\geq 1, 
\end{equation}
since $e^{\pm \ii n\theta} = \cos(n\theta) \pm \ii \sin(n\theta)$.  

Now we discuss the construction of $\Lbo_n$ and $\Lbe_n$. Let us use $\betao_n(\theta)$ and $\betae_n(\theta)$ to denote their weight function: 
\begin{equation}\label{eq:Lbnoe-1}
    \Lbo_n \veq \int_0^{2\pi}d\theta\betao_n(\theta)\Tcft(\theta),\quad \Lbe_n \veq \int_0^{2\pi}d\theta\betae_n(\theta)\Tcft(\theta). 
\end{equation}
We aim to choose $\betao_n(\theta),\betae_n(\theta)$ in Eq.~\eqref{eq:Lbnoe-1} so that they approximate $\sin(n\theta)$ and $\cos(n\theta)$. Based on Hypothesis~\ref{assump:chiral-CFT}, the weight function $\betao_n(\theta),\betae_n(\theta)$ shall be a linear combination of the coolness functions $\beta^{[\theta_1, \theta_2]}(\theta)$, introduced previously in Eq.~\eqref{eq:coolness-function}, on various intervals on a circle. Therefore, to obtain the desired $\Lbo_n,\Lbe_n$, one needs to answer the following two questions: (1) Which linear combination of coolness functions gives a good approximation of $\sin(n\theta),\cos(n\theta)$? (2) Which good modular flow generator can result in such a linear combination in its weight function? 

\begin{figure}[h!]
    \centering
    \includegraphics[width=0.95\columnwidth]{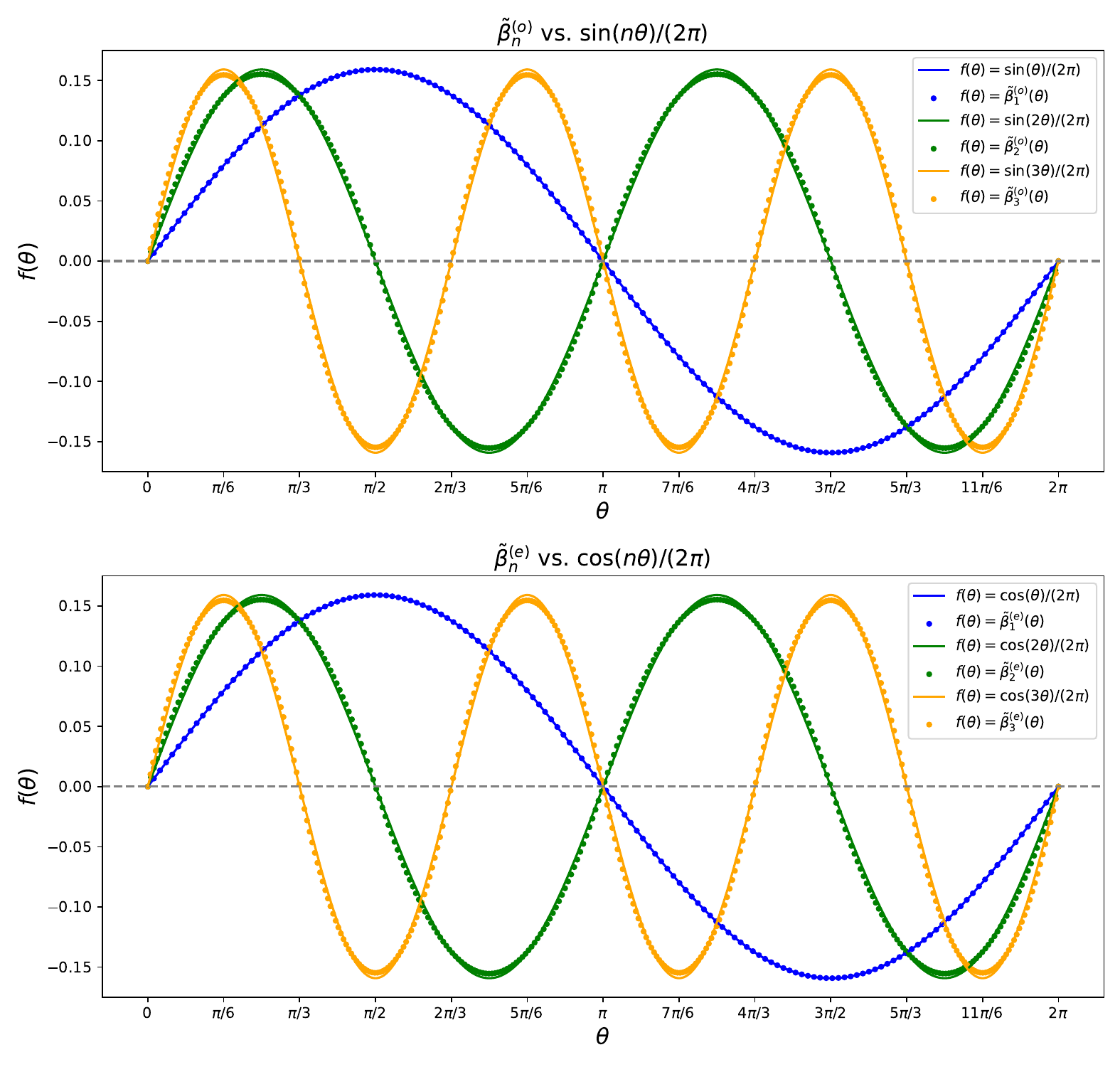}
    \caption{Comparison between $\beta_n^{(o,e)}(\theta)$ and $\sin(n\theta)/(2\pi),\cos(n\theta)/(2\pi)$ respectively.} 
    \label{fig:beta-vs-trig}
\end{figure}

For the first question, dividing a circle into $2n$ equal-size intervals 
$$\left[(k-1)\frac{2\pi}{2n}, k\frac{2\pi}{2n} \right],~k=1,\cdots,2n$$ 
and adding up the coolness function of these intervals with alternating signs will provide an approximation of $\sin(n\theta)$, up to a normalization factor (denoted as $A_n$). Similarly, the approximation of $\cos(n\theta)$ can be obtained by a shift of $\pi/(2n)$ angle along the circle.\footnote{The important part is the $\pi/(2n)$ phase difference between $\beta^{(o)}_n$ and $\beta^{(e)}_n$, It is also okay to call $\beta^{(o)}_n$ the approximation for $\cos(n\theta)$ and $\beta^{(e)}_n$ for $\sin(n\theta)$. That just amounts to a $\pi/(2n)$ overall coordinate translation.}  Explicitly,  
\begin{equation}\label{eq:betan}\begin{split}
    &\betao_n(\theta) = A_n\sum_{k=1}^{2n} (-1)^{k-1}\beta^{\[\frac{(k-1)\pi}{n}, \frac{k\pi}{n} \] }(\theta)\approx \frac{1}{2\pi}\sin(n\theta),\\ 
    &\betae_n(\theta)= \betao_n\(\theta+\frac{\pi}{2n}\) = A_n\sum_{k=1}^{2n} (-1)^{k-1}\beta^{\[\frac{(k-3/2)\pi}{n}, \frac{(k-1/2)\pi}{n} \] }(\theta)\approx  \frac{1}{2\pi}\cos(n\theta),
\end{split}
\end{equation}
where normalization factor $A_n$ is defined as
\begin{equation}\label{eq:An}
    A_n = \left\{ \begin{aligned}
        & \frac{1}{2\pi} &   &n = 1 \\ 
        &  \tan\( \frac{\pi}{2n} \)(n^2-1)/8,& &n\neq 1  
    \end{aligned}\right. ~.
\end{equation}
This factor $A_n$ in the equation above is obtained by demanding 
\begin{equation}
    \int_0^{2\pi} \beta^{(o)}_n(\theta)\sin(n\theta)=\frac{1}{2},\quad  \int_0^{2\pi} \beta^{(e)}_n(\theta)\cos(n\theta)=\frac{1}{2}.  
\end{equation}
See Figure~\ref{fig:beta-vs-trig} for a comparison between $\beta^{(o,e)}_n$ and $\sin(n\theta)/(2\pi),\cos(n\theta)/(2\pi)$ respectively. We will provide a quantitative comparison in Appendix~\ref{app:improvement}, discussing the error in this approximation in detail. 

Now, we need to identify a certain \emph{good} modular flow generator that results in such weight function Eq.~\eqref{eq:betan}. 
To that end, we utilize the following ``twisting trick" [Figure~\ref{fig:twist}]. For illustration purposes, let us consider the construction for $n=2$. We divide the chiral gapless edge into four intervals of equal size, labeled by $x_k \equiv [(k-1)\pi/2, k\pi/2],k=1,2,3,4$. We wish to obtain a good modular flow generator, denoted as $\Lbo_2$, whose weight function is $\beta^{x_1}(\theta)-\beta^{x_2}(\theta)+\beta^{x_3}(\theta)-\beta^{x_4}(\theta)$, that is: 
\begin{equation}\label{eq:L2}
    \Lbo_2 \stackrel{\mathbb{V}}{\propto} \Lcft = \Kcft_{x_1}^{\chi}-\Kcft_{x_2}^{\chi}+\Kcft_{x_3}^{\chi}-\Kcft_{x_4}^{\chi},
\end{equation}
where $\Kcft_a^{\chi}$ is the modular Hamiltonian of an interval $a$ for chiral CFT ground state, defined in Eq.~\eqref{eq:modH-chiCFT}.

One approach to achieve this would be to choose four regions 
$X_k$ anchored at the interval $x_k$, $k=1,2,3,4$ along the gapless edge shown in the left column of Figure~\ref{fig:twist}. The linear combination $L = K_{X_1}-K_{X_2}+K_{X_3}-K_{X_4}$, with alternating signs, acting on $\ket{\Psi}$ shall create the same edge excitations as $\Lcft$ in Eq.~\eqref{eq:L2} does on the chiral CFT ground state. However, this $L$ is not a good modular flow generator: $L$ will create bulk excitations along the entanglement cuts $b_k$ shown in the Figure~\ref{fig:twist} due to the action of $K_{X_k}$ on $\ket{\Psi}$. 
To cancel these bulk excitations, we introduce a \emph{twisting trick} to construct another linear combinations of modular Hamiltonian $L^{\rm twist} = K_{X'_1}-K_{X'_2}+K_{X'_3}-K_{X'_4}$. Each region $X'_k$ is obtained by a twisting of $X_k$ in the bulk, such that the boundary of $X'_k$ includes the entanglement cut $b_{k+1}$ from its neighboring untwisted region $X_{k+1}$, i.e. $b_{k+1}\subset \partial X'_{k}$; see the right column of Figure~\ref{fig:twist} for the illustration of the twisted regions. Notice $X'_k$ anchors at the same interval at the gapless edge as $X_k$. 
Therefore $L^{\rm twist}$ will create the same edge excitations as $L$. Now we can simply add up 
\be \Lbo_2 \equiv {A_2\over 2} \( L+L^{\rm twist} \),\ee
where $A_2$ is the prefactor defined in Eq.~\eqref{eq:An}.
Notice $\Lbo_2$ will create the same edge excitations as $L$ and $L^{\rm twist}$, but create no excitations in the bulk. Indeed, we can see that the excitations in $b_k$ created by $K_{X_k}$ is canceled by $K_{X'_{k-1}}$, as they have opposite sign in $\Lbo_2$. As a result, we obtain a good modular flow generator $\Lbo_2$ that satisfies Eq.~\eqref{eq:L2}. To obtain $\Lbe_2$, we can simply shift the whole setup by $\pi/4$ clockwise, such that $x_k \to y_k = [(k-3/2)\pi/2, (k-1/2)\pi/2],X_k \to Y_k, X'_k \to Y'_k$, and then use the same summation using $K_{Y_k},K_{Y'_k}$ as we did for $K_{X_k},K_{X'_k}$. 

The idea we described above works for the construction of $\Lbo_n,\Lbe_n$ with %can generalize easily to arbitrary 
$n\geq 1$, defined in Eq.~\eqref{eq:Lbnoe-1}. We now explicitly describe the algorithm using Figure~\ref{fig:Lno}: 
\begin{itemize}
    \item \emph{Step 1}: For a 2+1D purely-chiral state $\ket{\Psi}$, we first choose a gapless edge of length $L$ and divide it into $2n$ equal-size intervals, labeled by $x_j,j=1,\cdots,2n$. We remark that the starting point of the choice of intervals does not affect the construction due to the translation symmetry along the edge.
    \item \emph{Step 2}:  Then we make a set of twisted squares $X^{L}_{j}, X^{R}_j$ anchored at the interval $x_j$. They are twisted to the left $(L)$ and right $(R)$ such that $X^{L}_{j}$ meets $X^{R}_{j+1}$ at the top edge of the squares, see Figure~\ref{fig:Lno}. 
    \begin{figure}[!ht]
        \includegraphics[width=0.9\columnwidth]{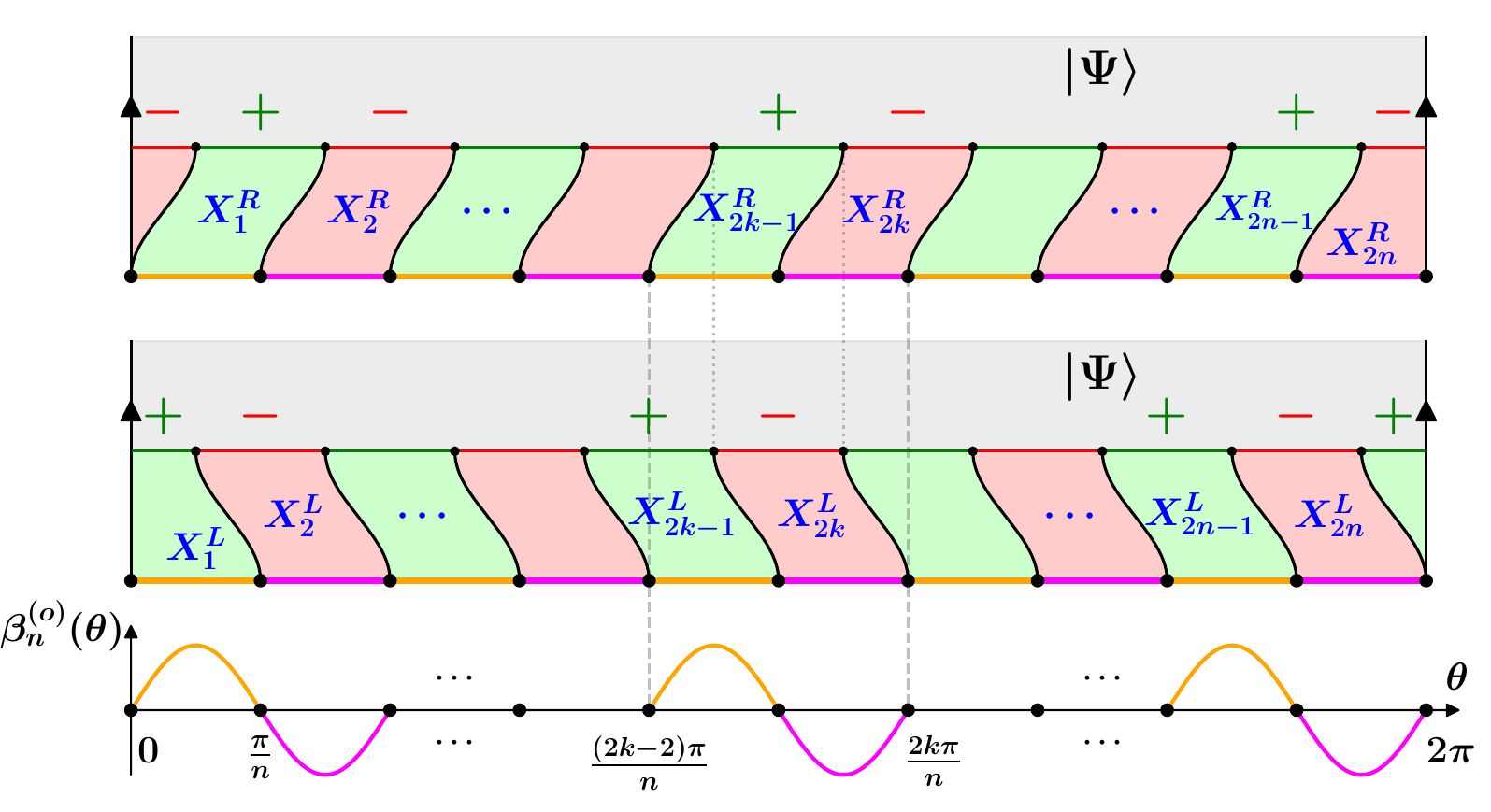}
        \caption{Construction of $\Lbo_n$. The color indicates the sign of the modular Hamiltonian of that region. Each of the figures depicts a connected component of a boundary.  Left and right sides of the figure are identified.  We can be agnostic about what happens farther into the bulk; for example, the whole geometry could be a disk. The regions used in the construction of $\Lbe_n$ are called $Y_k^R$ and $Y_k^L$; they are obtained by shifting $X_k^R$ and $X_k^L$ to the left by $\pi/2n$.} 
    \label{fig:Lno}
    \end{figure}
    \item \emph{Step 3:} With this setup, we construct 
    \begin{equation}\label{eq:step3}
    \Lbo_n = \frac{A_n}{2}\sum_{j=1}^{2n}(-1)^{j-1}\( K_{X_j^R} + K_{X_j^L} \).
    \end{equation}
    \item \emph{Step 4:} To obtain $\Lbe_n$, we can shift the intervals and squares by $L/(4n)$ distance to the left direction: $x_i \to y_i$, $(X^{L}_{j}, X^{R}_j)\to (Y^{L}_{j}, Y^{R}_j),j=1,\cdots,2n$. Then 
     \begin{equation}\label{eq:step4}
    \Lbe_n = \frac{A_n}{2}\sum_{j=1}^{2n}(-1)^{j-1}\( K_{Y_j^R} + K_{Y_j^L} \).
    \end{equation}
\end{itemize}

In summary, in this way, we obtain good modular flow generators $\Lbo_n,\Lbe_n$: 
\begin{equation}\label{eq:Lbnoe-2} \begin{split} 
    &\Lbo_n \veq \int_{0}^{2\pi} \beta_n^{(o)}(\theta) \Tcft(\theta)\approx \frac{1}{2\pi}\int_{0}^{2\pi} \sin(n\theta) \Tcft(\theta) \\ 
    &\Lbe_n \veq \int_{0}^{2\pi} \beta_n^{(e)}(\theta) \Tcft(\theta)\approx \frac{1}{2\pi}\int_{0}^{2\pi} \cos(n\theta) \Tcft(\theta)
\end{split}\end{equation}
At last, by making linear combinations of $\Lbo_n$ and $\Lbe_n$, 
\begin{equation}\label{eq:Leffn-eo}
    \Leff_{\pm n} = \Lbe_n \pm \ii \Lbo_n,
\end{equation}
we can approximate $e^{\pm \ii n\theta}/(2\pi)$ in the weight function, 
\begin{equation}\label{eq:Leffn}
    \begin{split}
        &\Leff_n  \veq \int_{0}^{2\pi} \(\beta^{(e)}_n+\ii \beta^{(o)}_n \) \Tcft(\theta) \approx  \frac{1}{2\pi}\int_{0}^{2\pi} e^{\ii n\theta} \Tcft(\theta) = \Lcft_n \\ 
        &\Leff_{-n}  \veq \int_{0}^{2\pi} \(\beta^{(e)}_n-\ii \beta^{(o)}_n \) \Tcft(\theta) \approx  \frac{1}{2\pi}\int_{0}^{2\pi} e^{-\ii n\theta} \Tcft(\theta) = \Lcft_{-n},
\end{split}
\end{equation}
and as a result we obtain the approximate Virasoro generators $\Leff_{\pm n}\vapprox \Lcft_{\pm n},n\geq 1$. 

\subsubsection{Expansion of $\Leff_n$ in Virasoro modes}
\label{subsubsec:ana-Leffn}
With the approximate $\Leff_n$ constructed, we now discuss the exact form of $\Leff_n$ in terms of Virasoro generators $L_m$. Such an expression of $\Leff_n$ is useful in analytically computing several quantities that are related to them, such as those discussed in Section~\ref{subsec:quant-properties}. The explicit expression of $\Leff_n$ also reveals the approximation ``error'' of $\Leff_n$ to $L_n$. 

For $n=0$, as we mentioned before in Section~\ref{subsubsec:L0}, the construction gives $\Leff_0 = L_0$. Therefore, we shall focus on $\Leff_{n\neq 0}$. We remark that even though the construction of $\Leff_{\pm 1}$ also gives exactly $L_{\pm 1}$, we still include $n=\pm 1$ into the discussion below, because the construction for all $n\neq 0$ follows the same procedure, so the analysis is applicable to all $n\neq 0$. 

Based on the construction Eq.~\eqref{eq:Leffn}, the approximate Virasoro generator takes the form
\begin{equation}\label{eq:Leff-beta}
    \Leff_{n} \veq \int_{0}^{2\pi}d\theta\beta_{n}(\theta) \Tcft(\theta), \quad n\neq 0
\end{equation}
where 
\begin{equation}
    \beta_{\pm n}(\theta) = \betae_n(\theta) \pm \ii \betao_n(\theta),\quad n\geq 1. 
\end{equation}
Employing $ \Tcft(\theta) = \sum_{m\in\mathbb{Z}} L_m e^{-\ii m\theta}$ in Eq.~\eqref{eq:Leff-beta}, one can obtain the exact form of $\Leff_n$ in terms of exact Virasoro generators $L_n$ 
\begin{equation}\label{eq:Leffn-exact}
    \Leff_n = \sum_{m\in\mathbb{Z}} \lambda_{n,m} L_m,
\end{equation}
where the $\lambda_{n,m}$ are the Fourier coefficients of $\beta_n(\theta)$: 
\begin{equation}\label{eq:lambda-nm}
    \lambda_{n,m} = \int_{0}^{2\pi}d\theta \beta_n(\theta)e^{-\ii m\theta}  
     = \left\{\begin{aligned}
         & 1, & & \text{ if } |n| = 1,m=n \\
         &\frac{n^3-n}{m^3-m},& &\text{ if } |n|\geq 2, m = (-1)^{k}(2k+1)n, k \in \mathbb{N} \\ 
         &0,& &\text{ otherwise}
     \end{aligned}
     \right.
\end{equation}
\begin{comment}
Utilizing the explicit result of $\lambda_{n,m}$, one can write $\Leff_n$ as 
\begin{equation}\label{eq:Leffn-exact}
    \Leff_n = L_n - \sum_{k\in 2\mathbb{N}+1}g_n(k) L_{-(2k+1)n} + \sum_{k\in 2\mathbb{N}+2} g_n(k) L_{(2k+1)n}, \quad n\neq 0
\end{equation}
where 
\begin{equation}
    g_n(k) = \frac{n^2-1}{(2k+1)^3n^2-(2k+1)}. 
\end{equation}
\end{comment}
We make several remarks regarding the exact form $\Leff_n,n\neq 0$. 
\begin{itemize}
    \item For $|n|=1$, we emphasize that $\Leff_n$ is exactly $L_n$, i.e. $\Leff_{\pm 1} = L_{\pm 1}$. This can be seen directly by checking $\beta_{\pm 1}(\theta) = e^{\pm \ii \theta}$.
    \item For $|n|\geq 2$, we can see that the Virasoro mode labels appearing in $\Leff_n$ are 
    $n$ multiplied by odd integers with alternating signs: $L_n, L_{-3n},L_{5n}$, etc. Among these modes, the leading order contribution is $L_n$, as $\lambda_{n,n}=1$. As $|m/n|$ increases, the magnitude of its coefficient $|\lambda_{n,m}|$ decay as $O(|m/n|^{-3})$, which ensures the contribution from these sub-leading terms are considerably small. 
\end{itemize}

Moreover, having obtained the exact form of $\Leff_n,n\neq 0$, we can also design a procedure to reduce the subleading terms in $\Leff_n$, so that it becomes a better approximation to $L_n$. The procedure is explicitly introduced in Appendix~\ref{app:improvement}. As a limit of such a procedure, one can obtain the \emph{exact} Virasoro generator $L_n$ as a linear combination of good modular flow generators. 

\subsection{Quantitative properties}
\label{subsec:quant-properties}
In this subsection, we discuss several quantitative properties related to $\Leff_n$, which will be tested in numerics [Section~\ref{sec:numerics}]. 
Based on Hypothesis~\ref{assump:chiral-CFT},
these proprieties are analytically expressed in terms of the Fourier coefficients $\Leff_n$ in Eq.~\eqref{eq:Leffn-exact}. Therefore the excellent agreement we shall see in Section~\ref{sec:numerics} provides support for the correctness of the exact form of $\Leff_n$ in Eq.~\eqref{eq:Leffn-exact}, as well as to the validity of Hypothesis~\ref{assump:chiral-CFT}. These quantities of $\Leff_n$ will also be compared with those from $L_n$. Such a comparison gives a measure of the quality of the approximation of $\Leff_n$ to $L_n$.

\subsubsection{Variance}
\label{subsubsec:variance}
The first quantity we consider is the variance of the good modular flow generators $\Lbo_n,\Lbe_n$ that are involved in the construction of $\Leff_n$. We will focus on $\Lbo_n$; the results for the other one $\Lbe_n$ can be obtained similarly. 

Based on Eq.~\eqref{eq:Leffn-eo}, $\Lbo_n,n\geq 1$ is of the form 
\begin{equation}
    \Lbo_n = \frac{-\ii }{2} (\Leff_n - \Leff_{-n}).
\end{equation}
Taking the exact form of $\Leff_{\pm n}$ from Eq.~\eqref{eq:Leffn-exact}, we can express $\Lbo_n$ in terms of Virasoro generators 
\begin{equation}
    \Lbo_n = \sum_{n\in\mathbb{Z}} \lambda^{(o)}_{n,m} L_m,
\end{equation}
where 
\begin{equation}\label{eq:lambdao-nm}
     \lambda^{(o)}_{n,m} = -\frac{\ii }{2}(\lambda_{n,m}-\lambda_{-n,m}),\quad n\geq 1. 
\end{equation}
As we mentioned in Section~\ref{subsubsec:variance}, the variance of a good modular flow generator can be expressed in terms of its Fourier coefficient as Eq.~\eqref{eq:sigma-Lb}. Therefore, for $\Lbo_n$, we obtain 
\begin{equation}
    \sigma(\Lbo_n)^2 = \frac{c}{12}\sum_{m\geq 0} |\lambda^{(o)}_{n,m}|^2 (m^3-m). 
\end{equation}
This quantity can be explicitly computed by making use of the analytic formula of $\lambda^{(o)}_{n,m}$ obtained from Eq.~\eqref{eq:lambdao-nm} and Eq.~\eqref{eq:lambda-nm}:
\begin{equation}\label{eq:sigma-Lb-analytic}
    \sigma(\Lbo_n)^2 = \frac{c}{192}n(n^2-1)^2 \left[ 2\psi\left( \frac{1}{2} \right)- \psi\left( \frac{n+1}{2n} \right)-\psi\left( \frac{n-1}{2n} \right)\right],
\end{equation}
where $\psi(z)$ is the polygamma function, defined as $\psi(z)\equiv \Gamma'(z)/\Gamma(z)$ from the gamma function $\Gamma(z)$ and its first derivative $\Gamma'(z)$. Thus, at large $n$, $\sigma(\Lbo_n)^2 \sim n^3$.  

\subsubsection{Expectation values of commutators}
\label{subsubsec:expt-comm}
We are also interested in expectation values of commutators among $\Leff_{n},n\neq 0$ since they encode information about the Virasoro algebra relations.  Explicitly, we shall consider $\bra{\Psi}[\Leff_n,\Leff_{-n}]\ket{\Psi},n\geq 1$. Recall $\Leff_{\pm n} = \Lbe_n \pm \ii \Lbo_n$, therefore the expectation value of the commutator can be expressed as 
\begin{equation}
    \bra{\Psi}[\Leff_n,\Leff_{-n}]\ket{\Psi} = 2\ii \bra{\Psi}[\Lbe_n,\Lbo_{n}]\ket{\Psi},\quad n\geq 1. 
\end{equation}

Before we explicitly compute $\ii \bra{\Psi}[\Lbe_n,\Lbo_{n}]\ket{\Psi},n\geq 1$, we first discuss the general case, namely, the expectation value $\ii \bra{\Psi} [\Lb^1,\Lb^2] \ket{\Psi}$ for any two good modular flow generators $\Lb^1,\Lb^2 \in \gc$. There are, in general, two methods to compute it. 

One way is to follow the same idea we used to compute the variance and use Fourier expansion of $\Lb^1,\Lb^2$ 
\begin{equation}
    \Lb^1 = \sum_{m\in\mathbb{Z}} \lambda^1_m L_m,\quad \Lb^2 = \sum_{m\in\mathbb{Z}} \lambda^2_m L_m,
\end{equation}
then the expectation value of their commutators is 
\begin{equation}
     \bra{\Psi}[\Lb^1,\Lb^2]\ket{\Psi} = \frac{c}{12}\sum_{m\in\mathbb{Z}} \lambda_m^1 \lambda_{-m}^2 (m^3-m). 
\end{equation}

The other way is to utilize the result in \cite{Zou2022}: The expectation value of the commutator between two chiral CFT modular Hamiltonian on interval $ab,bc$ is 
\begin{equation}\label{eq:cft-commutator}
    \ii{\bra{\Omega}} [ \Kcft_{ab}^{\chi}, \Kcft_{bc}^{\chi}]\ket{\Omega} = \frac{\pi c_{-}}{6}(2\eta_{a,b,c}-1), 
\end{equation}
where $\eta_{a,b,c} = \frac{|a|\cdot|c|}{|ab|\cdot |bc|}$ is the cross-ratio of the successive intervals $(a,b,c)$, and $c_{-}$ is the chiral central charge. With Eq.~\eqref{eq:cft-commutator} at hand, we can, in general, compute expectation values between any two good modular flow generators
\begin{equation}
    \begin{aligned}
        &\gc \ni \Lb^1 = \sum_{A^1} \lambda_A^1 K_{A^1} \veq \sum_{a^1} \lambda_{A^1}\Kcft^{\chi}_{a^1},\\ 
        &\gc \ni \Lb^2 = \sum_{A^2} \lambda_A^2 K_{A^2} \veq \sum_{a^2} \lambda_{A^2}\Kcft^{\chi}_{a^2}.
    \end{aligned}
\end{equation}
In the equation above, $a^1,a^2$ are the intervals where $A^1,A^2$ intersect with the gapless edge. Using the isometry relation in Hypothesis~\ref{assump:chiral-CFT}, the expectation value of the commutator $[\Lb^1,\Lb^2]$ is  
\begin{equation}
    \ii \bra{\Psi} [\Lb^1,\Lb^2] \ket{\Psi} = \sum_{A^1,A^2}\lambda_{A^1}\lambda_{A^2} \left( \ii \bra{\Omega}[\Kcft_{a^1},\Kcft_{a^2}]\ket{\Omega}\right),
\end{equation}
where each commutator $\ii \bra{\Omega}[\Kcft_{a^1},\Kcft_{a^2}]\ket{\Omega}$ in the right hand side can be computed via Eq.~\eqref{eq:cft-commutator}. 

We shall use the second method to compute $\ii \bra{\Psi} [\Lbe_n,\Lbo_{n}]\ket{\Psi}$, as it provides a succinct analytic formula of the result. Based on the construction of $\Lbe_n,\Lbo_n$ in Eqs.~\eqref{eq:step3} and \eqref{eq:step4},
\begin{equation}
    \Lbo_n \veq A_n\sum_k (-1)^{k-1} \Kcft_{x_k},\quad \Lbe_n \veq A_n\sum_k (-1)^{k-1} \Kcft_{y_k},
\end{equation}
where $x_k = \left[ \frac{(k-1)\pi}{n}, \frac{k\pi}{n} \right]$ and $y_k = \left[ \frac{(k-3/2)\pi}{n}, \frac{(k-1/2)\pi}{n} \right]$ are sequences of intervals on a circle, as explained in Section~\ref{subsubsec:Leffn}. By counting the intersection of intervals, we can obtain
\begin{equation}
    \ii \bra{\Psi} [\Lbe_n,\Lbo_{n}]\ket{\Psi} = A_n^2\cdot 4n\cdot \frac{\pi c_{-}}{6}(1-2\eta_n), 
\end{equation}
where 
\begin{equation}
    \eta_n = \frac{\sin(\pi/(4n))^2}{\sin(\pi/(2n))^2}. 
\end{equation}
Therefore, plugging in the expression of $A_n$ in Eq.~\eqref{eq:An}, we can obtain the final expression of $\bra{\Psi} [\Leff_n,\Leff_{-n}]\ket{\Psi}$. 
\begin{equation}\label{eq:Lnn-analytic}
    \bra{\Psi} [\Leff_n,\Leff_{-n}]\ket{\Psi} = \frac{\pi c_{-}}{24} n(n^2-1)^2 \frac{\sin^2(\pi/(4n))}{\cos(\pi/(2n))}, \quad c_{-} = c. 
\end{equation}
where the chiral central charge $c_{-}$ is equal to the holomorphic central charge $c$, since we are working with purely chiral CFT\footnote{$\bar{c}=0$, therefore $c_{-} = c-\bar{c} = c$.}. 
%\XL{Is this derivation too detailed or too brief?}\JM{It looks good to me.}

Later we will numerically compute the commutators $\bra{\Psi} [\Leff_n,\Leff_{-n}]\ket{\Psi}$ and compare them with Eq.~\eqref{eq:Lnn-analytic}. Such tests examine the validity of the Hypothesis~\ref{assump:chiral-CFT}. We are also going to compare the numerical results with the results obtained by Virasoro algebra generators. These tests demonstrate the quality of $\Leff_n$ as an approximation to Virasoro generator $L_n$.

As a mathematical aside, from the two ways of computing $\bra{\Psi}[\Lb^1,\Lb^2]\ket{\Psi}$ we are able to show a number of interesting relations between their Fourier coefficients and the set of cross-ratios. For example, in the computation of $\bra{\Psi}[\Lbe_n,\Lbo_{n}]\ket{\Psi}$, we have shown 
\be 
\sum_{m\in\mathbb{Z}} |\lambda_{n,m}|^2 (m^3-m) = \frac{\pi }{2} n(n^2-1)^2 \frac{\sin^2(\pi/(4n))}{\cos(\pi/(2n))},\quad n\geq 1
\ee
where $\lambda_{n,m}$ is given in Eq.~\eqref{eq:lambda-nm}. 

\subsubsection{Expectation value of double commutators}
\label{subsubsec:expt-double-comm}
Besides the expectation values of commutators, we shall also consider expectation values of double commutators $\bra{\Psi}[[\Leff_m,\Leff_n],\Leff_k]\ket{\Psi}$. The results can be explicitly computed by utilizing the Fourier expansion of $\Leff_n$ (Eq.~\eqref{eq:Leffn-exact}) in terms of Virasoro generators. 
Such quantities will allow us to test further the closure of our realization of the Virasoro algebra.

One particular double commutator that we will compute is $ \bra{\Psi}[[\Leff_{n},L_0],\Leff_{-n}]\ket{\Psi}$, $n\geq 1$. By making use of the Fourier expansion of $\Leff_{\pm n}$ as follows 
\begin{equation}
    \Leff_n = \sum_{m\in\mathbb{Z}} \lambda_{n,m} L_m,\quad \Leff_{-n} = (\Leff_n)^{\dagger} = \sum_{m\in\mathbb{Z}} \lambda_{n,m}^* L_{-m},\quad n\geq 1,
\end{equation}
and the Virasoro algebraic relations, we can derive %\BS{in the following equation, we have $m \in \mathbb{Z}$ right?} \XL{Yeah, changed to make it explicit}
\begin{equation}\label{eq:Ln0n}
    \bra{\Psi}[[\Leff_{n},L_0],\Leff_{-n}]\ket{\Psi} = \frac{c}{12}\sum_{m\in\mathbb{Z}} |\lambda_{n,m}|^2 (m^4-m^2).
\end{equation}
With the analytic formula of $\lambda_{n,m}$ given in Eq.~\eqref{eq:lambda-nm}, we are able to explicitly compute Eq.~\eqref{eq:Ln0n} and the result is
\begin{equation}\label{eq:Ln0n-analytic}
    \bra{\Psi}[[\Leff_{n},L_0],\Leff_{-n}]\ket{\Psi} = \frac{\pi c}{48} n (n^2-1)^2 \tan\left( \frac{\pi}{2n} \right). 
\end{equation}
The result shall be compared with numerical results. Such a test examines not only the Fourier expansion of $\Leff_{\pm n}$ but also the construction of $L_0$.  

Note that Eq.~\eqref{eq:Ln0n} is the $\CO(t^2)$ term in the quantity 
$ \bra{\Psi(t)} L_0 \ket{\Psi(t)}$, with $\ket{\Psi(t)} = e^{ \ii t \Leff_{-n}} \ket{\Psi} $, and therefore as explained in  
Section~\ref{subsec:schwarzian} is the leading contribution to a Schwarzian functional.
%==================================
% end of sec:virasoro-construction
%==================================

% ==============================
% sec:numerics
% ===============================
\section{Numerical tests}
\label{sec:numerics} 
In the previous section, we explicitly described the procedures to construct approximate Virasoro algebra $\Leff_n$ and made several predictions based on Hypothesis~\ref{assump:chiral-CFT}. In this section, we study two specific reference states, namely $p+\ii p$ superconductor (SC) ground state and semion ground state, to test the construction of $\Leff_n$ via the predictions we discussed in the previous section. We will also compare results obtained by $\Leff_n$ and $L_n$, to test the quality of the approximation $\Leff_n \approx L_n$. 

$p+\ii p$ SC is a free fermion system. By the means introduced by \cite{Eisler2009}, we are able to reach pretty large system sizes. Therefore, we will perform a thorough test in $p+\ii p$ SC.  One may wonder, however, whether the conclusions rely on the free fermion property.   Therefore, we also perform tests on a ground state with chiral semion topological order, which has no free-fermion representation.  For the semion ground state, due to the limitation of the computation power, we will not do as many tests as in $p+\ii p$ SC, but the results nevertheless indicate that the construction works and the predictions are valid. 

\subsection{Tests in $p+\ii p$ superconductor}
We choose the $p+\ii p$ SC Hamiltonian as Eq.~\eqref{eq:pipH-2}, with the same parameters used in Section~\ref{sec:fidelity_tests}. Again, we choose the anti-periodic boundary condition in the $x$-direction and the open boundary condition in the $y$-direction so that there is no flux threaded inside the cylinder, and full-boundary {\bf A0}, {\bf A1} introduced in Section~\ref{subsec:cylinder-argument} are satisfied. We will specify the size of the cylinder $L_x\times L_y$ later. 

The ground state of $p+\ii p$ SC for this choice of parameters is in its chiral gapped phase. In the bulk, its entanglement entropy satisfies the area law and therefore, the entanglement bootstrap axioms {\bf A0} and {\bf A1} are approximately satisfied.
It has chiral central charge $c_{-}=1/2$ in the bulk and has chiral gapless edge, which is robust against any local perturbations. 
Next we check these expectations.

\subsubsection{Tests of the ground state} \label{sec:axioms-test-p+ip}
Before we test the construction of the approximate Virasoro generators, we first check whether $p+\ii p$ ground state is indeed a suitable state to do the constructions. Therefore, we will check the following: (1) The entanglement bootstrap axioms {\bf A0} and {\bf A1} in the bulk as well as the full boundary {\bf A0} and {\bf A1}. Recall this property is crucial in our argument that the bulk reduced density matrices are invariant under good modular flow. (2) Modular commutators. This quantity indicates the chirality of the state and computes the chiral central charge $c_{-} = c - \bar c$ from the bulk. (3) Vector fixed-point equation: 
As argued in~\cite{Lin2023, strength-of-FPE},
the satisfaction of the vector fixed-point equation near the edge is a strong indication that the edge is described by a CFT. Moreover, the proportionality factor in the vector fixed-point equation contains the total central charge $\ctot = c + \bar c$ of the edge CFT. If the total central charge $\ctot$ is equal to the absolute value of chiral central charge $c_{-}$, then this indicates the edge CFT is purely chiral. Therefore, this test provides strong evidence that Hypothesis~\ref{assump:chiral-CFT} is satisfied by the state. 

\begin{figure}[!h]
    \centering
    \includegraphics[width=0.90\columnwidth]{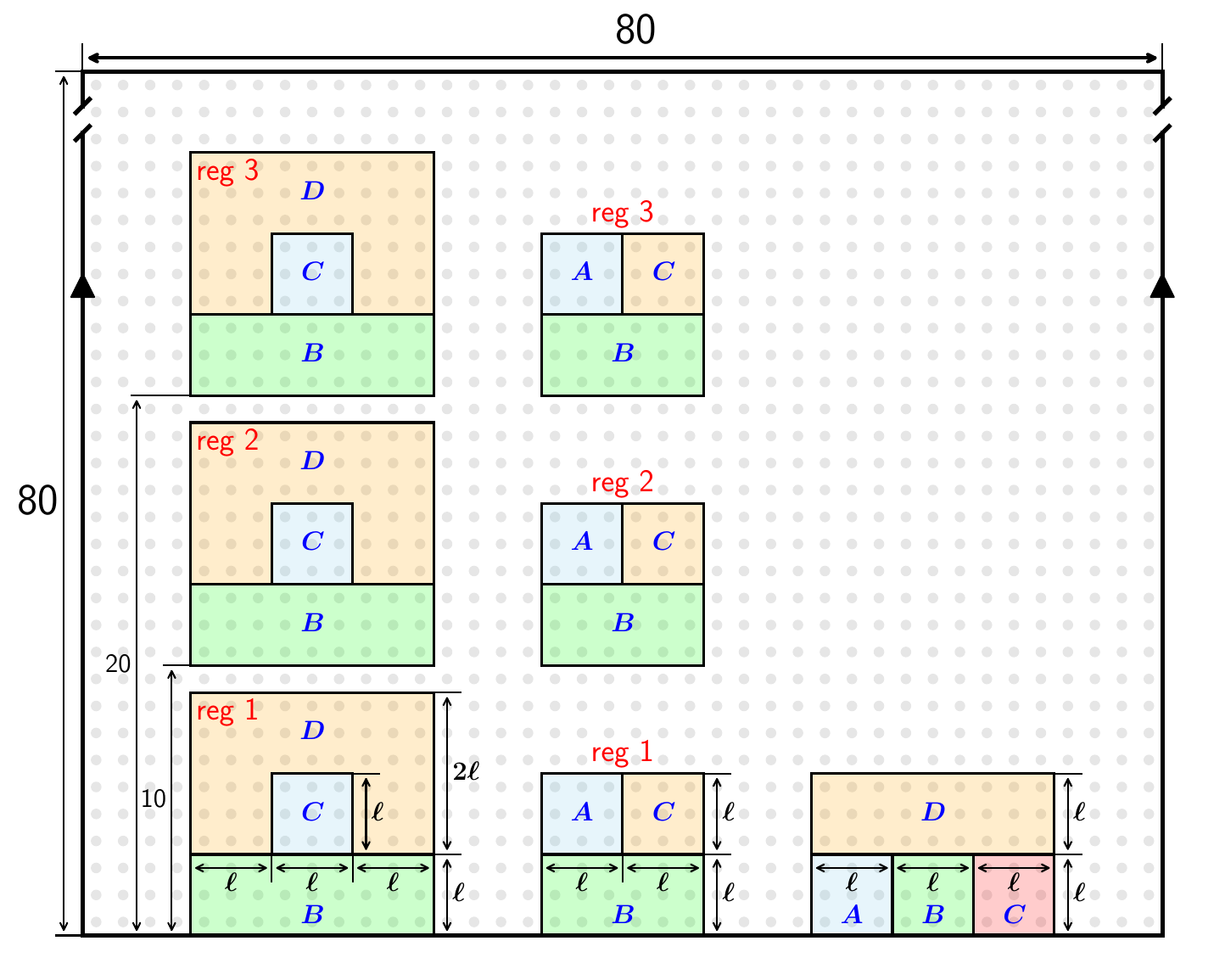}
    \caption{Subsystems used in numerical tests. $BCD$ in regions 1, 2, and 3 in the left column are used to test bulk entanglement bootstrap axioms {\bf A0} and {\bf A1}. $ABC$ in regions 1, 2, 3 of the middle column are used to test the modular commutator  formula. The region $ABCD$ in the right column, near the edge, is used to test the vector fixed-point equation. The location of the subsystems can be inferred from the plot. We omit the $x$-coordinate of each region because the ground state is translational invariant in that direction.}
    \label{fig:axioms-regions}
\end{figure}

We choose the subsystems shown in Figure~\ref{fig:axioms-regions} on a cylinder with size $L_x = 80, L_y = 80$. The tests are explicitly enumerated as follows:  
\begin{enumerate}
\item For \textbf{A0} and \textbf{A1}, we compute 
\begin{equation}
    \Delta(BD,C,\emptyset) = S_{BCD}+S_{C}-S_{BD},\quad \Delta(B,C,D) = S_{BD}+S_{CD}-S_B-S_D 
\end{equation}
respectively, and see if they are close to zero. The subsystems are regions 1,2,3 shown in the left column in in Figure~\ref{fig:axioms-regions}, which are progressively farther away from the gapless edge. We also compute the full-boundary {\bf A0} and {\bf A1}, using the setup shown in Figure~\ref{fig:boundary-axioms-regions}. 
\begin{figure}[!h]
    \centering
    \includegraphics[width=0.8\columnwidth]{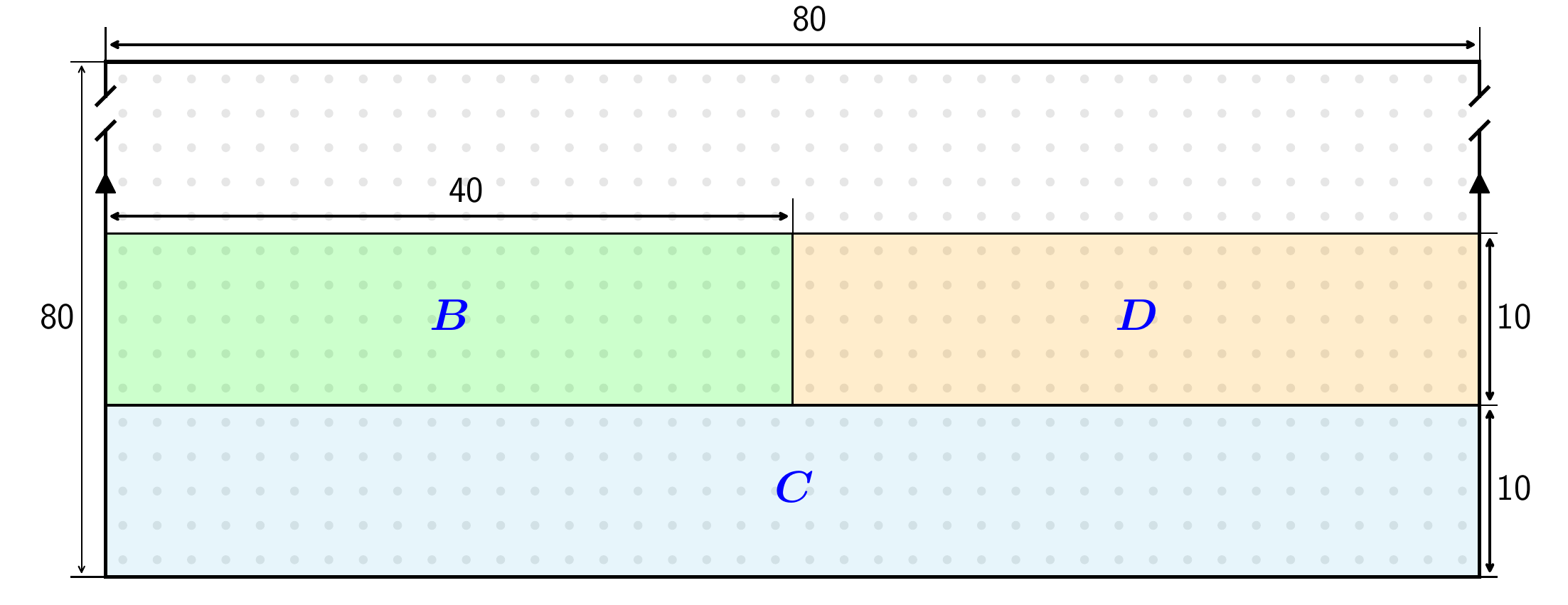}
    \caption{Subsystems used in the test of full boundary axioms {\bf A0} and {\bf A1}.}
    \label{fig:boundary-axioms-regions}
\end{figure}
\item For the modular commutator, we computed 
\begin{equation}
    J(A,B,C)= \ii \bra{\Psi}[K_{AB},K_{BC}]\ket{\Psi}
\end{equation}
for regions 1,2,3 shown in the middle column, which are also progressively farther away from the gapless edge. We compare the results with $c_{-}\pi/3$ with $c_{-}=1/2$.
\item For the vector fixed-point equation \eqref{eq:bdy-vector-fixed-point-equation}, the setup is shown in the right column in Figure~\ref{fig:axioms-regions}. We construct $\hat{\Delta}(AD,B,C) = K_{ABD}+K_{BC}-K_{AD}-K_C$ and $\hat{I}(A:C|B) = K_{AB}+K_{BC}-K_{ABC}-K_B$ to obtain 
\begin{equation}
    \KD(\eta) = \eta\hat{\Delta}(AD,B,C)+(1-\eta)\hat{I}(A:C|B),
\end{equation}
with $\eta$ being the cross-ratio of the intervals $(a,b,c)$
\begin{equation}
    \eta = \frac{\sin(\pi|a|/L_x)\sin(\pi|c|/L_x)}{\sin(\pi|bc|/L_x)\sin(\pi|ab|/L_x)}.
\end{equation} 
To test $\KD(\eta)\ket{\Psi}\propto\ket{\Psi}$, we compute the square root of its variance: 
\begin{equation}
    \sigma(\KD(\eta)) = \sqrt{\bra{\Psi} \KD(\eta)^2\ket{\Psi} - \bra{\Psi}\KD(\eta)\ket{\Psi}^2 }
\end{equation}
and see if it vanishes. We also compute the proportionality factor 
\begin{equation}
    \bra{\Psi}\KD(\eta)\ket{\Psi} = \frac{\ctot}{6}h(\eta),
\end{equation}
with $h(\eta) = -\eta\ln(\eta) - (1-\eta)\ln(1-\eta)$, and test whether $\ctot = 1/2$, which is an indication that the edge CFT is purely chiral. 
\end{enumerate}

\begin{figure}[!h]
    \centering
    \includegraphics[width=0.95\columnwidth]{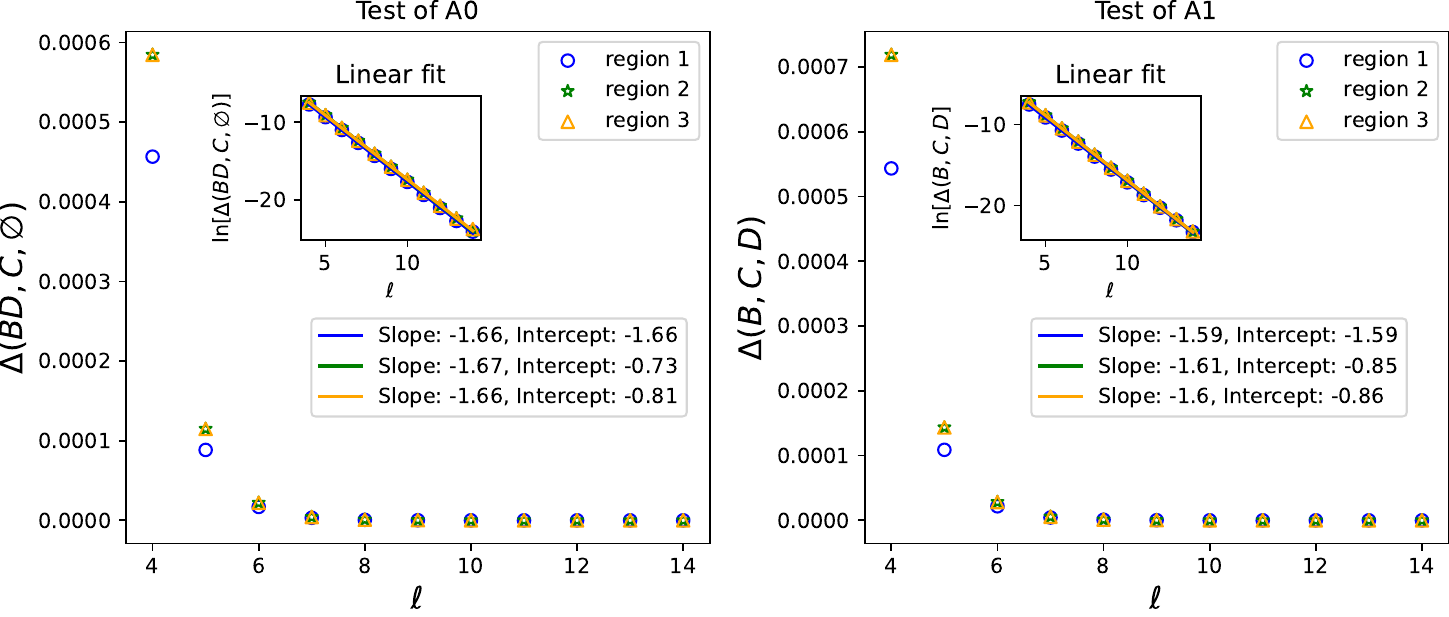}
    \caption{Numerical results for the violation of entanglement bootstrap axioms {\bf A0} and {\bf A1} in the $p+ \ii p$ SC ground state as a function of the linear size of the subsystem $\ell$, as shown in Figure~\ref{fig:axioms-regions}. From the slope of the linear fit result of {\bf A0}, one can extract an upper bound of the correlation length, as explained in Eq.~\eqref{eq:corr-length}}
    \label{fig:EB-axioms}
\end{figure}

The linear sizes of the subsystems for these tests are labeled by $\ell$, as shown in Figure~\ref{fig:EB-axioms}. 
We now describe the numerical results. 

{\it (1) Entanglement bootstrap axioms:} The results for axioms {\bf A0, A1} are shown in Figure~\ref{fig:EB-axioms}. 
For the parameters we chose, both $\Delta(BD,C,\emptyset)$ and $\Delta(B,C,D)$ decay to zero exponentially as the linear size $\ell$ of the subsystem increases. For some different choices of parameters,\footnote{One instance of such parameter is $\Delta=0.9,\mu=1.3$ in Eq.~\eqref{eq:pipH-2}. Such an algebraic decay feature persists if one varies these parameters around this choice.} as far as we observed, \textbf{A1} and modular commutator could decay algebraically, while \textbf{A0} still decays exponentially. It is interesting that the means of approach to the fixed point seems not to be a universal property of the phase.  The reason for such a distinction is a topic for further study.

As an aside, the rate of exponential decay of $\Delta(BD,C,\emptyset)$ (i.e.~the entanglement entropy combination that appears in \textbf{A0}) can be used as a characterization of correlation length, as it upper bounds any correlation functions in the following way:
\begin{equation}
    \Delta(BD,C,\emptyset)\geq I(A:C)\geq \frac{1}{2}\left( \frac{\langle \calO_A \calO_C\rangle_c}{\| \calO_A \| \cdot \|\calO_C \|} \right)^2, 
\end{equation}
where $A$ is any region outside $BCD$, for $BCD$ shown in the left column in Figure~\ref{fig:axioms-regions}, and $\calO_A,\calO_C$ are any two operators supported on $A$ and $C$ respectively. The first inequality follows from SSA. 
The second inequality is a well-known result in quantum information theory \cite{Wolf:2007tdq}, where $\langle \calO_A\calO_C\rangle_c \equiv \langle \calO_A \calO_C\rangle - \langle \calO_A\rangle \langle \calO_C\rangle $ stands for the connected correlation function and $\|\calO\|$ is defined as  $ \mathrm{sup}\left\{ \left. \sqrt{\langle \psi| \calO^{\dagger} \calO|\psi\rangle}\right|\forall \ket{\psi} \text{ w.t. }|\langle \psi |\psi\rangle| = 1 \right\}$. Therefore, from the exponential decaying violation of \textbf{A0}, i.e. $\Delta(BD,C,\emptyset) \propto e^{-\ell/\xi_*}$, one can obtain an upper bound on the correlation length $\xi$ of the correlation function $\langle \calO_A\calO_C\rangle_c \propto e^{-\ell/\xi}$: 
\begin{equation}\label{eq:corr-length}
    \Delta(BD,C,\emptyset) = \alpha e^{-\ell/\xi_*} \geq \frac{1}{2}\left( \frac{\langle \calO_A \calO_C\rangle_c}{\| \calO_A \| \cdot \|\calO_C \|} \right)^2 = \beta e^{-2\ell/\xi} \quad \Rightarrow \quad 
    \xi\leq 2\xi_*,  
\end{equation}
where $\alpha,\beta$ are two prefactors which are not important in the current discussion. As a result, from the linear fit of $\ln(\Delta(BD,C,\emptyset))$ versus $\ell$ shown in the first plot in Figure~\ref{fig:EB-axioms}, we can conclude that the correlation length $\xi$ of any correlation function satisfies $ \xi \le 2/1.66 \approx 1.2$ lattice spacings, for the choice of $p+\ii p$ Hamiltonian parameters used in the numerics.    

We also computed the violation of full-boundary {\bf A0, A1} in the setup shown in Figure~\ref{fig:boundary-axioms-regions}. 
The results are of the order $O(10^{-8})$. Recall such full-boundary axioms play an important role in the cylinder argument [Section~\ref{subsec:cylinder-argument}], as well as in the discussion of relation between state overlap and fidelity of reduced density matrices on the edge region for comparison between the reference state and coherent states [Section~\ref{subsec:state-difference}]. The latter shall be tested in the following subsubsection~\ref{subsubsec:coherent-state-tests}. 

\begin{figure}[!h]
    \centering
    \includegraphics[width=0.5\columnwidth]{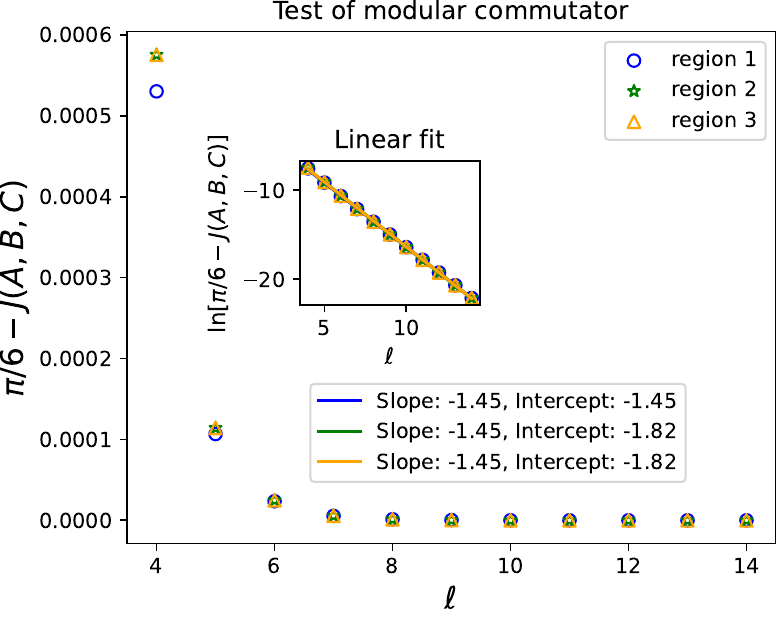}
    \caption{Difference of the numerical results of the modular commutator from its theoretical expectation in the $p+\ii p$ SC ground state, for various regions, as a function of the linear size of the region $\ell$ shown in Figure~\ref{fig:axioms-regions}.}
    \label{fig:modular-commutator}
\end{figure}

{\it (2) Modular commutators:} The results for modular commutators are shown in Figure~\ref{fig:modular-commutator}. With this choice of parameters, the modular commutators $J(A,B,C)$ approach the expected value $\frac{\pi c_{-}}{3}$ with $c_{-} = 1/2$ exponentially, which agrees with the fact that the non-trivial phase of the $p+\ii p$ superconductor 
has a chiral central charge $1/2$. As for the test of {\bf A1}, we also observe that $J(A,B,C)$ can approach $\pi/6$ algebraically for certain other choices of parameters. The fact that the data for regions 1, 2 and 3 largely overlap corroborates the fact that the state satisfies bulk \textbf{A1} with pretty small violation, because bulk \textbf{A1} enables one to deform or move the regions of the subsystems used in the modular commutator in the bulk without changing its value \cite{Kim2021}\footnote{One might be concerned that region 1 for modular commutator test touches the gapless edge. We comment that the proof using {\bf A1} still goes through, as long as the triple intersection point is far away (compared with the correlation length) from the edge and $A,B,C$ are not all intersecting with the gapless edge.}. 

\begin{figure}[!h]
    \centering
    \includegraphics[width=0.95\columnwidth]{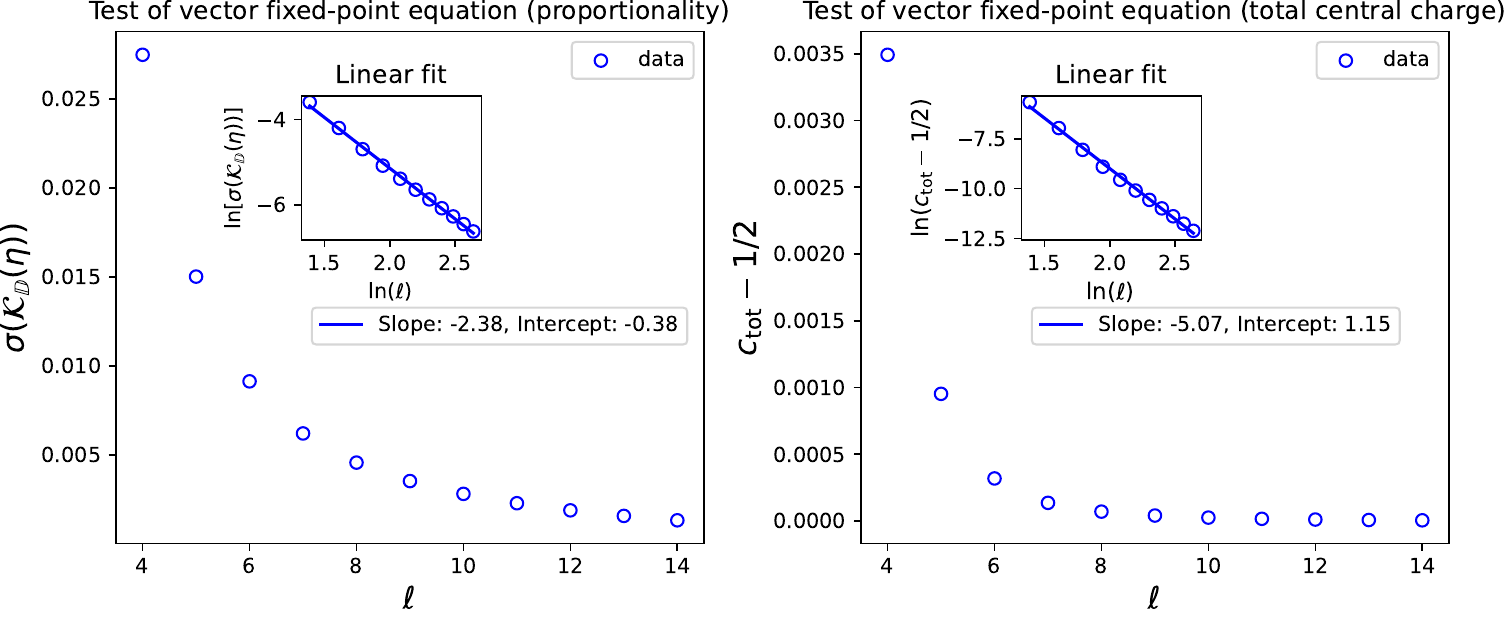}
    \caption{Left: The amount of violation of the vector fixed-point equation in the $p + \ii p $ superconductor ground state, as a function of linear region size $\ell$, as shown in Fig.~\ref{fig:axioms-regions}.   The $\eta$ used in the fixed point equation is the geometrical cross-ratio.  Right: The deviation of the central charge inferred from the RHS from its expected value.  The fact that we find $c_\text{tot} = 1/2$ is evidence that the state is indeed purely chiral.}
    \label{fig:vector-id}
\end{figure}

{\it (3) Vector fixed-point equation:} The results of tests of the vector fixed-point equation are shown in Figure~\ref{fig:vector-id}. The left column shows the results for testing the proportionality via $\sigma(\KD(\eta))$, and the right column shows the results for the total central charge computed from the proportionality factor. Both results satisfy our expectations with algebraically decreasing discrepancy. As the system size grows, $\sigma(\KD(\eta))\to 0$ indicates the CFT nature of the gapless edge. And $\ctot$ approaches  $1/2$, the expected chiral central charge, indicating that the CFT is purely chiral. 

In summary, the numerical data indicates that: (1) The $p+\ii p$ SC ground state $\ket{\Psi}$ from the Hamiltonian Eq.~\eqref{eq:pipH-2} with such choice of parameters satisfies EB axioms {\bf A0}, {\bf A1} (both bulk version and full boundary version) and is in a non-trivial chiral phase. (2) It has a purely chiral CFT on the gapless edge, whose total central charge agrees with the chiral central charge computed from the bulk wavefunction. Therefore, such a state is ideal for the construction of approximate Virasoro generators.  

\subsubsection{Tests of constructions of Virasoro generators}
\label{subsec:virasoro_tests}
In the previous subsubsection, we concluded that the $p+\ii p$ SC ground state from the Hamiltonian Eq.~\eqref{eq:pipH-2} with such a choice of parameters listed in [Section~\ref{sec:fidelity_tests}] is suitable for the construction of approximate Virasoro generators $\Leff_n$. We now put the system on a cylinder of size $L_x=240, L_y=24$ and test the construction of $\Leff_n$ from its ground state $\ket{\Psi}$. Such a choice of total system size is large enough so that the linear size of the smallest subsystem used in the construction of $\Leff_n$ is much larger than the correlation length, as discovered in the previous subsubsection. The gapless edge is along the $x$-direction of the cylinder. In order to construct $\Leff_n$ with several choices of $n$, we choose a pretty large $L_x$. This is because the construction of $\Leff_n$ with large $n$ requires dividing the gapless edge into many intervals, and a large $L_x$ can make sure that the length of the smallest interval is much larger than the correlation length. In the $y$-direction, the construction does not demand many lattice sites, so $L_y$ does not have to be very large. 

\begin{figure}[!ht]
    \centering
    \includegraphics[width=0.9\columnwidth]{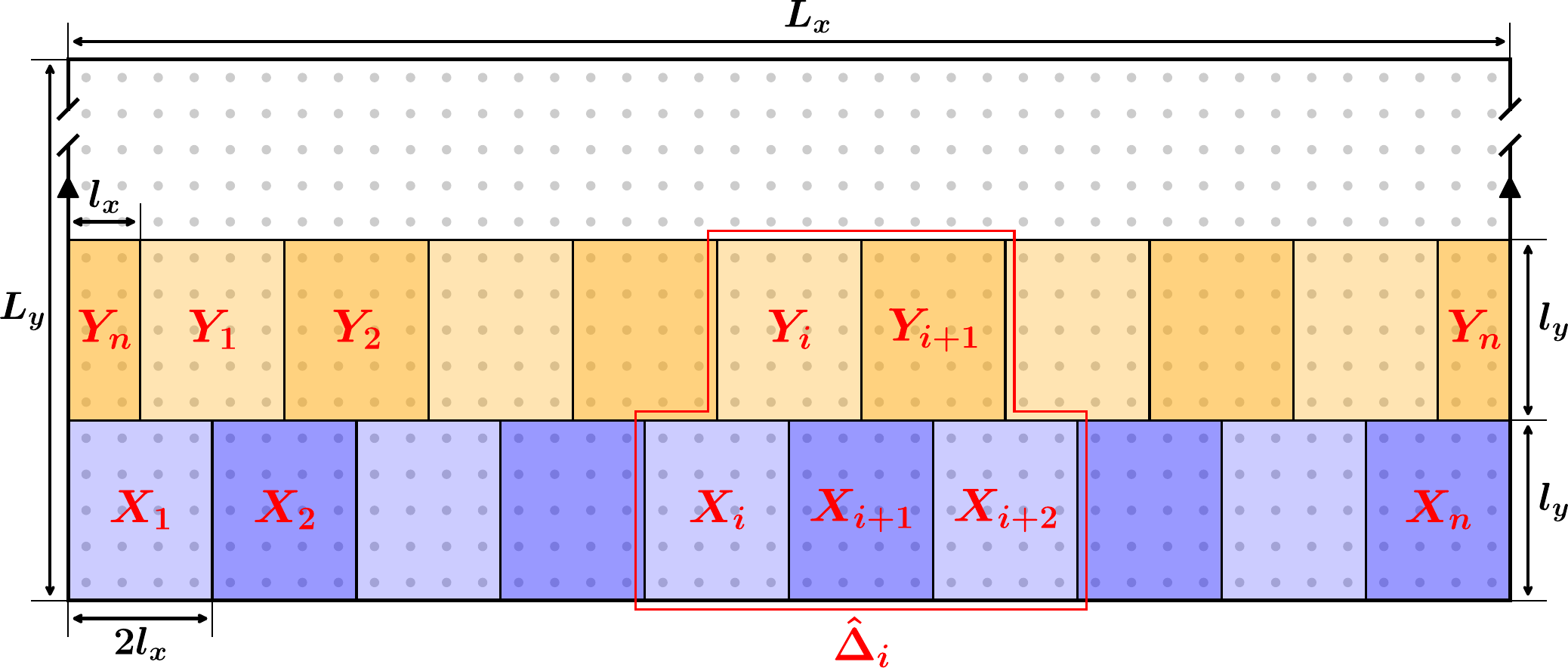}
    \caption{ Regions used in the construction of $\Leff_0$. The locations of equal-sized ($2l_x \times l_y$) regions $X_i, Y_i$ can be inferred from the figure. In the numerical tests, we used $L_x = 240, L_y = 24,l_x = 20,l_y = 6$.   Each gray dot in the picture is not quite a coarse-grained site for a literal representation of our numerical setup.}
    \label{fig:L0-num}
\end{figure}

\begin{figure}[!ht]
    \centering
    \includegraphics[width=0.9\columnwidth]{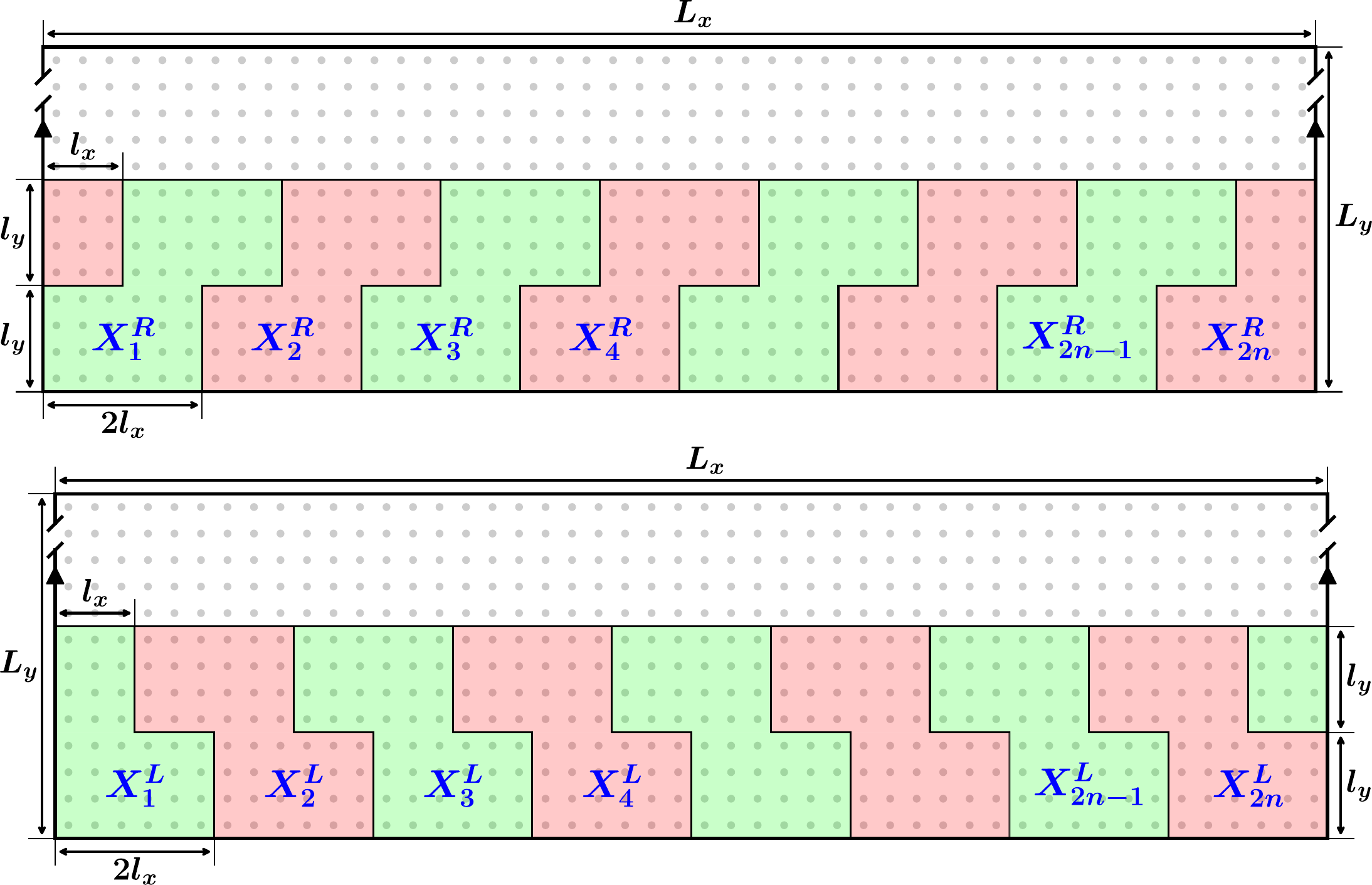}
    \caption{Construction of $\Lbo_n$ (and $\Lbe_n$). The location and size of $X_j^R, X_j^L$ can be inferred from the diagram. The color indicates the sign of the modular Hamiltonian of that region used in $\Lbo_n$: Green ($+$), Red ($-$). In the numerical tests, we used $L_x = 240, L_y = 24,l_x = \frac{N_x}{4n},l_y = 6$.
    The regions used in the construction of $\Lbe_n$ are called $Y_k^R$ and $Y_k^L$, which are regions obtained from $X_k^R$ and $X_k^L$ by shifting to the left by $l_x$.}  
    \label{fig:Lno-num}
\end{figure}

Following the instructions given in Section~\ref{sec:virasoro-construction}, we construct 
\begin{equation}\begin{split}
    &\Leff_0 = \frac{1}{4\pi \tan\(\frac{\pi}{N}\)} \sum_{i=1}^{N} \left( \hat{\Delta}_i - \bra{\Psi}\hat{\Delta}_i\ket{\Psi}\right) \\ \textrm{where} \quad  & \hat{\Delta}_i = K_{X_{i}X_{i+1}Y_i} + K_{X_{i+2}X_{i+1}Y_{i+1}} - K_{X_{i}Y_i} - K_{X_{i+2}Y_{i+1}} \\  
\end{split}
\end{equation}
for regions $X_i, Y_i$, $i = 1, \cdots, N$ shown in Figure~\ref{fig:L0-num}.
For $n \ge 1$, we construct approximate Virasoro generators $\Leff_n$ and $\Leff_{-n}$ as
\begin{equation}\begin{split}
    &\Leff_n = \Lbe_n +i \Lbo_n , \quad \Leff_{-n} = \Lbe_n - i \Lbo_n \\ 
    \textrm{with} \quad &\Lbo_n = \frac{A_n}{2}\sum_{j=1}^{2n} (-1)^{j-1} \( K_{X_j^R} + K_{X_j^L}\) \\
    \textrm{and} \quad &\Lbe_n = \frac{A_n}{2}\sum_{j=1}^{2n} (-1)^{j-1} \( K_{Y_j^R} + K_{Y_j^L} \), 
\end{split}
\end{equation} 
where regions $X_j^R, X_j^L, Y_j^R, Y_j^L$ are shown in Figure~\ref{fig:Lno-num} and 
\begin{equation}%\label{eq:An}
    A_n = \left\{ \begin{aligned}
        & \frac{1}{2\pi} &   &n = 1 \\ 
        &  \tan\( \frac{\pi}{2n} \)(n^2-1)/8,& &n\neq 1  
    \end{aligned}\right. . 
\end{equation}

To test the construction, we will numerically compute several quantities involving $\Leff_n$, and make the following two comparisons: 
\begin{itemize}
    \item We compare the numerical results with theoretical predictions discussed in Section~\ref{subsec:quant-properties}. The theoretical prediction is made under the assumption that $\ket{\Psi}$ satisfies Hypothesis~\ref{assump:chiral-CFT} is exactly, so that one can do the computation under the CFT ground state $\ket{\Omega}$. We shall call these theoretical results as ``CFT prediction''. This comparison tests the validity of the Hypothesis~\ref{assump:chiral-CFT}. 
    \item We compare the numerical results with those quantities from exact Virasoro generator $L_n$. The latter is computed also under the assumption that Hypothesis~\ref{assump:chiral-CFT} is satisfied\footnote{When we talk about $L_n$, we should assume Hypothesis~\ref{assump:chiral-CFT} is satisfied, as it is defined by $L_n \veq \Lcft_n$ which utilize $\mathbb{V}$ in the hypothesis. Note we are not constructing $L_n$ in the lattice system and computing the results from them.}. These results are titled with ``Virasoro algebra'' (or ``Vir'' for short) when we list the data. Such a comparison estimates the quality of approximation of $\Leff_n$ to $L_n$. 
\end{itemize}
Before we present the results, we emphasize that by ``\emph{comparison}'' of $x$ with $y$, we explicitly mean $\left|{x - y \over y }\right|$ when we list the data below.

We first examine some selected commutation relations $\bra{\Psi} [\Leff_n, \Leff_{-n}] \ket{\Psi}$ for $n = 1,2,\cdots 6$, and make comparisons with the CFT predictions
\begin{equation}\label{eq:Lnn-cft}
    \text{CFT prediction: }\bra{\Psi} [\Leff_n, \Leff_{-n}] \ket{\Psi} = \frac{\pi c}{24} n(n^2-1)^2 \frac{\sin^2(\pi/(4n))}{\cos(\pi/(2n))}  
\end{equation}
with the central charge $c=1/2$, which is computed in Section~\eqref{eq:Lnn-analytic} in Section~\ref{subsubsec:expt-comm} under the Hypothesis~\ref{assump:chiral-CFT}. The results are listed in Table~\ref{tab:commutator-expectations}. The pretty small discrepancy ($10^{-5}$ in magnitude) between the numerical results and CFT expectation indicates the reference state $\ket{\Psi}$ we used in numerics satisfies the hypothesis pretty well on the length scale we chose. We also compare the data with results from exact Virasoro generators  
\begin{equation}\label{eq:Lnn-vir}
    \text{Vir: }  \bra{\Psi} [L_n,L_{-n}] \ket{\Psi}  = \bra{\Omega} [\Lcft_n,\Lcft_{-n}] \ket{\Omega} = \frac{c}{12} (n^3-n), \quad c=\frac{1}{2},
\end{equation}
which are computed under Hypothesis~\ref{assump:chiral-CFT}. The dominating source for the deviation is because we ignore the higher harmonic terms in $\Leff_n$ even though we know how to compute them. The purpose of such a comparison is to test the quality of the approximation of $\Leff_n$ to $L_n$. The fact that the ``error'' is still small, indicates that higher harmonics only contribute a small correction and supports the name ``approximate Virasoro generators'' for $\Leff_n$. Later in appendix~\ref{app:improvement}, we shall improve the construction and do similar comparisons. As we shall see the ``approximation error'' is reduced as expected. 

\begin{table}[!h]
\begin{tabular}{|l|l|l|l|l|l|}
\hline
$n$ & $\langle   \Psi|[\Leff_n, \Leff_{-n}] \Psi\rangle$ & \makecell{CFT \\ predictions} & \makecell{Virasoro algebra \\ $\bra{\Omega}[\Lcft_{n},\Lcft_{-n}]\ket{\Omega}$} & \makecell{\small Comparison \\ \small with \\ \small CFT prediction} & \makecell{\small  Comparison \\ \small with  \small Vir} \\ \hline
1   & $O(10^{-8})$                                       & 0                                & 0                            & NA                               & NA                                         \\ \hline
2   & 0.24399045                                         & 0.24399193                       & 0.25                         & $6.05\times   10^{-6}$           & 2.404\%                                    \\ \hline
3   & 0.97200578                                         & 0.97201215                       & 1                            & $6.55\times   10^{-6}$           & 2.799\%                                    \\ \hline
4   & 2.42663026                                         & 2.42665060                        & 2.5                          & $8.38\times   10^{-6}$           & 2.935\%                                    \\ \hline
5   & 4.85014000                                         & 4.85020037                       & 5                            & $1.245\times   10^{-5}$          & 2.997\%                                    \\ \hline
6   & 8.48474652                                         & 8.48491574                       & 8.75                         & $1.994\times   10^{-5}$          & 3.031\%                                    \\ \hline
\end{tabular}
\caption{Results for expectation values of some selected commutators among approximate Virasoro generators $\Leff_n$, and their comparison with CFT predictions and with results from exact Virasoro generator.}
\label{tab:commutator-expectations}
\end{table}

Secondly, we test the expectation values of double commutators $\bra{\Psi} [[\Leff_m,\Leff_n],\Leff_k]\ket{\Psi}$ among the approximate Virasoro generators. As above, the tests are summarized into two groups. The first group is the computation of $\bra{\Psi} [[\Leff_n,\Leff_0],\Leff_{-n}]\ket{\Psi}, n=1,\cdots,6$. The results are listed in Table~\ref{tab:Ln0n}. We first compare the results with the CFT prediction, as computed in Eq.~\eqref{eq:Ln0n-analytic} in Section~\ref{subsubsec:expt-double-comm}:
\begin{equation}\label{eq:Ln0n-cft-num}
    \text{CFT prediction: } \bra{\Psi} [[\Leff_n,\Leff_0],\Leff_{-n}]\ket{\Psi} =  \frac{\pi c}{48} n (n^2-1)^2 \tan\left( \frac{\pi}{2n} \right),
\end{equation}
as well as the results from the exact Virasoro generators: 
\begin{equation}\label{eq:Ln0n-vir}
    \text{Vir: } \bra{\Psi} [[L_n,L_0],L_{-n}]\ket{\Psi} = \bra{\Omega} [[\Lcft_n,\Lcft_0],\Lcft_{-n}]\ket{\Omega} = \frac{c}{12}(n^4-n^2). 
\end{equation}
The central charge in Eq.~\eqref{eq:Ln0n-analytic} and Eq.~\eqref{eq:Ln0n-vir} are both $c=1/2$. The results are listed in Table~\ref{tab:Ln0n}. The sources of the discrepancy shown in both comparisons are similar to those we discussed in the commutator test above. 
The second group is $\bra{\Psi} [[\Leff_m,\Leff_n],\Leff_k]\ket{\Psi}$ of some selected non-zero $m,n,k$, and we compare the results with those obtained by Virasoro algebra:
\begin{equation}
    \text{Vir: }\bra{\Psi} [[L_m,L_n],L_k]\ket{\Psi} = \bra{\Omega} [[\Lcft_m,\Lcft_n],\Lcft_k]\ket{\Omega} = \frac{c}{12}(m-n)(k^3-k)\delta_{m+n,-k},
\end{equation}
with $c = 1/2$. The results are listed in Table~\ref{tab:Lmnk}
We also computed the results using ``improved'' approximate Virasoro algebra in Appendix~\ref{app:improvement}, and the approximation error is reduced as expected. 

\begin{table}[]
\begin{tabular}{|l|l|l|l|l|l|}
\hline
$n$ & {\small $\bra{\Psi}[[\Leff_n, \Leff_0],\Leff_{-n}]\ket{\Psi}$} & \makecell{\small CFT \\ prediction} & \makecell{\small $\bra{\Omega}[[\Lcft_n,\Lcft_0],$  \\ $ \Lcft_{-n}]\ket{\Omega}$} & \makecell{\small Comparison \\ \small with \\ \small CFT prediction} & \makecell{\small  Comparison \\ \small with  \small Vir} \\  \hline
1   & $O(10^{-5})$                                          & 0              & 0                                                         & NA                              & NA                                       \\ \hline
2   & 0.58734                                               & 0.58905        & 0.5                                                       & 0.290\%                         & 14.981\%                                 \\ \hline
3   & 3.60489                                               & 3.62776       & 3                                                         & 0.626\%                         & 15.834\%                                 \\ \hline
4   & 12.07894                                              & 12.19960       & 10                                                        & 0.989\%                         & 15.006\%                                 \\ \hline
5   & 30.25392                                              & 30.62296       & 25                                                        & 1.205\%                         & 13.150\%                                 \\ \hline
6   & 63.53049                                              & 64.44933       & 52.5                                                      & 1.426\%                         & 11.298\%                                 \\ \hline
\end{tabular}
\caption{Results of $\bra{\Psi}[[\Leff_n, \Leff_0],\Leff_{-n}]\ket{\Psi}$ and comparison with CFT prediction and results of Virasoro algebra generators.} 
\label{tab:Ln0n}
\end{table}

\begin{table}[!h]
\begin{tabular}{|l|l|l|l|}
\hline
$(m,n,k)$ & $\bra{\Psi}[[\Leff_m, \Leff_n],\Leff_{k}]\ket{\Psi}$ & $\bra{\Omega}[[\Lcft_m, \Lcft_n],\Lcft_{k}]\ket{\Omega}$ & \makecell{ Comparison with \\ Virasoro algebra} \\ \hline
$(2,1,-3) $ & $1.00376+2.0\times   10^{-5} \ii$                      & 1                                                        & 0.376\%                                   \\ \hline
$(3,1,-4)$  & $4.95632+1.3\times   10^{-4} \ii$                      & 5                                                        & 0.874\%                                   \\ \hline
$(3,2,-5)$  & $4.98279   + 0.0 \ii$                                  & 5                                                        & 0.344\%                                   \\ \hline
$(4,2,-6)$  & $17.40158-6\times   10^{-4} \ii $                      & 17.5                                                     & 0.562\%                                   \\ \hline
\end{tabular}
\caption{Tests of $\bra{\Psi}[[\Leff_m, \Leff_n],\Leff_{k}]\ket{\Psi}$ and comparison with the Virasoro generators.}
\label{tab:Lmnk}
\end{table}

\subsubsection{Tests about coherent states}
\label{subsubsec:coherent-state-tests}

After testing the constructions of approximate Virasoro generators, we now test some selected properties of coherent states, defined as $\ket{\Psi(t)} = e^{i\Lb t} \ket{\Psi}$ for some good modular flow generator $\Lb\in\gc$ [Section~\ref{sec:edge-subspace-coherent}]. 

To be more explicit, we use $p+\ii p$ SC with the same parameters as we used before and put the system on a cylinder with $L_x = 72$, $L_y=18$. The boundary condition is the same as before: open boundary condition in $y$ direction and anti-periodic boundary condition in $x$-direction. We construct $\Lbo_2$ and $\Lbo_3$ to generate the good modular flow acting on the reference state. The construction is similar to what we did in the previous subsubsection shown in Figure~\ref{fig:Lno-num}, in which the size parameters for the subsystems used in the construction are $l_x = L_x/(4n),l_y = 6$. 

We focus on testing the ``difference'' between the coherent state $\ket{\Psi(t)}$ and the reference state $\ket{\Psi} = \ket{\Psi(0)}$, which is characterized by the fidelity $F(\rho_{X}(0),\rho_{X}(t))$ between density matrices $\rho_X(0),\rho_X(t)$ from $\ket{\Psi}$ and $\ket{\Psi(t)}$. We choose the region $X$ as the entire system and the entire edge region. Explicitly, the edge region in the test is a cylinder that includes the entire gapless edge on which the modular flows are acting and extends 5 lattice spacings into the bulk. 

Based on the idea that good modular flow preserves the bulk state, we expect the two choices of $X$ to give the same results [Section \ref{subsec:state-difference}]\footnote{Since the system is on a cylinder, the proof of this statement requires full boundary {\bf A0}. As we showed in Section~\ref{sec:axioms-test-p+ip}, indeed, with the current choice of Hamiltonian parameters and boundary conditions, the full boundary {\bf A0} is satisfied with considerably small errors. Therefore, the proof is applicable here.}. 

\begin{figure}[!h]
    \centering
    \includegraphics[width=0.8\columnwidth]{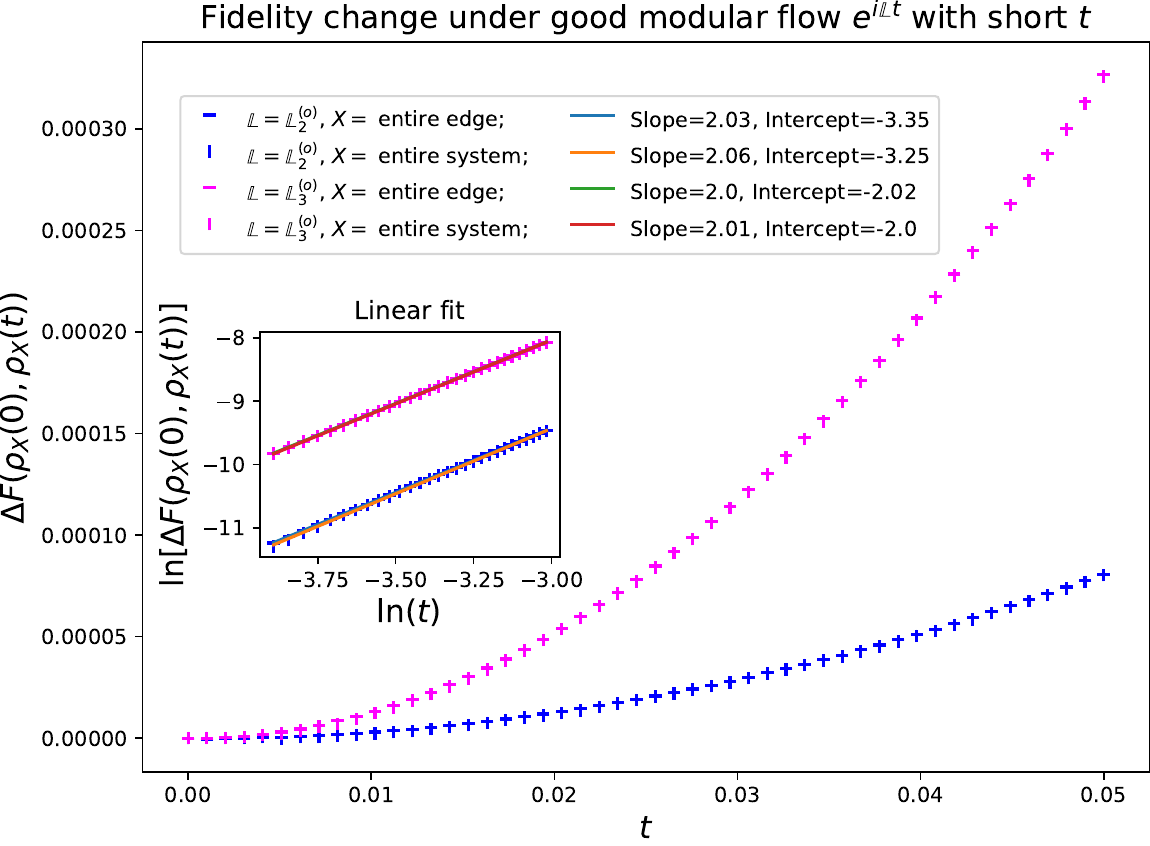}
    \caption{Change of fidelity $\Delta F(\rho_X(0),\rho_X(t))\equiv F(\rho_X(0),\rho_X(0))-F(\rho_X(0),\rho_X(t))$ between reduced density matrices $\rho_X(0)$ and $\rho_X(t)$ under good modular flow $e^{\ii \Lb t}$, with flow generator $\Lb$ taken to be $\Lbo_2,\Lbo_3$. $\rho_X(t)$ is the reduced density matrix of $\ket{\Psi} = e^{\ii \Lb t}\ket{\Psi}$ of region $X$. The region $X$ is chosen to be the entire system and the entire edge.} 
    \label{fig:fidelity-alpha}
\end{figure}

The change of fidelity with the evolution time $\Delta F(\rho_X(0),\rho_X(t)) \equiv F(\rho_X(0),\rho_X(0)) - F(\rho_X(0),\rho_X(t))$ is plotted in Figure~\ref{fig:fidelity-alpha}. We also compute $\alpha(\Lbo_2,X),\alpha(\Lb_3,X)$ defined in Eq.~\eqref{eq:alpha} in Section~\ref{sec:fid-general}, by fitting the data of 
\begin{equation}
    \Delta F(\rho_X(0),\rho_X(t)) \approx \frac{1}{2}\alpha(\Lb,X)t^2,
\end{equation}
for the modular flow generated by $\Lb=\Lbo_2,\Lbo_3$ as a function of $t$ in the small $t$ regions. Based on the argument in Section~\ref{subsec:state-difference}, we shall expect they are equal to $\sigma(\Lbo_2)^2$, $\sigma(\Lbo_3)^2$. Recall from Section~\ref{subsubsec:variance} that we can analytically compute $\sigma(\Lbo_n)^2$ using Eq.~\eqref{eq:sigma-Lb-analytic} under the Hypothesis~\ref{assump:chiral-CFT}. Therefore, we shall compare the results of $\alpha(\Lbo_2),\alpha(\Lbo_3)$ with the CFT prediction of $\sigma(\Lbo_2)^2,\sigma(\Lbo_3)^2$ from Eq.~\eqref{eq:sigma-Lb-analytic}. Moreover, we also compare the numerical evaluation of $\sigma(\Lbo_2)^2,\sigma(\Lbo_3)^2$ with the CFT prediction, which estimates how well the reference state from the lattice system satisfies the Hypothesis~\ref{assump:chiral-CFT}. The data and comparison results are listed in Table~\ref{tab:alpha}. 

\begin{table}[!h]
\begin{tabular}{|l|l|l|l|l|}
\hline
$\Lb$          & \makecell{$\sigma(\Lb)^2$ CFT \\ prediction} & \makecell{$\sigma(\Lb)^2$ numerical \\ results} & $\alpha(\Lb,\text{entire system})$ & $\alpha(\Lb,\text{entire edge})$ \\ \hline
$\Lb = \Lbo_2$ & 0.065322                       & 0.065013 (0.05\%)                  & 0.077222 (18.22\%)                 & 0.070044 (7.23\%)                \\ \hline
$\Lb = \Lbo_3$   & 0.262994                       & 0.261825 (0.04\%)                  & 0.271779 (3.34\%)                  & 0.264401 (0.53\%)                \\ \hline
\end{tabular}
\caption{Results of $\sigma(\Lb)^2$ and $\alpha(\Lb,X)$. $\Lb$ is chosen to be $\Lbo_2,\Lbo_3$ and $X$ is chosen to be the entire system and entire edge. The data in the second row is the CFT prediction. The rest of the data are numerical results. The comparison between numerical results with CFT prediction is given in the parenthesis after each data.}
\label{tab:alpha}
\end{table}

The agreement in Fig.~\ref{fig:fidelity-alpha} and Table~\ref{tab:alpha} supports the theoretical prediction that the action of good modular flow generators is only on the gapless edge.  
Furthermore, the agreement between the numerical results for the variance and the CFT prediction further supports the validity of Hypothesis~\ref{assump:chiral-CFT}.  

%%%%%%%%%%%%%%%%%%
% Semion test starts here
%%%%%%%%%%%%%%%%%%

\subsection{Tests in chiral semion wavefunction}\label{sec:semion}

We also test the construction of approximate Virasoro generators starting from a chiral semion wavefunction. The chiral semion model describes a chiral topologically ordered phase with two abelian anyons $\mathbf{1}$ (vacuum) and $s$ (semion). The wavefunction we use is the discrete version of $\nu = 1/2$ bosonic Laughlin ground state introduced by Nielsen {\it et al} \cite{Nielsen2012}, with no semion excitations inserted. Such a wavefunction is expected to have chiral central charge $c_{-}=1$ in the bulk and a purely-chiral CFT ground state on the edge with total central charge $c_\text{tot}=1$. Therefore, it will satisfy Hypothesis~\ref{assump:chiral-CFT} and be suitable to apply our construction of Virasoro generators. 

The wavefunction is on a cylinder and the underlying degrees of freedom are arranged as ``ladders'' of qubits [Figure~\ref{fig:semion-region}]. The explicit expression of the wavefunction $\ket{\Psi}$ is introduced in Appendix~\ref{app:semion}. 

\begin{figure}[!h]
    \centering
    \includegraphics[width=0.76\columnwidth]{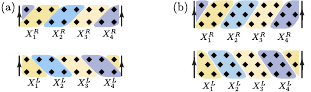}
    \caption{A finite cylinder on which we define the semion wavefunction. The horizontal direction is glued periodically. (a) on a 16-qubit 2-leg ladder, and (b) on a 24-qubit 3-leg ladder. See Appendix~\ref{app:semion} for the details of the wavefunction. We shall also need regions $Y$s, which are obtained from $X$s by a shift by one site to the right.} 
    \label{fig:semion-region}
\end{figure}

Following the same method as above, we construct 
\begin{equation}
    \Leff_{\pm 2} = \Lbe_2 \pm \ii \Lbo_2,
\end{equation}
where 
\begin{equation}
    \Lbo_2 = \frac{A_2}{2}\sum_{k=1}^{4}(-1)^k(K_{X_k^R}+K_{X_k^L}) ,\quad \Lbe_2 =  \frac{A_2}{2}\sum_{k=1}^{4}(-1)^k(K_{Y_k^R}+K_{Y_k^L}).
\end{equation}
The prefactor $A_2 = 3/8$ follows from Eq.~\eqref{eq:An}. Due to the limitation of system size allowed by the single value decomposition (SVD) method, we cannot construct approximate Virasoro generators of higher mode numbers. 

With the construction $\Lbo_2,\Lbe_2$ and $\Leff_{\pm 2}$, we compute the expectation values of commutators $\bra{\Psi}[\Leff_2,\Leff_{-2}]\ket{\Psi}$ and the variance $\sigma(\Lbo_2)^2$, and make comparisons with the theoretical prediction based on Hypothesis~\ref{assump:chiral-CFT}. We also compare the commutators of approximate Virasoro generators $\Leff_{\pm 2}$ with the results from exact Virasoro generators $L_{\pm 2}$ to test the quality of the approximation. 

The results are given in Table~\ref{tab:semion-L22} and Table~\ref{tab:semion-sigma}. We see that the finite-size corrections to the approximate Virasoro algebra are large for the 16-site system.  However, the agreement with CFT predictions is already quite good for the system with 24 sites. 

\begin{table}[!h]
\begin{tabular}{|l|l|l|l|l|l|}
\hline
Tests            & {\small $\bra{\Psi}[\Leff_2, \Leff_{-2}]\ket{\Psi}$} & \makecell{\small CFT\\ prediction} & {\small $\bra{\Omega}[\Lcft_2, \Lcft_{-2}]\ket{\Omega}$} & \makecell{\small Comparison \\ with CFT \\ prediction} & \makecell{\small Comparison \\ with Vir 
%\\ Virasoro algebra
} \\ \hline
16 sites         & 0.242078                                    & 0.487984       & 0.5                                             & 50.39\%                        & 51.58\%                          \\ \hline
24 sites & 0.442392                                    & 0.487984       & 0.5                                             & 9.34\%                         & 11.5\%                           \\ \hline
\end{tabular}
\caption{Same as Table~\ref{tab:commutator-expectations}, but for the semion wavefunction.}
\label{tab:semion-L22}
\end{table}

\begin{table}[!h]
\begin{tabular}{|l|l|l|l|}
\hline
Tests    & $\sigma(\Lbo_2)^2$ CFT prediction & $\sigma(\Lbo_2)^2$ numerical results & Comparison \\ \hline
16 sites & 0.129965                          & 0.110759                             & 14.78\%    \\ \hline
24 sites & 0.129965                          & 0.122377                             & 5.84\%     \\ \hline
\end{tabular}
\caption{Comparison of the variance of second Virasoro mode in the semion wavefunction with CFT predictions from Hypothesis~\ref{assump:chiral-CFT}.} 
\label{tab:semion-sigma}
\end{table}
% ==============================
% end of sec:numerics
% ===============================

\section{Discussion}\label{sec:discussion}
\subsection{Summary}

We have described a way to construct Virasoro generators from a purely-chiral wavefunction. The argument is based on entanglement bootstrap axioms within the bulk and a hypothesis [Hypothesis \ref{assump:chiral-CFT}] about the chiral CFT description of the physical edge. The theoretical predictions are tested against numerical data. A notable concept we introduce is the good modular flow; the idea is that such flows preserve the bulk density matrix, but create edge excitations that form a representation of the Virasoro algebra. 

We have provided theoretical arguments (based on a hypothesis) and numerical evidence that our method to extract the Virasoro algebra works for both invertible phases and chiral topological orders. In particular, we have tested $p+\ii p$ superconductor ground states and found an excellent match with the theoretical prediction. We also tested our prediction on a semion wavefunction, and the error is consistent with what we should expect on systems of small sizes, thus supporting our prediction.

We give a concrete summary of the results in the following list: 
\begin{itemize}
    \item Assumptions: (1) Entanglement bootstrap axioms {\bf A0}, {\bf A1} \cite{shi2020fusion} in the bulk, which give a sharp implementation of area law of entanglement. (2) Hypothesis~\ref{assump:chiral-CFT}, which asserts the existence of a map between certain modular flow generators and CFT operators.
    \item Definitions: Good modular flow [Definition~\ref{definition:good_modular_flow_succinct}] and [Definition~\ref{def:gdot}]. An important property for a good modular flow generator is that it does not change the bulk reduced density matrices. The theoretical argument, which is based on the bulk {\bf A1} assumption, is given in 
        [Section~\ref{sec:def-good-modular-flows}] and numerical verification is given in [Section~\ref{sec:fidelity_tests}].
    \item List of results regarding chiral edge: 
    \begin{itemize}
        \item Vector fixed-point equations are satisfied near the edge [Eq.~\eqref{eq:bdy-vector-fixed-point-equation}]. 
        \item $\calH_{\mathrm{edge}}^{\bullet}$ is a carrier space for a representation of the Virasoro algebra [Section~\ref{sec:edge-subspace-coherent}]. 
        \item We give a construction of $L_0$ [Eq.~\eqref{eq:L0}] and approximate Virasoro generators $\Tilde{L}_{n\neq 0}$ based on the twisting trick Figure~\ref{fig:twist} [Eq.~\eqref{eq:step3}, Eq.~\eqref{eq:step4}, Eq.~\eqref{eq:Leffn}]. Some quantitative analysis of $\Tilde{L}_{n\neq 0}$ is provided. 
        \item Our construction of the approximate Virasoro algebra passes a number of numerical tests [Section~\ref{sec:numerics}]. 
    \end{itemize}
\end{itemize}

\subsection{Further remarks}

Our construction of Virasoro generators is effective for purely-chiral systems but not for systems with both left and right movers.
This is because the vector fixed point equation~\eqref{eq:bdy-vector-fixed-point-equation} holds only when the cross-ratios associated with the left- and right-moving modes are the same $\eta_L = \eta_R$.

This issue can be resolved by separating the left- and right-moving modes using the commutators of the modular Hamiltonians, similar to the construction of the global conformal generators in \cite{Lin2023}. Alternatively, another approach involves performing a Fourier transformation in space followed by a Fourier transformation in time.

Another difference between the purely-chiral system and the systems with both left and right movers is how they change under a modular flow.
We start from a state that satisfies the vector fixed-point equation~\eqref{eq:bdy-vector-fixed-point-equation} and apply a finite time modular flow.
For a system with both left and right movers, since the modular flow acts differently to the left- and right-movers, the cross-ratios become different $\eta_L (t) \ne \eta_R(t)$ and the vector fixed-point equation no longer holds.

On the other hand, for a purely-chiral theory, where there is only one set of modes and one corresponding cross-ratio, we can always set $\eta$ in the vector fixed-point equation~\eqref{eq:bdy-vector-fixed-point-equation} to be the cross-ratio. 
This ensures that the vector fixed-point equation holds, which enables us to generate the family of states utilized in the section on coherent states, \S\ref{sec:edge-subspace-coherent}.

\subsection{Future directions}

One might think that good modular flows only exhibit nontrivial features near the edge. Nevertheless, we believe that similar results can also be extracted directly from the bulk. In pursuit of this, we will introduce the notion of an ``instantaneous modular flow'' and generalize the vector fixed-point equation, Eq.~\eqref{eq:bdy-vector-fixed-point-equation}.

We would like to further explore the relation between the conformal data of the 1+1D chiral edge and the anyon data of the 2+1D bulk under the framework of entanglement bootstrap. Such results would support the axioms widely believed in the category theory approach to this problem~\cite{2018NuPhB.927..140K, Kong:2019byq, Kong:2019cuu}.

One may hope to extend our formalism to include Kac-Moody algebra, which arises in CFT with continuous global symmetry.
Presumably, this involves replacing the entanglement Hamiltonian with the logarithm of the flux-insertion operator $U_A(\theta) = \prod_{x \in A } u_x(\theta) \veq e^{ \ii \theta \int_{b(A)} dx j_0(x) } $, where $\mathbb{V}$ is a generalization of the isometry in Hypothesis~\ref{assump:chiral-CFT} to include other degrees of freedom.  

Our definition of good modular flow allows the inclusion of regions with disconnected components (this is the distinction between 
$\g$ and $\gc$). The modular Hamiltonians of such regions in 1+1D CFT are no longer local~\cite{Casini2009,Cardy:2016fqc,Klich2017,Arias2018}, and depend on the detailed operator content of the CFT (not just $c$ and the Hamiltonian density).  
We thus expect that such generators involve operators beyond the Virasoro generators, but still preserve the bulk density matrix. This provides a possible route towards generalizing the isometry $\mathbb{V}$ in Hypothesis~\ref{assump:chiral-CFT} to access other CFT operators besides the stress tensor.
Note that, in contrast to the good modular flow studied in this paper, this type of modular flow with disconnected components likely does not preserve the vector fixed-point equation.

We have expressed the elements of the Virasoro algebra as quantum information quantities from the chiral ground state.
It is known that for a system with CFT behaviors (e.g.~Hypothesis~\ref{assump:chiral-CFT}), the Virasoro algebra forms a Lie algebra, which is a vector space with commutator relations.
However, the current construction only implies that the constructed Virasoro algebra forms a vector space.
Can we find a checkable condition on the chiral state which implies these commutator relations?

\subsection{Preview of results about irregular edges}

In Nature, many chiral edges appear to be irregular, such as those in experiments (e.g.~\cite{saminadayar1997observation,zhang2009distinct,ofek2010role,nakamura2020direct,Bartolomei2020,2023arXiv230711054S, veillon2024observation}). 
The theoretical reason for the persistence of edge modes in such situations is that chiral edges are robust and cannot be gapped out if we add disorder (locally perturbing the Hamiltonian) to the edge. However, as we shall discuss in \cite{q-cross-ratio}, with some general considerations, such irregular edges are secretly regular in a broader sense: they have a well-defined cross-ratio, measured using information-theoretic quantities. We believe that our way of extracting the Virasoro algebra can also work in the context of irregular edges and give evidence for this in the forthcoming paper.

\vfill\eject
{\bf Acknowledgments.} 
This work was supported in part by
funds provided by the U.S. Department of Energy
(D.O.E.) under cooperative research agreement 
DE-SC0009919, 
by the University of California Laboratory Fees Research Program, grant LFR-20-653926,
and by the Simons Collaboration on Ultra-Quantum Matter, which is a grant from the Simons Foundation (652264, JM). IK acknowledges supports from NSF under award number PHY-2337931.
JM received travel reimbursement from the Simons Foundation;
the terms of this arrangement have been reviewed and approved by the University of California, San Diego, in accordance with its conflict of interest policies.  The bureaucrats there are very proud of this fact and insist that we tell you about their accomplishments.

\appendix
\renewcommand{\theequation}{\Alph{section}.\arabic{equation}}

\section{Notation}
\begin{center}
\begin{tabular}{|c|c|}
\hline 
Notations & Meanings  \\ \hline \hline
$c$ & Holomorphic central charge \\ \hline
$\bar c$ & Anti-holomorphic central charge \\ \hline
$\ctot$ & Total central charge: $\ctot=c+\bar{c}$ \\ \hline
$c_{-}$ & Chiral central charge: $c_{-} = c-\bar c$ \\ \hline \hline
$K_A$ & Modular Hamiltonian of region $A$ in 2+1D state $\ket{\Psi}$ \\ \hline
$\Kcft_a$ & Modular Hamiltonian of an interval $a$ in 1+1D CFT ground state \\ \hline 
$\Kcft_a^{\chi}$ & Holomorphic component of $\Kcft_a$ \\ \hline \hline 
$\mathbb{V}$ & An isometry defined in Hypothesis~\ref{assump:chiral-CFT} \\ \hline
$\Lb$ & Good modular flow generator [Definition~\ref{definition:good_modular_flow_succinct}] \\ \hline
$\Lcft$ & \makecell{Linear combination of chiral CFT modular Hamiltonians,\\ with $\Lb\mathbb{V} = \Lcft\mathbb{V}$} \\ \hline
$\Lcft_n$ & Virasoro generator from a purely chiral CFT \\ \hline 
$L_n$ & \makecell{Representation of Virasoro generators on 2+1D edge, \\ with $L_n\mathbb{V} = \Lcft_n\mathbb{V}$} \\ \hline 
$\Lbo_n,\Lbe_n$ & \makecell{Two particular good modular flow generators, \\ defined in Eq.~\eqref{eq:step3} and Eq.~\eqref{eq:step4}} \\ \hline 
$\Leff_n$ & \makecell{Approximate Virasoro generators,\\ defined in Eq.~\eqref{eq:Leffn-eo}} \\ \hline 
\end{tabular}
\end{center}
Some remarks on these notations: 
\begin{itemize}
    \item If a CFT is purely chiral, then $\bar c=0$, therefore $\ctot = c = |c_{-}|$. 
    \item Usually, the curly symbols, such as $\Kcft, \Lcft$ are operators for a CFT, while others are on the Hilbert space of the 2+1D state.    
\end{itemize}

%==============
% smoothness and fourier coefficients 
%==============
\section{Smoothness and decay of Fourier coefficients}
\label{sec:smooth-fourier}
Here, we prove a statement that relates the smoothness of a function and the asymptotic decay properties of its Fourier coefficients.

Let $f(x)$ be a real-valued function and $f^{(m)}(x)$ denote its $m$-th derivative. We call a function $f(x)$ `piecewise $C^m$' if $|f^{(m)}(x)|$ 
is bounded and piecewise continuous with a finite number of finite jump discontinuities, while $f^{(m+1)}(x)$ is unbounded. By $f(x)$ having a finite jump discontinuity at $x_{*}$, we mean $0<|\lim_{x\to x_{*}^{+}}f(x)-\lim_{x\to x_{*}^{-}}f(x)|<\infty$.

\begin{lemma}\label{lemma:fourier-discontinuity}
    Let $f(\theta)$ be a real-valued $2\pi$-periodic function. If $f(\theta)$ is piecewise $C^m$, then its Fourier coefficient 
    \begin{equation}
        \Tilde{f}_n \equiv \frac{1}{2\pi}\int_0^{2\pi}d\theta f(\theta) e^{-\ii n\theta},
    \end{equation}
    has the property $|\Tilde{f}_n|\sim O(1/n^{m+1})$. That is, to be more explicit, $\lim_{n\to\infty}n^{m+1}|\Tilde{f}_n|$ is bounded. 
\end{lemma}
\begin{proof}
    Since $f(\theta) = \sum_n \Tilde{f}_n e^{\ii n\theta}$, 
    \begin{equation}\label{eq:intfm}
        f^{(k)}(\theta) = \sum_{n} {\ii}^k n^{k} \Tilde{f}_n e^{\ii \theta n}\quad \Rightarrow \quad \Tilde{f}_n \cdot n^{k} =\frac{(-\ii)^k}{2\pi } \int_0^{2\pi} d\theta f^{(k)}(\theta)e^{-\ii n\theta}.
    \end{equation}
    $f(\theta)$ is piecewise $C^m$. Therefore, let $\theta_1<\theta_2<\cdots <\theta_N$ be the discontinuous points of $f^{(m)}(\theta)$, then $f^{(m)}(\theta)$ can be written as 
    \begin{equation}
        f^{(m)}(\theta) = \sum_{k=1}^{N} \Theta(\theta-\theta_{k})\Theta(\theta_{k+1}-\theta)F_k(\theta),
    \end{equation}
    where $\theta_{N+1}\equiv \theta_1$, $\Theta(x)$ is the Heaviside step function. $F_k(\theta)$ is some smooth function, which is identical to $f^{(m)}(\theta)$ on the interval $[\theta_{k},\theta_{k+1}]$. The derivative of $f^{(m)}(\theta)$ involves various $\delta$ functions: 
    \begin{equation}\begin{split}
        f^{(m+1)}(\theta) = \sum_{k=1}^N \Big[& \delta(\theta-\theta_k)\Theta(\theta_{k+1}-\theta)F_k(\theta) - \Theta(\theta-\theta_{k})\delta(\theta_{k+1}-\theta)F_k(\theta) \\ 
        &+ \Theta(\theta-\theta_{k})\Theta(\theta_{k+1}-\theta)\frac{dF_k(\theta)}{d\theta}\Big]. 
    \end{split}
    \end{equation}
    Therefore, even though the function $f^{(m+1)}(\theta)$ itself is not upper bounded, the absolute value of its integral  
    \begin{equation}
       \left|  \frac{1}{2\pi} \int_0^{2\pi}d\theta f^{(m+1)}(\theta) e^{-\ii n\theta}  \right|
    \end{equation}
    shall be a upper bounded by some finite number $M$. This is because there is a finite number of $\delta$ functions, and each $\delta$ function only contributes a finite number to the result of the integral. 

    Then utilizing Eq.~\eqref{eq:intfm}, one can see that 
    \begin{equation}
        |\Tilde{f}_n| \cdot|n^{m+1}| =   \left| \frac{1}{2\pi} \int_0^{2\pi}d\theta f^{(m+1)}(\theta) e^{-\ii n\theta}  \right| \leq M,
    \end{equation}
    Hence $|\Tilde{f}_n|\sim O(1/n^{m+1})$.\footnote{Some careful readers might wonder, is it possible $|\Tilde{f}_n|\sim O(n^{-(m+2)})$? The answer is no because if so, $f^{(m)}(\theta) = \sum_{n} i^m n^{m} \Tilde{f}_n e^{i\theta n}$ will be absolutely summable, and as a result, it will absolutely and uniformly converge to a continuous function, so the function would actually be piecewise $C^{m+1}$.}
\end{proof}

We can then apply this lemma to the weight function $f(\theta)$ of a good modular flow generator 
\begin{equation}\label{eq:L-T-A1}
    \gc \ni \Lb \veq \int_{0}^{2\pi} f(\theta) \Tcft(\theta). 
\end{equation}
As $f(\theta)$ is piecewise $C^2$, with a finite number of jump discontinuities, we can conclude 
\begin{equation}
    \tilde{f}_n = \frac{1}{2\pi}\int_{0}^{2\pi}d\theta f(\theta) e^{-\ii n\theta} \sim O(n^{-3}). 
\end{equation}

For bad modular flow generators, if we assume its action on the edge is of a similar form as Eq.~\eqref{eq:L-T-A1}, the weight function will only be piecewise $C^1$, and therefore $|\tilde{f}_n|\sim O(|n|^{-2})$. 

%==============
% analysis of the construction and improvement
%==============

\section{Improvement of the approximate Virasoro generators}
\label{app:improvement}
We introduced the construction of the approximate Virasoro generators $\Leff_n$ and their expansions in terms of Virasoro generators 
\begin{equation}
     \Leff_n = \sum_{m\in\mathbb{Z}}\lambda_{n,m}L_m,
\end{equation}
with $\lambda_{n,m}$ given in Eq.~\eqref{eq:lambda-nm}. In this section, we introduce a procedure to reduce the contribution of the higher harmonic terms in $\Leff_n$. As a result of the procedure, we shall obtain the Virasoro generator as a linear combination of the approximate Virasoro generators $\Leff_n$. Furthermore, this indicates that $\{\Leff_n\ket{\Psi}\}_{n\in\mathbb{Z}}$ forms a (non-orthorgonal) basis of the vector space spanned by $\{L_n\ket{\Psi}\}_{n\in\mathbb{Z}}$. 

\subsection{Improvement procedure}
We first discuss the improvement of $\Leff_n$ to produce a better approximation to $L_n$. Throughout this section, we shall focus on $|n|\geq 2$ cases, as $\Leff_{0},\Leff_{\pm 1}$ are already equal to $L_0,L_{\pm 1}$ by their constructions. 

Upon a close examination of the Fourier coefficients $\lambda_{n,m}$ in Eq.~\eqref{eq:lambda-nm}, we can see that the mode numbers of Virasoro generators in $\Leff_n$ are $(-1)^j(2j+1),j\in\mathbb{N}$. To be more explicit, 
\begin{equation}\label{eq:Leffn-expand-2}
    \Leff_n = L_n + \sum_{j\in\mathbb{N}^*}\lambda_{n,(-1)^j(2j+1)n}L_{(-1)^j(2j+1)n},
\end{equation}
where $\mathbb{N}^*$ is defined as $\mathbb{N}\setminus\{0\}$, and 
\begin{equation}
    \lambda_{n,(-1)^j(2j+1)n} = (-1)^j \frac{n^2-1}{(2j+1)^3n^2 - (2j+1)}.
\end{equation}
Moreover, it is also these modes $\left\{ L_{(-1)^j(2j+1)n}\right\}_{j\in\mathbb{N}^*}$ that contribute in the Virasoro mode expansion of $\Leff_{(-1)^{k}(2k+1)n},k\in\mathbb{N}^*$. This can be seen by 
\begin{equation}\begin{split}
    &(-1)^{j}(2j+1)n\cdot (-1)^{k}(2k+1) \\ 
    = &(-1)^{j+k}[2(2jk+j+k)+1 ]n \\ 
    = &(-1)^{m}(2m+1)n,\quad \text{with }m=2jk+j+k. 
\end{split}
\end{equation}
Therefore, it is plausible that one can use the approximate Virasoro generators $\left\{ \Leff_{(-1)^j(2j+1)n}\right\}_{j\in\mathbb{N}^*}$ themselves to cancel the contribution of $\left\{ L_{(-1)^j(2j+1)n}\right\}_{j\in\mathbb{N}^*}$ in $\Leff_n$, as it will not introduce new Virasoro modes. 

The exact expression for $L_n$ in terms of $\left\{ \Leff_{(-1)^j(2j+1)n}\right\}_{j\in\mathbb{N}}$ can be obtained by the following procedure. Starting from $\Leff_n$, we can see the first sub-leading order contribution is from $L_{-3n}$. This term can be canceled by the replacement
\begin{equation}
    \Leff_{n} \equiv(\Leff_{n})^{(0)} \to (\Leff_{n})^{(1)}
    \equiv \Leff_{n} - \lambda_{-3n}\Leff_{-3n}. 
\end{equation}
Then we can cancel the first sub-leading order contribution in $(\Leff_n)^{(1)}$ (which is from $L_{5n}$) by $\Leff_{5n}$, and so on. Denoting the improved approximate Virasoro generator after the $k$-th correction as $(\Leff_n)^{(k)}$, we can summarize the procedure as the following two steps: (1) Find the first sub-leading order $L_m$ in $(\Leff_n)^{(k)}$ and its coefficient $\lambda^{(k)}_{n,m}$. (2) Cancel this term by 
\begin{equation}
    (\Leff_n)^{(k)} \to (\Leff_n)^{(k+1)} = (\Leff_n)^{(k)} - \lambda^{(k)}_{n,m}\Leff_m. 
\end{equation}
Then the Virasoro generator $L_n$ can be obtained as the limit
\begin{equation}\label{eq:Ln-Leffn}
    L_n = \lim_{k\to\infty}(\Leff_n)^{(k)}, 
\end{equation}
in the sense that 
\begin{equation}\label{eq:lim-lambda-nm}
    \lim_{k\to\infty} \lambda_{n,m}^{(k)} = 0,\quad \forall m\neq n. 
\end{equation}

\subsection{Numerical tests of the improvement}
We can test the success of the procedure elaborated in the previous subsection numerically in the $p+\ii p$ SC ground state.  
In the construction of $\Leff_{\pm n}$, the next leading order contribution is $L_{\mp 3n}$ [Eq.~\eqref{eq:Leffn-expand-2}]. We can thus subtract this term, yielding the following construction: 
\begin{equation}
    \Leff_{\pm n} \to (\Leff_{\pm n})' = \Leff_{\pm n} + \frac{n^2-1}{27n^2-3}\Leff_{\mp 3n} \\ 
\end{equation}
Although there are additional terms being introduced from $\Leff_{\pm 3n}$, we find the total error is reduced [Table~\ref{table:vir-improvement}]. One can repeat the process to further improve the approximation. 

For example, we can improve $\Leff_{\pm 2}$ and $\Leff_{\pm 4}$ as follows, using the setups and construction in Section~\ref{subsec:virasoro_tests}. For $\Leff_{\pm 2}$, the next leading order is $\Lcft_{\pm 6}$. Then we use $\Leff_{\pm 6}$ to subtract off $L_{\pm 6}$. For $\Leff_{\pm 4}$, we use $\Leff_{\pm 12}$ to cancel the next leading term $L_{\pm 12}$: 
\begin{equation}\begin{split}
    &\Leff_{\pm 2} \to (\Leff_{\pm 2})' = \Leff_{\pm 2} + \frac{2^2-1}{27\times 2^2-3}\Leff_{\mp 6} \\ 
    &\Leff_{\pm 4} \to (\Leff_{\pm 4})' = \Leff_{\pm 4} + \frac{4^2-1}{27\times 4^2-3}\Leff_{\mp 12}. 
\end{split}
\end{equation}
As expected, we found improvements; see Table~\ref{table:vir-improvement}

\begin{table}[!h]
\setlength\extrarowheight{3pt}
\begin{tabular}{|l|ll|ll|}
\hline
\multirow{2}{*}{Tests}                                & \multicolumn{2}{l|}{Original data with $\Leff_{\pm n}$}                            & \multicolumn{2}{l|}{Replacing $\Leff_{\pm n}$ with $(\Leff_{\pm n})'$}                         \\ \cline{2-5} 
                                                      & \multicolumn{1}{l|}{Results} & \makecell{Comparison with \\ Virasoro algebra} & \multicolumn{1}{l|}{Results} & \makecell{Comparison with \\ Virasoro algebra} \\ \hline
$\bra{\Psi} [\Leff_2, \Leff_{-2}]\ket{\Psi}$          & \multicolumn{1}{l|}{0.2440}  & 2.404\%                          & \multicolumn{1}{l|}{0.2509}  & 0.367\%                          \\ \hline
$\bra{\Psi} [\Leff_4, \Leff_{-4}]\ket{\Psi}$          & \multicolumn{1}{l|}{2.4266}  & 2.935\%                          & \multicolumn{1}{l|}{2.5113}  & 0.454\%                          \\ \hline
$\bra{\Psi} [[\Leff_2,\Leff_0] \Leff_{-2}]\ket{\Psi}$ & \multicolumn{1}{l|}{0.5749}  & 14.98\%                          & \multicolumn{1}{l|}{0.5272}  & 5.44\%                           \\ \hline
$\bra{\Psi} [[\Leff_4,\Leff_0] \Leff_{-4}]\ket{\Psi}$ & \multicolumn{1}{l|}{11.5006} & 15.01\%                          & \multicolumn{1}{l|}{10.4140} & 4.14\%                           \\ \hline
\end{tabular}
\caption{Commutator tests for the constructed Virasoro generators, with and without improvements. }
\label{table:vir-improvement}
\end{table}

\section{Semion Setup}\label{app:semion} 

In this appendix, we explain the detailed setup of the semion wavefunction used in Section~\ref{sec:semion}.  This model wavefunction, introduced by Nielsen \emph{et al}~\cite{Nielsen2012}, describes a chiral topological order with semion statistics. It is also known as a discrete version of the $\nu=\frac{1}{2}$ bosonic Laughlin state~\cite{Laughlin1983,Kalmeyer1987}. The semion topological order has  $c_-=1$ and two species of Abelian anyons $\{1, s\}$. The central charge of the chiral edge is expected to be $c=1$, which should be the same as its total central charge since the state is purely chiral.

The construction of the semion wavefunction in Ref.~\cite{Nielsen2012} is as follows.  Consider a pair $(N, \{z_j\}_{j=1}^N)$, where $N$ is an even integer denoting the number of qubits in the system, and $z_j\in \mathbb{C}$ for $j= 1, \cdots, N$ are distinct complex numbers. For each  $z_j$ on the complex plane, we assign a qubit. 
The system is bosonic, and thus the total Hilbert space is a tensor product of these qubits. 
The wavefunction (not normalized) is
\begin{equation}
\vert \Psi(N, \{z_j\}_{j=1}^N) \rangle = \sum_{\{s_i\}_{i=1}^N} w(\{s_i\}_{i=1}^N) \vert \{s_i\}_{i=1}^N \rangle, \quad s_i= \pm 1.
\end{equation}
The complex coefficients $w(\{s_i\}_{i=1}^N)$ are given by 
\begin{equation}
w(\{s_i\}_{i=1}^N)= \delta_{\{s_i\}} \prod_{n<m}^N (z_n-z_m)^{\frac{1}{2} s_n s_m},
\end{equation}
where $\delta_{\{s_i\}}=1$ for $\sum_i s_i =0$ and $\delta_{\{s_i\}}=0$ otherwise. 

\begin{figure}
    \centering
    \includegraphics[width = 0.88 \columnwidth]{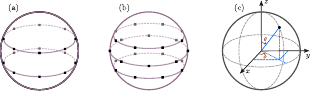}
    \caption{The sphere coordinates for the semion 2 and 3-leg ladders used in Section~\ref{sec:semion}.}
    \label{fig:semion-sphere}
\end{figure}

It is convenient to map the coordinates $z_j$ to the sphere via the stereographic projection.
Namely, we assign a sphere coordinate $(\theta_j, \phi_j)$ to $z_j$ according to
\begin{equation}
z_j= \frac{\sin\theta_j}{1+ \cos\theta_j} \exp\left(\ii \phi_j\right), \qquad  \theta_j\in [0,\pi] \textrm{ and } \phi_j\in [0,2\pi).
\end{equation}
The unit vector in 3-dimensional Euclidean space, corresponding to a sphere coordinate $(\theta,\phi)$ is $\vec{n}= (\sin\theta \cos\phi, \sin\theta \sin\phi,\cos\theta)$, as explained in Fig.~\ref{fig:semion-sphere}(c). 
As an aside, the sphere coordinate is convenient in this problem because sphere rotation keeps the wavefunction invariant \cite{Nielsen2012}. 

Now, we are ready to specify the wavefunctions used in Section~\ref{sec:semion}. This is done by specifying the coordinates on the sphere. 
\begin{itemize}
    \item For the 2-leg ladder in Figure~\ref{fig:semion-sphere}(a), the 16 points correspond to the coordinates
    \begin{equation}
      \text{ladder 1:} \left( \frac{\pi}{2}-\frac{\pi}{8}, \frac{k\pi}{4} \right), \quad \text{ladder 2:} \left( \frac{\pi}{2}+\frac{\pi}{8}, \frac{k\pi}{4}\right),
    \end{equation}
    with $k=1,2,\cdots,8$.
    \item For the 3-leg ladder in Figure~\ref{fig:semion-sphere}(b), the 24 points correspond to the coordinates
    \begin{equation}
      \text{ladder 1:} \left( \frac{\pi}{2}-\frac{\pi}{8}, \frac{k\pi}{4}\right), \quad \text{ladder 2:} \left( \frac{\pi}{2}, \frac{k\pi}{4}+\frac{\pi}{8} \right), \quad  \text{ladder 3:} \left( \frac{\pi}{2}+\frac{\pi}{8}, \frac{k\pi}{4} \right ),
    \end{equation}
    with $k=1,2,\cdots,8$.
\end{itemize}

%==============
% cylinder argument
%==============

\section{The cylinder argument}
\label{appendix:cylinder-argument}
 
\subsection{Review of 1+1D CFT fixed-point equation}
 
Here, we briefly review a local condition satisfied by 1+1D conformal field theory ground states $\ket\Omega$ on a real line in \cite{Lin2023}, which will play an important role in our analysis. The entanglement entropy of an interval of length $\ell$ in the ground state, $S(\ell) = { \ctot \over 6 } \ln  { \ell \over \epsilon }$ \cite{Holzhey:1994we, Calabrese:2009qy}, is by itself not strong enough to imply that the state is a CFT ground state. For the ground state of a CFT, the modular Hamiltonian $K_A$, $ \rho_A  = e^{ - K_A}$, for a single interval on the line is \cite{Cardy:2016fqc}: 
\be\label{eq:K-CFT} 
\begin{aligned} \Kcft_{[x_1, x_2]} =  \int_{x_1}^{x_2} dx \beta_{[x_1, x_2]}(x) h(x) + S_{[x_1, x_2]} \Ione , \text{ where }\\
\beta_{[x_1, x_2]}(x) = 2 \Theta(x-x_1)\Theta(x_2-x) { (x-x_1)(x_2-x) \over x_2 - x_1} .
\end{aligned}\ee
Here $T(x)$ is the Hamiltonian density of the CFT, and the $\Theta(x)$ is the Heaviside step function. 

A certain linear combination of these locally-quadratic functions is identically zero.  
Therefore the corresponding combination of modular Hamiltonians is proportional to the identity. For any three contiguous intervals $A, B$ and $C$,
\begin{align}   \mathcal{K}_{\Delta} &  \equiv   \Kcft_{AB} + \Kcft_{BC} - \eta (\Kcft_A + \Kcft_C) - (1-\eta) (\Kcft_B + \Kcft_{ABC}) = {\ctot \over 6 } h(\eta) \Ione 
\label{eq:Kdelta_temp}
\\ & \equiv \eta \hat \Delta(A,B,C) + (1-\eta) \hat I(A:C|B) ~. 
\end{align}
Here 
\be \hat \Delta(A,B,C) \equiv \Kcft_{AB} + \Kcft_{BC} -\Kcft_{A} - \Kcft_{C},
~~~\hat I(A:C|B) \equiv \Kcft_{AB} + \Kcft_{BC} - \Kcft_B - \Kcft_{ABC}  ~.\ee
The function $h(\eta)$ on the right-hand side of Eq.~\eqref{eq:Kdelta_temp} is determined by taking expectation values to be 
the binary entropy function 
$h(\eta) = - \eta \ln  \eta - (1-\eta) \ln (1-\eta)$.  In a lattice wavefunction (as opposed to continuum field theory), the density matrix may have a kernel, and so $\Kcft$ may be singular; 
as a result, the operator form of this equation is less robust than the following {\it vector fixed-point equation}:
\be \mathcal{K}_\Delta \ket\Omega = {\ctot \over 6 } h(\eta)  \ket\Omega .   \label{eq:star} \ee

\subsection{Vector fixed-point equation for chiral edge}
\label{subsec:chiral-edge-fixed-point-equation}

In this Section, we derive several nontrivial identities for the modular Hamiltonians localized near the edge. Let us first recall the setup. We shall consider a finite cylinder hosting a 2+1D reference state $\ket\Psi$ with chiral gapless edges on both the top and the bottom edge; see Figure~\ref{fig:cylinder_decomposition}. It can be overall viewed as a 1+1D non-chiral CFT [Figure~\ref{fig:cylinder_argument}] by the \emph{cylinder assumption}: if we dimensionally reduce along the direction normal to the boundaries, we obtain a 1+1D field theory with conformal symmetry. Thus for the regions in Figure~\ref{fig:cylinder_argument}, we get
\be  \( \eta \hat \Delta(A,B,C) + (1-\eta) \hat I(A:C|B) \) \ket{\Psi } 
= { \ctot \over  6 } h(\eta) \ket{\Psi}, \ee
where $\eta$ is the cross-ratio computed form the circle and $h(\eta)$ is the binary entropy function as above. The total central charge is $\ctot = c_L + c_R$, with $c_L,c_R$ being the total central charge of the top and bottom edge. For purely chiral state $c_L = c_R = |c_{-}|$, where $c_{-}$ is the chiral central charge of the state. 

Our goal is to derive Eq.~\eqref{eq:bdy-vector-fixed-point-equation}, a version of the vector fixed-point equation near the chiral edge. For the reader's convenience, we restate this below. For the regions shown in Figure~\ref{fig:KD_eta}, the fixed-point equation is
\be  
\(  \eta \hat \Delta(AA', B, CC') + (1 - \eta) \hat I(A:C|B) \)
\ket{\Psi} \propto \ket{\Psi}.
~~  \label{eq:chiral-fixed-point-equation}\ee
The assumptions behind this derivation are the cylinder assumption [Section~\ref{subsec:cylinder-argument}], axioms of the entanglement bootstrap in the bulk [Section~\ref{sec:primer_eb}], and the full boundary axioms introduced in Section~\ref{subsec:cylinder-argument} [Figure~\ref{fig:full_boundary_axioms}]. 

\cylindertovectorfixedpointeqn*

\begin{proof}[Proof of \ref{prop:cylinder-to-vector-fixed-point-eqn}]
We wish to show:
\begin{align}
\KD^L \ket{\psi}
  = \Gamma^L \ket{\psi}  \label{eq:conclusion}
\quad \text{and} \quad 
 \KD^R \ket\psi 
 & = \Gamma^R \ket{\psi} ,
\end{align}
where, for $\alpha = L,R$,
$\KD^\alpha \equiv \( \eta \Delta(A_\up^\alpha,B^\alpha,C_\up^\alpha) + (1-\eta) I(A^\alpha:C^\alpha |B^\alpha )\) $, with 
$\eta$ is the geometric cross-ratio for the regions $abc$ on the boundary, 
and $C_\up^L \equiv C^LC', A_\up^L \equiv A^LA'$, 
$C_\up^R \equiv C^RC', A_\up^R \equiv A^RA'$ [Figure~\ref{fig:cylinder_decomposition}].  
Furthermore $\Gamma^L + \Gamma^R = { \ctot \over 6 }h(\eta)$.

\begin{figure}[t]
    \centering
    \begin{tikzpicture}
    \draw[fill=black!30!white] (1.2,0) -- ++ (1.2,0) -- ++ (0, 4) -- ++ (-1.2, 0) -- cycle;
    \draw[fill=blue!30!white] (2.4,0) -- ++ (1.2,0) -- ++ (0, 4) -- ++ (-1.2, 0) -- cycle;
    \draw[fill=orange!30!white] (3.6,0) -- ++ (1.2,0) -- ++ (0, 4) -- ++ (-1.2, 0) -- cycle;
    \draw[] (1.2, 4/3) -- ++ (3.6, 0);
    \draw[] (1.2, 8/3) -- ++ (3.6, 0);
    \draw[very thick] (0,0) -- (6,0);
    \draw[very thick] (0,4) -- (6,4);
    \node[] () at (1.8, 4/6) {$A^R$};
    \node[] () at (1.8, 12/6) {$A^M$};
    \node[] () at (1.8, 20/6) {$A^L$};
    \node[] () at (3, 4/6) {$B^R$};
    \node[] () at (3, 12/6) {$B^M$};
    \node[] () at (3, 20/6) {$B^L$};
    \node[] () at (4.2, 4/6) {$C^R$};
    \node[] () at (4.2, 12/6) {$C^M$};
    \node[] () at (4.2, 20/6) {$C^L$};

    \draw[line width=3pt, red] (1.2, 0) -- ++ (1.2, 0);
    \draw[line width=3pt, green] (2.4, 0) -- ++ (1.2, 0);
    \draw[line width=3pt, blue] (3.6, 0) -- ++ (1.2, 0);
    \node[below] () at (1.8, 0) {$a$};
    \node[below] () at (3.0, 0) {$b$};
    \node[below] () at (4.2, 0) {$c$};

    \begin{scope}[xshift=8cm]
    \draw[fill=black!30!white] (1.2,0) -- ++ (1.2,0) -- ++ (0, 16/6) -- ++ (-1.2, 0) -- cycle;
    \draw[fill=blue!30!white] (2.4,0) -- ++ (1.2,0) -- ++ (0, 8/6) -- ++ (-1.2, 0) -- cycle;
    \draw[fill=orange!30!white] (3.6,0) -- ++ (1.2,0) -- ++ (0, 16/6) -- ++ (-2.4, 0) -- ++ (0, -8/6) -- ++ (1.2, 0)-- cycle;
    \draw[] (1.2, 16/6) -- ++ (0, 8/6);
    \draw[] (2.4, 16/6) -- ++ (0, 8/6);
    \draw[] (3.6, 16/6) -- ++ (0, 8/6);
    \draw[] (4.8, 16/6) -- ++ (0, 8/6);
    \draw[] (1.2, 4/3) -- ++ (3.6, 0);
    \draw[] (1.2, 8/3) -- ++ (3.6, 0);
    \draw[very thick] (0,0) -- (6,0);
    \draw[very thick] (0,4) -- (6,4);
    \node[] () at (1.8, 4/6) {$A^R$};
    \node[] () at (1.8, 12/6) {$A'$};
    \node[] () at (1.8, 20/6) {$A^L$};
    \node[] () at (3, 4/6) {$B^R$};
    \node[] () at (3, 20/6) {$B^L$};
    \node[] () at (4.2, 4/6) {$C^R$};
    \node[] () at (3.6, 12/6) {$C'$};
    \node[] () at (4.2, 20/6) {$C^L$};

    \draw[line width=3pt, red] (1.2, 0) -- ++ (1.2, 0);
    \draw[line width=3pt, green] (2.4, 0) -- ++ (1.2, 0);
    \draw[line width=3pt, blue] (3.6, 0) -- ++ (1.2, 0);
    \node[below] () at (1.8, 0) {$a$};
    \node[below] () at (3.0, 0) {$b$};
    \node[below] () at (4.2, 0) {$c$};
        
    \end{scope}
    \end{tikzpicture}
    \caption{Two decompositions of a strip connecting two physical edges.}
    \label{fig:cylinder_decomposition}
\end{figure}
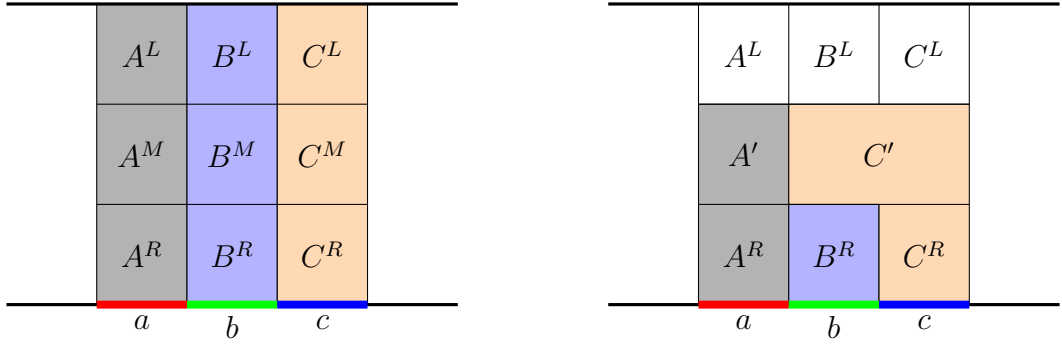

Here is the argument, in two steps. First, we can disconnect the two boundaries in the following sense. We show that 
the LHS of the equation \eqref{eq:chiral-fixed-point-equation} is
\begin{equation}
\KD \ket{\Psi} = ( \KD^L(\eta) + \KD^R(\eta) ) \ket{\Psi}
\label{eq:fixed_point_disconnected}
\end{equation}
where $\KD^{L/R}$ is made from modular Hamiltonians of regions that only touch the $L/R$ boundary.

To show Eq.~\eqref{eq:fixed_point_disconnected}, we use the full-boundary \textbf{A1} [Figure~\ref{fig:full_boundary_axioms}] to conclude that $ I(A^L:A^R|A^M)_{\ket\Psi} = 0 $; see Figure~\ref{fig:cylinder_decomposition} for the description of these regions. Therefore, 
\be\label{eq:vertical-disconnection}  K_{A} \ket{\Psi}  = ( K_{A^{ML} } + K_{A^{MR} } - K_{A^M}  ) \ket\Psi ,\ee
and similarly for the other vertical strips.  
Therefore 
\begin{align} \hat I(A:C|B) \ket{\psi} 
& = \( \hat I(A^{LM}:C^{LM}|B^{LM})  \nonumber
+ \hat I(A^{RM}:C^{RM}|B^{RM})   \right. 
\\ & \left. -  \hat I(A^M:C^M|B^M)   \) \ket\psi  ~.\end{align}
But bulk \textbf{A1} says  $\hat I(A^M:C^M|B^M)   \ket\psi   = 0 $.  Therefore, using bulk \textbf{A1} to deform the regions,  we get
\be\label{eq:part-one}
\hat I(A:C|B) \ket{\Psi} 
= \( \hat I(A^{L}:C^{L}|B^{L}) 
+ \hat I(A^{R}:C^{R}|B^{R})   \) \ket\Psi.
\ee

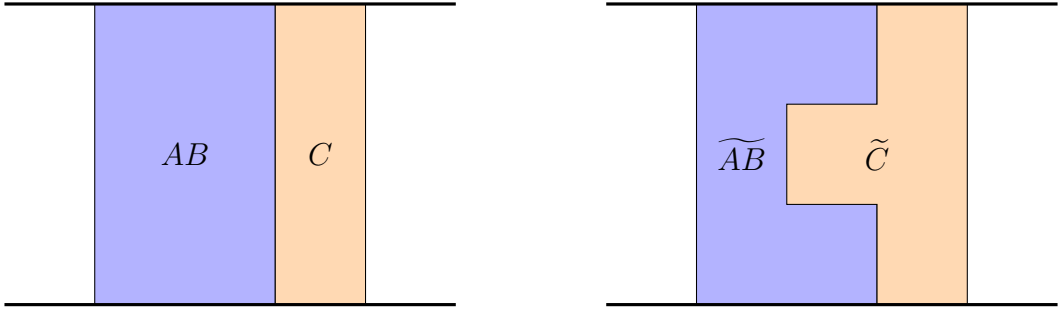
\begin{figure}[t]
    \centering
    \begin{tikzpicture}
    \draw[fill=blue!30!white] (1.2,0) -- ++ (2.4,0) -- ++ (0, 4) -- ++ (-2.4, 0) -- cycle;
    \draw[fill=orange!30!white] (3.6,0) -- ++ (1.2,0) -- ++ (0, 4) -- ++ (-1.2, 0) -- cycle;
    \draw[very thick] (0,0) -- (6,0);
    \draw[very thick] (0,4) -- (6,4);
    \node[] () at (2.4, 12/6) {$AB$};
    \node[] () at (4.2, 12/6) {$C$};

    \begin{scope}[xshift=8cm]
    \draw[fill=blue!30!white] (1.2,0) -- ++ (2.4,0) -- ++ (0, 4) -- ++ (-2.4, 0) -- cycle;
    \draw[fill=orange!30!white] (3.6,0) -- ++ (1.2,0) -- ++ (0, 24/6) -- ++ (-1.2, 0) -- ++ (0, -8/6) -- ++ (-1.2, 0)-- ++ (0, -8/6) -- ++ (1.2, 0) -- cycle;
    \draw[very thick] (0,0) -- (6,0);
    \draw[very thick] (0,4) -- (6,4);
    
    \node[] () at (1.8, 12/6) {$\widetilde{AB}$};
    
    \node[] () at (3.6, 12/6) {$\widetilde{C}$};
        
    \end{scope}
    \end{tikzpicture}
    \caption{Because of bulk \textbf{A1}, $(K_{AB} - K_C)\ket\Psi= (K_{\widetilde{AB}} - K_{\widetilde{C}})\ket\Psi$. Comparing this figure with Figure~\ref{fig:cylinder_decomposition}, we find $\widetilde{AB} = AB \setminus B^M$ and $\widetilde{C} = C\cup B^M$.}
    \label{fig:delta_bulk_deform}
\end{figure}

Now we use bulk \textbf{A1} to deform the regions 
as in Figure~\ref{fig:delta_bulk_deform}:
\be \begin{split} \hat \Delta(A,B,C)\ket\Psi
&= \( K_{AB} - K_C \) + \( K_{BC} - K_A \) \ket\Psi \\
&= \( K_{\widetilde{AB}} - K_{\widetilde{C}} \) + \( K_{BC} - K_A \) \ket\Psi 
\end{split}~.
\ee
Then using \eqref{eq:vertical-disconnection} to decompose each of these pieces vertically, we have 
\be \hat \Delta(A,B,C) \ket{\Psi} 
=  \(  \hat \Delta(A_\up^L, B^L, C^L_\up)  +  \hat \Delta(A_\up^R, B^R, C^R_\up) \)   \ket{\Psi}  . 
\label{eq:part-two}
\ee
Combining \eqref{eq:part-one} with \eqref{eq:part-two}
gives the sum of the criticality conditions for the two boundaries:
\be\label{eq:sum-of-conditions} \( \KD^L + \KD^R\) \ket\Psi = {\ctot \over 6 } h(\eta)\ket\Psi . \ee

Note that \eqref{eq:sum-of-conditions} implies 
\be \( \KD^L + \KD^R\)\rho^{LR} = {\ctot h(\eta)\over 6 }  \rho^{LR},  \label{eq:sum-on-rhos}\ee  
where $\rho^{LR}$ is the reduced density matrix for $\psi$ reduced to the
union of the regions $L, R$ of support of $\KD^L$ and $\KD^R$ respectively.

Using full-boundary A0, we conclude that $I(L:R) = 0 $ and therefore 
$\rho^{LR} = \rho^L \otimes \rho^R$.
Since the state factorizes, the contributions of $\KD^L$ and $\KD^R$ 
to \eqref{eq:sum-of-conditions} are independent.
More precisely, we can take the trace of \eqref{eq:sum-on-rhos} over $R$ to get 
\be   \KD^L  \rho^{L} +  \tr_R( \KD^R \rho^R) \rho^{L} = {\ctot h(\eta)\over 6 }  \rho^{L},  \ee  
which says 
\be   \KD^L  \rho^{L} \propto  \rho^{L}.  \ee  
Therefore, each of them acting on $\ket\Psi$ gives a c-number, which is denoted as $\Gamma^L,\Gamma^R$ respectively. One can purify the equation above to obtain 
\begin{equation}
    \KD^L \ket{\Psi} = \Gamma^L \ket{\Psi}.
\end{equation}
The one for $R$ can be obtained similarly. 
The two c-numbers add up to 
\begin{equation}
    \Gamma^L +\Gamma^R = \frac{\ctot}{6}h(\eta).
\end{equation}
\end{proof}
If we assume the two edges contribute equally to the proportionality factor, then we obtain 
\begin{equation}
    \Gamma^L = \frac{c_L}{6}h(\eta),\quad \Gamma^R = \frac{c_R}{6}h(\eta)
\end{equation}
where $c_L = c_R$ and $\ctot = c_L + c_R$. $c_R,c_L$ can be interpreted as the holomorphic and anti-holomorphic central charge of the 1+1D CFT.

We note that the assumptions in Proposition~\ref{prop:cylinder-to-vector-fixed-point-eqn} (i.e., cylinder assumption, bulk \textbf{A1}, full-boundary \textbf{A0}, and full-boundary \textbf{A1}), together with the extra assumption that the bulk is purely chiral, leads to the following equation: 
\be \label{eq:tempabc}
\( \hat \Delta(A_\up,B,C_\up) -  \Delta(A_\up,B,C_\up) \) \ket\Psi 
 \veq \int dx f_{A,B,C}(x)  \hhh(x)\ket\Omega \ee 
with $A_\up \equiv AA', C_\up \equiv CC'$ as in the figure in \eqref{eq:chiral-fixed-point-equation}.
Here $\hhh(x)$ is the Hamiltonian density for the edge CFT and 
$f_{A,B,C}(x)$ is the $C^1$ bump function implied by Hypothesis \ref{assump:chiral-CFT}. This is an equation that we use throughout the body of results that follow from Hypothesis~\ref{assump:chiral-CFT}. Thus, our argument below can be yet another justification for these derivations that do not rely on Hypothesis~\ref{assump:chiral-CFT}.

The argument for Eq.~\eqref{eq:tempabc} follows a similar chain of reasoning as in Proposition~\ref{prop:cylinder-to-vector-fixed-point-eqn}. Consider again the regions $ABC$ in Fig.~\ref{fig:cylinder_argument}.  First, we have \eqref{eq:part-two}. On the other hand, the cylinder assumption implies that 
\be 
\( \hat \Delta(A,B,C) - \Delta(A,B,C) \) \ket\Psi 
\veq \int dx f_{A,B,C}(x) \( \hhh_L(x) + \hhh_R(x) \) \ket{\Omega} \ee
where $f_{A,B,C}(x)$ is the $C^1$ bump function implied by Hypothesis~\ref{assump:chiral-CFT}, and $\hhh_L$ and $\hhh_R$ are 
the right-moving and left-moving parts of the CFT Hamiltonian.  In the case where the CFT is purely chiral, these contributions are localized at the two ends of the cylinder.  
That is, full-boundary A0 implies that 
the two terms on the RHS of \eqref{eq:part-two} are independent, and therefore
\begin{align}
\( \hat \Delta(A_\up^L,B^L,C_\up^L) -\Delta(A_\up^L,B^L,C_\up^L) \)  \ket\Psi 
& \veq \int dx f_{A,B,C}(x)  \hhh_L(x)\ket\Omega , ~~~ \nonumber
\\ 
\(\hat \Delta(A_\up^R,B^R,C_\up^R) -\Delta(A_\up^R,B^R,C_\up^R) \)\ket\Psi 
& \veq \int dx f_{A,B,C}(x)  \hhh_R(x)\ket\Omega .
\end{align}
%==================================
% end of sec:cylinder-argument
%==================================

\section{Fidelity near the edge under good modular flow}
\label{app:fidelity}

\begin{figure}[!h]
    \centering
    \begin{tikzpicture}
    \draw[thick,fill=green!20] (0,0) circle (3 cm);
    \draw (-2.75 cm,0) node {$C$};
    \draw[thick,fill=blue!20] (0,0) circle (2.5 cm);
    \draw (-2.25 cm,0) node {$B$};
    \draw[thick,fill=red!20] (0,0) circle (2 cm);
    \draw (0,0) node {$A$};
\end{tikzpicture}
    \caption{The partition used in the proof of Eq.~\eqref{eq:edge-Fid}. Here we do not assume the topology of $A$, which could have holes that are not shown. }
    \label{fig:ABC-for-fid}
\end{figure}
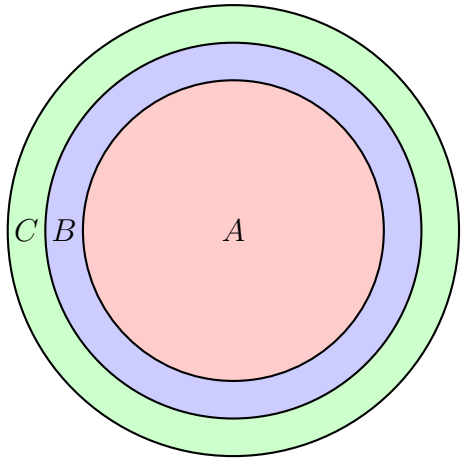

This appendix explains the relation $F(\rho_{\Omega}(0),\rho_{\Omega}(t)) = F(\rho_{\text{edge}}(0), \rho_{\text{edge}}(t))$ based on the full-boundary {\bf A0} condition. This result is used in Section~\ref{subsec:state-difference}.
Consider a good modular flow generated by $\Lb$, intersecting with only one gapless edge\footnote{If a good modular flow generator covers more than one edge, then one can first decompose the generator and then deform it to all the edges, similar as we did in the cylinder argument in \ref{appendix:cylinder-argument}.}. We partition the entire system $\Omega$ into $ABC$ shown in Figure~\ref{fig:ABC-for-fid}. We let $C$ be the annulus that covers the particular gapless edge on which $\Lb$ has nontrivial action, and $B$ be another annulus that surrounds $C$ and separates $C$ from the remaining region $A$\footnote{We do not make any further assumption about the topology of $A$}. Without loss of generality, one can think the support of $\Lb$ is within $C$, as its support can be deformed in the bulk such that it is localized near the edge [Section~\ref{sec:good-modular-flows}]. We claim that 
\begin{equation}\label{eq:edge-Fid}
    F(\rho_{ABC}(0),\rho_{ABC}(t)) = F(\rho_{BC}(0),\rho_{BC}(t)).
\end{equation}

This is plausible because the good modular flow does not change the state in the bulk $AB$. 

The proof of Eq.~\eqref{eq:edge-Fid} follows from a well-known fact: that fidelity between any two states $\rho$ and $\sigma$ is non-increasing under any quantum channel $\calE$, i.e., $F(\rho,\sigma)\leq F(\calE(\rho),\calE(\sigma))$. First, note that $F(\rho_{ABC}(0),\rho_{ABC}(t))\leq F(\rho_{BC}(0),\rho_{BC}(t))$ by choosing the quantum channel to be the partial trace over $A$. On the other hand, we have $ F(\rho_{BC}(0),\rho_{BC}(t))\leq F(\rho_{ABC}(0),\rho_{ABC}(t))$. The reason is as follows: Firstly, due to full-boundary {\bf A0}, 
\begin{equation}
    0\leq I(A:C|B)_{\ket{\Psi}}\leq \Delta(B,C,\emptyset)_{\ket{\Psi}}=0 \quad \Rightarrow\quad I(A:C|B)_{\ket{\Psi}}=0,
\end{equation}
where the first two ``$\leq$'' signs are due to strong subadditivity~\cite{Lieb1973}. Secondly, under the good modular flow $\ket{\Psi(t)} = e^{i\Lb t}\ket{\Psi}$ still satisfies the Markov condition $I(A:C|B)_{\ket{\Psi(t)}}$. To see this, notice $\ket{\Psi(t)}$ still satisfies the full-boundary {\bf A0}: 
\begin{equation}
    \Delta(B,C,\emptyset)_{\ket{\Psi(t)}} = (S_A + S_{AB}-S_{B})_{\ket{\Psi(t)}} = (S_A + S_{AB}-S_{B})_{\ket{\Psi}}=0, 
\end{equation}
where the first equal sign is due to the purity of $\ket{\Psi(t)}$ and the second equal sign is due to the support of $\Lb$ is within $C$ and as a result $\rho_{AB}(0) = \rho_{AB}(t)$. As a result of Markov structure of $\rho_{ABC}(0)$ and $\rho_{ABC}(t)$, one can use Petz recovery map $\calE_{B\to AB}^{\rho_{AB}(0)}$\footnote{The same channel for recovery because $\rho_{AB}(0) = \rho_{AB}(t)$.} \cite{Petz2003} to recover both $\rho_{ABC}(t)$ from $\rho_{BC}(t)$ for any $t \in \mathbb{R}$. Therefore, we can conclude $F(\rho_{ABC}(0),\rho_{ABC}(t))= F(\rho_{BC}(0),\rho_{BC}(t))$. 

\phantomsection\addcontentsline{toc}{section}{References}
\bibliographystyle{ucsd}
\bibliography{revised-ref} 

\begingroup\raggedright\begin{thebibliography}{10}

\bibitem{Chamon:2004lew}
C.~Chamon, ``{Quantum Glassiness},'' {\em Phys. Rev. Lett.} {\bf 94} (2005),
  no.~4 040402, \href{http://arxiv.org/abs/cond-mat/0404182}{{\tt
  cond-mat/0404182}}.

\bibitem{Haah2011}
J.~Haah, ``Local stabilizer codes in three dimensions without string logical
  operators,'' {\em Phys. Rev. A} {\bf 83} (Apr, 2011) 042330,
  \href{https://link.aps.org/doi/10.1103/PhysRevA.83.042330}{https://link.aps.org/doi/10.1103/PhysRevA.83.042330}.

\bibitem{nandkishore2019fractons}
R.~M. Nandkishore and M.~Hermele, ``Fractons,'' {\em Annual Review of Condensed
  Matter Physics} {\bf 10} (2019) 295--313,
  \href{http://arxiv.org/abs/1803.11196}{{\tt 1803.11196}}.

\bibitem{pretko2020fracton}
M.~Pretko, X.~Chen, and Y.~You, ``Fracton phases of matter,'' {\em
  International Journal of Modern Physics A} {\bf 35} (2020), no.~06 2030003,
  \href{http://arxiv.org/abs/2001.01722}{{\tt 2001.01722}}.

\bibitem{Witten1988}
E.~Witten, ``{Topological quantum field theory},'' {\em Communications in
  Mathematical Physics} {\bf 117} (1988), no.~3 353 -- 386.

\bibitem{Wen1990}
X.~G. Wen and Q.~Niu, ``Ground-state degeneracy of the fractional quantum Hall
  states in the presence of a random potential and on high-genus Riemann
  surfaces,'' {\em Phys. Rev. B} {\bf 41} (May, 1990) 9377--9396,
  \href{https://link.aps.org/doi/10.1103/PhysRevB.41.9377}{https://link.aps.org/doi/10.1103/PhysRevB.41.9377}.

\bibitem{Wilczek1982}
F.~Wilczek, ``Quantum Mechanics of Fractional-Spin Particles,'' {\em Phys. Rev.
  Lett.} {\bf 49} (Oct, 1982) 957--959,
  \href{https://link.aps.org/doi/10.1103/PhysRevLett.49.957}{https://link.aps.org/doi/10.1103/PhysRevLett.49.957}.

\bibitem{Kitaev2006}
A.~{Kitaev} and J.~{Preskill}, ``{Topological Entanglement Entropy},'' {\em
  \prl} {\bf 96} (Mar., 2006) 110404,
  \href{http://arxiv.org/abs/hep-th/0510092}{{\tt hep-th/0510092}}.

\bibitem{Levin2006}
M.~{Levin} and X.-G. {Wen}, ``{Detecting Topological Order in a Ground State
  Wave Function},'' {\em \prl} {\bf 96} (Mar., 2006) 110405,
  \href{http://arxiv.org/abs/cond-mat/0510613}{{\tt cond-mat/0510613}}.

\bibitem{Calabrese:2009qy}
P.~Calabrese and J.~Cardy, ``{Entanglement entropy and conformal field
  theory},'' {\em J. Phys. A} {\bf 42} (2009) 504005,
  \href{http://arxiv.org/abs/0905.4013}{{\tt 0905.4013}}.

\bibitem{Ryu2006}
S.~Ryu and T.~Takayanagi, ``Holographic Derivation of Entanglement Entropy from
  the anti--de Sitter Space/Conformal Field Theory Correspondence,'' {\em Phys.
  Rev. Lett.} {\bf 96} (May, 2006) 181602,
  \href{https://link.aps.org/doi/10.1103/PhysRevLett.96.181602}{https://link.aps.org/doi/10.1103/PhysRevLett.96.181602}.

\bibitem{shi2020fusion}
B.~Shi, K.~Kato, and I.~H. Kim, ``Fusion rules from entanglement,'' {\em Annals
  of Physics} {\bf 418} (2020) 168164,
  \href{http://arxiv.org/abs/1906.09376}{{\tt 1906.09376}}.

\bibitem{Shi:2019ngw}
B.~Shi, ``{Verlinde formula from entanglement},'' {\em Phys. Rev. Res.} {\bf 2}
  (2020), no.~2 023132, \href{http://arxiv.org/abs/1911.01470}{{\tt
  1911.01470}}.

\bibitem{Shi:2020rne}
B.~Shi and I.~H. Kim, ``{Entanglement bootstrap approach for gapped domain
  walls},'' {\em Phys. Rev. B} {\bf 103} (2021), no.~11 115150,
  \href{http://arxiv.org/abs/2008.11793}{{\tt 2008.11793}}.

\bibitem{knots-paper}
J.-L. Huang, J.~McGreevy, and B.~Shi, ``{Knots and entanglement},'' {\em
  SciPost Phys.} {\bf 14} (2023) 141,
  \href{https://scipost.org/10.21468/SciPostPhys.14.6.141}{https://scipost.org/10.21468/SciPostPhys.14.6.141}.

\bibitem{Shi:2023kwr}
B.~Shi, J.-L. Huang, and J.~McGreevy, ``{Remote detectability from entanglement
  bootstrap I: Kirby's torus trick},''
  \href{http://arxiv.org/abs/2301.07119}{{\tt 2301.07119}}.

\bibitem{Kim2021}
I.~H. {Kim}, B.~{Shi}, K.~{Kato}, and V.~V. {Albert}, ``{Chiral Central Charge
  from a Single Bulk Wave Function},'' {\em \prl} {\bf 128} (Apr., 2022)
  176402, \href{http://arxiv.org/abs/2110.06932}{{\tt 2110.06932}}.

\bibitem{Fan:2022ipl}
R.~Fan, R.~Sahay, and A.~Vishwanath, ``Extracting the quantum hall conductance
  from a single bulk wave function,'' {\em Phys. Rev. Lett.} {\bf 131} (2023)
  186301, \href{http://arxiv.org/abs/2208.11710}{{\tt 2208.11710}}.

\bibitem{Hastings2005-decay-cor}
M.~B. {Hastings} and T.~{Koma}, ``{Spectral Gap and Exponential Decay of
  Correlations},'' {\em Communications in Mathematical Physics} {\bf 265}
  (Aug., 2006) 781--804, \href{http://arxiv.org/abs/math-ph/0507008}{{\tt
  math-ph/0507008}}.

\bibitem{Holzhey:1994we}
C.~Holzhey, F.~Larsen, and F.~Wilczek, ``{Geometric and renormalized entropy in
  conformal field theory},'' {\em Nucl. Phys. B} {\bf 424} (1994) 443--467,
  \href{http://arxiv.org/abs/hep-th/9403108}{{\tt hep-th/9403108}}.

\bibitem{Myers:2010tj}
R.~C. Myers and A.~Sinha, ``{Holographic c-theorems in arbitrary dimensions},''
  {\em JHEP} {\bf 01} (2011) 125, \href{http://arxiv.org/abs/1011.5819}{{\tt
  1011.5819}}.

\bibitem{Koo:1993wz}
W.~M. Koo and H.~Saleur, ``{Representations of the Virasoro algebra from
  lattice models},'' {\em Nucl. Phys. B} {\bf 426} (1994) 459--504,
  \href{http://arxiv.org/abs/hep-th/9312156}{{\tt hep-th/9312156}}.

\bibitem{Levin2013}
M.~{Levin}, ``{Protected Edge Modes without Symmetry},'' {\em Physical Review
  X} {\bf 3} (Apr., 2013) 021009, \href{http://arxiv.org/abs/1301.7355}{{\tt
  1301.7355}}.

\bibitem{Chatterjee2022}
A.~{Chatterjee} and X.-G. {Wen}, ``{Holographic theory for continuous phase
  transitions: Emergence and symmetry protection of gaplessness},'' {\em \prb}
  {\bf 108} (Aug., 2023) 075105, \href{http://arxiv.org/abs/2205.06244}{{\tt
  2205.06244}}.

\bibitem{Laughlin1983}
R.~B. Laughlin, ``Anomalous Quantum Hall Effect: An Incompressible Quantum
  Fluid with Fractionally Charged Excitations,'' {\em Phys. Rev. Lett.} {\bf
  50} (May, 1983) 1395--1398,
  \href{https://link.aps.org/doi/10.1103/PhysRevLett.50.1395}{https://link.aps.org/doi/10.1103/PhysRevLett.50.1395}.

\bibitem{Kitaev2005}
A.~{Kitaev}, ``{Anyons in an exactly solved model and beyond},'' {\em Annals of
  Physics} {\bf 321} (Jan., 2006) 2--111,
  \href{http://arxiv.org/abs/cond-mat/0506438}{{\tt cond-mat/0506438}}.

\bibitem{Lin2023}
T.-C. {Lin} and J.~{McGreevy}, ``{Conformal Field Theory Ground States as
  Critical Points of an Entropy Function},'' {\em Phys. Rev. Lett.} {\bf 131}
  (2023) 251602, \href{http://arxiv.org/abs/2303.05444}{{\tt 2303.05444}}.

\bibitem{Hu:2020suv}
Q.~Hu, A.~Franco-Rubio, and G.~Vidal, ``{Emergent universality in critical
  quantum spin chains: entanglement Virasoro algebra},''
  \href{http://arxiv.org/abs/2009.11383}{{\tt 2009.11383}}.

\bibitem{Osborne:2021ppp}
T.~J. Osborne and A.~Stottmeister, ``{Conformal Field Theory from Lattice
  Fermions},'' {\em Commun. Math. Phys.} {\bf 398} (2023), no.~1 219--289,
  \href{http://arxiv.org/abs/2107.13834}{{\tt 2107.13834}}.

\bibitem{Wang:2022qxf}
R.~Wang, X.~Zeng, C.~Shen, and L.-Y. Hung, ``{Virasoro and Kac-Moody algebra in
  generic tensor network representations of two-dimensional critical lattice
  partition functions},'' {\em Phys. Rev. B} {\bf 106} (2022), no.~11 115116,
  \href{http://arxiv.org/abs/2205.04500}{{\tt 2205.04500}}.

\bibitem{Zeng:2022swq}
X.~Zeng, R.~Wang, C.~Shen, and L.-Y. Hung, ``{Virasoro generators in the
  Fibonacci model tensor network: Tackling finite-size effects},'' {\em Phys.
  Rev. B} {\bf 107} (2023), no.~24 245146,
  \href{http://arxiv.org/abs/2212.02937}{{\tt 2212.02937}}.

\bibitem{Casini2008}
H.~{Casini}, ``{Relative entropy and the Bekenstein bound},'' {\em Classical
  and Quantum Gravity} {\bf 25} (Oct., 2008) 205021,
  \href{http://arxiv.org/abs/0804.2182}{{\tt 0804.2182}}.

\bibitem{Li2008}
H.~{Li} and F.~D.~M. {Haldane}, ``{Entanglement Spectrum as a Generalization of
  Entanglement Entropy: Identification of Topological Order in Non-Abelian
  Fractional Quantum Hall Effect States},'' {\em \prl} {\bf 101} (July, 2008)
  010504, \href{http://arxiv.org/abs/0805.0332}{{\tt 0805.0332}}.

\bibitem{Kim:2021tse}
I.~H. Kim, B.~Shi, K.~Kato, and V.~V. Albert, ``{Modular commutator in gapped
  quantum many-body systems},'' {\em Phys. Rev. B} {\bf 106} (2022), no.~7
  075147, \href{http://arxiv.org/abs/2110.10400}{{\tt 2110.10400}}.

\bibitem{Lin2022-op}
T.-C. {Lin}, I.~H. {Kim}, and M.-H. {Hsieh}, ``{A new operator extension of
  strong subadditivity of quantum entropy},'' {\em Letters in Mathematical
  Physics} {\bf 113} (June, 2023) 68,
  \href{http://arxiv.org/abs/2211.13372}{{\tt 2211.13372}}.

\bibitem{Lieb1973}
E.~H. Lieb and M.~B. Ruskai, ``{Proof of the strong subadditivity of
  quantum‐mechanical entropy},'' {\em Journal of Mathematical Physics} {\bf
  14} (11, 1973) 1938--1941,
  \href{http://arxiv.org/abs/https://pubs.aip.org/aip/jmp/article-pdf/14/12/1938/8146113/1938\_1\_online.pdf}{{\tt
  https://pubs.aip.org/aip/jmp/article-pdf/14/12/1938/8146113/1938\_1\_online.pdf}},
  \href{https://doi.org/10.1063/1.1666274}{https://doi.org/10.1063/1.1666274}.

\bibitem{Fidelity-2008}
P.~E.~M.~F. {Mendon{\c{c}}a}, R.~D.~J. {Napolitano}, M.~A. {Marchiolli}, C.~J.
  {Foster}, and Y.-C. {Liang}, ``{Alternative fidelity measure between quantum
  states},'' {\em \pra} {\bf 78} (Nov., 2008) 052330,
  \href{http://arxiv.org/abs/0806.1150}{{\tt 0806.1150}}.

\bibitem{Bures1969}
D.~Bures, ``An Extension of Kakutani's Theorem on Infinite Product Measures to
  the Tensor Product of Semifinite w*-Algebras,'' {\em Transactions of the
  American Mathematical Society} {\bf 135} (1969) 199--212,
  \href{http://www.jstor.org/stable/1995012}{http://www.jstor.org/stable/1995012}.

\bibitem{Cardy:2016fqc}
J.~Cardy and E.~Tonni, ``{Entanglement hamiltonians in two-dimensional
  conformal field theory},'' {\em J. Stat. Mech.} {\bf 1612} (2016), no.~12
  123103, \href{http://arxiv.org/abs/1608.01283}{{\tt 1608.01283}}.

\bibitem{Casini2009}
H.~Casini and M.~Huerta, ``{Reduced density matrix and internal dynamics for
  multicomponent regions},'' {\em Class. Quant. Grav.} {\bf 26} (2009) 185005,
  \href{http://arxiv.org/abs/0903.5284}{{\tt 0903.5284}}.

\bibitem{Klich2017}
I.~Klich, D.~Vaman, and G.~Wong, ``Entanglement Hamiltonians for Chiral
  Fermions with Zero Modes,'' {\em Phys. Rev. Lett.} {\bf 119} (Sep, 2017)
  120401,
  \href{https://link.aps.org/doi/10.1103/PhysRevLett.119.120401}{https://link.aps.org/doi/10.1103/PhysRevLett.119.120401}.

\bibitem{Arias2018}
R.~E. Arias, H.~Casini, M.~Huerta, and D.~Pontello, ``Entropy and modular
  Hamiltonian for a free chiral scalar in two intervals,'' {\em Phys. Rev. D}
  {\bf 98} (Dec, 2018) 125008,
  \href{https://link.aps.org/doi/10.1103/PhysRevD.98.125008}{https://link.aps.org/doi/10.1103/PhysRevD.98.125008}.

\bibitem{Halperin1982}
B.~I. Halperin, ``Quantized Hall conductance, current-carrying edge states, and
  the existence of extended states in a two-dimensional disordered potential,''
  {\em Phys. Rev. B} {\bf 25} (Feb, 1982) 2185--2190,
  \href{https://link.aps.org/doi/10.1103/PhysRevB.25.2185}{https://link.aps.org/doi/10.1103/PhysRevB.25.2185}.

\bibitem{Witten:1988hf}
E.~Witten, ``{Quantum Field Theory and the Jones Polynomial},'' {\em Commun.
  Math. Phys.} {\bf 121} (1989) 351--399.

\bibitem{Wen1991}
X.~G. Wen, ``Gapless boundary excitations in the quantum Hall states and in the
  chiral spin states,'' {\em Phys. Rev. B} {\bf 43} (May, 1991) 11025--11036,
  \href{https://link.aps.org/doi/10.1103/PhysRevB.43.11025}{https://link.aps.org/doi/10.1103/PhysRevB.43.11025}.

\bibitem{Hellerman:2021fla}
S.~Hellerman, D.~Orlando, and M.~Watanabe, ``{Quantum Information Theory of the
  Gravitational Anomaly},'' \href{http://arxiv.org/abs/2101.03320}{{\tt
  2101.03320}}.

\bibitem{Callan:1984sa}
C.~G. Callan, Jr. and J.~A. Harvey, ``{Anomalies and Fermion Zero Modes on
  Strings and Domain Walls},'' {\em Nucl. Phys. B} {\bf 250} (1985) 427--436.

\bibitem{Stinespring_1955}
W.~F. Stinespring, ``Positive Functions on C*-Algebras,'' {\em Proceedings of
  the American Mathematical Society} {\bf 6} (1955), no.~2 211--216,
  \href{http://www.jstor.org/stable/2032342}{http://www.jstor.org/stable/2032342}.

\bibitem{Uhlmann1976}
A.~{Uhlmann}, ``{The ``transition probability'' in the state space of a
  {\ensuremath{*}}-algebra},'' {\em Reports on Mathematical Physics} {\bf 9}
  (Apr., 1976) 273--279.

\bibitem{q-cross-ratio}
I.~H. {Kim}, X.~{Li}, T.-C. {Lin}, J.~{McGreevy}, and B.~{Shi}, ``{Conformal
  geometry from entanglement},'' {\em arXiv e-prints} (Apr., 2024)
  arXiv:2404.03725, \href{http://arxiv.org/abs/2404.03725}{{\tt 2404.03725}}.

\bibitem{Bombin2010}
H.~Bombin, ``Topological Order with a Twist: Ising Anyons from an Abelian
  Model,'' {\em Phys. Rev. Lett.} {\bf 105} (Jul, 2010) 030403,
  \href{https://link.aps.org/doi/10.1103/PhysRevLett.105.030403}{https://link.aps.org/doi/10.1103/PhysRevLett.105.030403}.

\bibitem{Read1999}
N.~{Read} and D.~{Green}, ``{Paired states of fermions in two dimensions with
  breaking of parity and time-reversal symmetries and the fractional quantum
  Hall effect},'' {\em \prb} {\bf 61} (Apr., 2000) 10267--10297,
  \href{http://arxiv.org/abs/cond-mat/9906453}{{\tt cond-mat/9906453}}.

\bibitem{Brown2013}
B.~J. Brown, S.~D. Bartlett, A.~C. Doherty, and S.~D. Barrett, ``Topological
  Entanglement Entropy with a Twist,'' {\em Phys. Rev. Lett.} {\bf 111} (Nov,
  2013) 220402,
  \href{https://link.aps.org/doi/10.1103/PhysRevLett.111.220402}{https://link.aps.org/doi/10.1103/PhysRevLett.111.220402}.

\bibitem{Kim2014-1D}
I.~H. {Kim}, ``{Entropic topological invariant for a gapped one-dimensional
  system},'' {\em \prb} {\bf 89} (June, 2014) 235120,
  \href{http://arxiv.org/abs/1306.4771}{{\tt 1306.4771}}.

\bibitem{strength-of-FPE}
in~progress (2024).

\bibitem{francesco2012conformal}
P.~Francesco, P.~Mathieu, and D.~S{\'e}n{\'e}chal, {\em Conformal field
  theory}.
\newblock Springer Science \& Business Media, 2012.

\bibitem{Hollands2019}
S.~{Hollands}, ``{Relative entropy for coherent states in chiral CFT},'' {\em
  Letters in Mathematical Physics} {\bf 110} (Dec., 2019) 713--733,
  \href{http://arxiv.org/abs/1903.07508}{{\tt 1903.07508}}.

\bibitem{Stanford:2017thb}
D.~Stanford and E.~Witten, ``{Fermionic Localization of the Schwarzian
  Theory},'' {\em JHEP} {\bf 10} (2017) 008,
  \href{http://arxiv.org/abs/1703.04612}{{\tt 1703.04612}}.

\bibitem{Zou2022}
Y.~{Zou}, B.~{Shi}, J.~{Sorce}, I.~T. {Lim}, and I.~H. {Kim}, ``{Modular
  Commutators in Conformal Field Theory},'' {\em \prl} {\bf 129} (Dec., 2022)
  260402, \href{http://arxiv.org/abs/2206.00027}{{\tt 2206.00027}}.

\bibitem{Eisler2009}
I.~{Peschel} and V.~{Eisler}, ``{Reduced density matrices and entanglement
  entropy in free lattice models},'' {\em Journal of Physics A Mathematical
  General} {\bf 42} (Dec., 2009) 504003,
  \href{http://arxiv.org/abs/0906.1663}{{\tt 0906.1663}}.

\bibitem{Wolf:2007tdq}
M.~M. Wolf, F.~Verstraete, M.~B. Hastings, and J.~I. Cirac, ``{Area Laws in
  Quantum Systems: Mutual Information and Correlations},'' {\em Phys. Rev.
  Lett.} {\bf 100} (2008), no.~7 070502,
  \href{http://arxiv.org/abs/0704.3906}{{\tt 0704.3906}}.

\bibitem{Nielsen2012}
A.~E.~B. {Nielsen}, J.~I. {Cirac}, and G.~{Sierra}, ``{Laughlin Spin-Liquid
  States on Lattices Obtained from Conformal Field Theory},'' {\em \prl} {\bf
  108} (June, 2012) 257206, \href{http://arxiv.org/abs/1201.3096}{{\tt
  1201.3096}}.

\bibitem{2018NuPhB.927..140K}
L.~{Kong} and H.~{Zheng}, ``{Gapless edges of 2d topological orders and
  enriched monoidal categories},'' {\em Nuclear Physics B} {\bf 927} (Feb.,
  2018) 140--165, \href{http://arxiv.org/abs/1705.01087}{{\tt 1705.01087}}.

\bibitem{Kong:2019byq}
L.~Kong and H.~Zheng, ``{A mathematical theory of gapless edges of 2d
  topological orders. Part I},'' {\em JHEP} {\bf 02} (2020) 150,
  \href{http://arxiv.org/abs/1905.04924}{{\tt 1905.04924}}.

\bibitem{Kong:2019cuu}
L.~Kong and H.~Zheng, ``{A mathematical theory of gapless edges of 2d
  topological orders. Part II},'' {\em Nucl. Phys. B} {\bf 966} (2021) 115384,
  \href{http://arxiv.org/abs/1912.01760}{{\tt 1912.01760}}.

\bibitem{saminadayar1997observation}
L.~Saminadayar, D.~Glattli, Y.~Jin, and B.~c.-m. Etienne, ``Observation of the
  e/3 fractionally charged Laughlin quasiparticle,'' {\em Physical Review
  Letters} {\bf 79} (1997), no.~13 2526,
  \href{http://arxiv.org/abs/cond-mat/9706307}{{\tt cond-mat/9706307}}.

\bibitem{zhang2009distinct}
Y.~Zhang, D.~T. McClure, E.~M. Levenson-Falk, C.~M. Marcus, L.~N. Pfeiffer, and
  K.~W. West, ``Distinct signatures for Coulomb blockade and Aharonov-Bohm
  interference in electronic Fabry-Perot interferometers,'' {\em Physical
  Review B} {\bf 79} (2009), no.~24 241304,
  \href{http://arxiv.org/abs/0901.0127}{{\tt 0901.0127}}.

\bibitem{ofek2010role}
N.~Ofek, A.~Bid, M.~Heiblum, A.~Stern, V.~Umansky, and D.~Mahalu, ``Role of
  interactions in an electronic Fabry--Perot interferometer operating in the
  quantum Hall effect regime,'' {\em Proceedings of the National Academy of
  Sciences} {\bf 107} (2010), no.~12 5276--5281,
  \href{http://arxiv.org/abs/0911.0794}{{\tt 0911.0794}}.

\bibitem{nakamura2020direct}
J.~Nakamura, S.~Liang, G.~C. Gardner, and M.~J. Manfra, ``Direct observation of
  anyonic braiding statistics,'' {\em Nature Physics} {\bf 16} (2020), no.~9
  931--936, \href{http://arxiv.org/abs/2006.14115}{{\tt 2006.14115}}.

\bibitem{Bartolomei2020}
H.~{Bartolomei}, M.~{Kumar}, R.~{Bisognin}, A.~{Marguerite}, J.~M. {Berroir},
  E.~{Bocquillon}, B.~{Pla{\c{c}}ais}, A.~{Cavanna}, Q.~{Dong}, U.~{Gennser},
  Y.~{Jin}, and G.~{F{\`e}ve}, ``{Fractional statistics in anyon collisions},''
  {\em Science} {\bf 368} (Apr., 2020) 173--177,
  \href{http://arxiv.org/abs/2006.13157}{{\tt 2006.13157}}.

\bibitem{2023arXiv230711054S}
J.~{Schirmann}, S.~{Franca}, F.~{Flicker}, and A.~G. {Grushin}, ``{Physical
  properties of the Hat aperiodic monotile: Graphene-like features, chirality
  and zero-modes},'' {\em arXiv e-prints} (July, 2023) arXiv:2307.11054,
  \href{http://arxiv.org/abs/2307.11054}{{\tt 2307.11054}}.

\bibitem{veillon2024observation}
A.~Veillon, C.~Piquard, P.~Glidic, Y.~Sato, A.~Aassime, A.~Cavanna, Y.~Jin,
  U.~Gennser, A.~Anthore, and F.~Pierre, ``Observation of the scaling dimension
  of fractional quantum Hall anyons,''
  \href{http://arxiv.org/abs/2401.18044}{{\tt 2401.18044}}.

\bibitem{Kalmeyer1987}
V.~Kalmeyer and R.~Laughlin, ``Equivalence of the resonating-valence-bond and
  fractional quantum Hall states,'' {\em Physical review letters} {\bf 59}
  (1987), no.~18 2095.

\bibitem{Petz2003}
D.~{Petz}, ``{Monotonicity of Quantum Relative Entropy Revisited},'' {\em Rev.
  Math. Phys.} {\bf 15} (2003) 79--91,
  \href{http://arxiv.org/abs/quant-ph/0209053}{{\tt quant-ph/0209053}}.

\end{thebibliography}\endgroup
\end{document}